\RequirePackage{fix-cm}
\documentclass[pdflatex,sn-nature,iicol]{sn-jnl}

\usepackage{graphicx}
\usepackage{amsmath,amssymb,amsfonts}
\usepackage{amsthm}
\usepackage{upgreek}
\usepackage{physics}
\usepackage{bm}
\usepackage{xcolor}
\usepackage{booktabs}
\usepackage{microtype}
\usepackage{placeins}
\usepackage{float}
\usepackage{tikz}
\usepackage{cleveref}
\hypersetup{hypertexnames=false,allcolors=black,bookmarksdepth=3}

\renewcommand{\addressfont}{\normalfont\fontsize{9bp}{11bp}\selectfont\raggedcenter}

\newcommand{\eqlab}[1]{\label{eq:#1}}
\renewcommand{\eqref}[1]{Eq.~(\ref{eq:#1})}

\newcommand{\figref}[1]{Fig.~\ref{fig:#1}}

\newcommand{\figlab}[1]{\label{fig:#1}}

\theoremstyle{plain}
\newtheorem{theorem}{Theorem}

\unnumbered
\begin{document}

\title[Multiport nanocavity photonic routing]{\centering\fontsize{20bp}{25bp}\selectfont
  Multiport photonic routing with a single\\
  cavity-based electro-optic phase shifter per port}
\author*[1]{\fnm{Alessandro} \sur{Buzzi}}\email{abuzzi@mit.edu}

\author[1,2]{\fnm{Bjarke A.} \sur{Frederiksen}}

\author*[1,3]{\fnm{Dashiell L.\,P.} \sur{Vitullo}}\email{dashiell.l.vitullo.civ@army.mil}

\author*[2]{\fnm{Mikkel} \sur{Heuck}}\email{mheu@dtu.dk}\equalcont{These authors jointly supervised this work.}

\author*[1]{\fnm{Dirk R.} \sur{Englund}}\email{englund@mit.edu}\equalcont{These authors jointly supervised this work.}

\affil[1]{\orgdiv{Research Laboratory of Electronics},
           \orgname{Massachusetts Institute of Technology},
           \orgaddress{\city{Cambridge}, \state{MA}, \postcode{02139}, \country{USA}}}

\affil[2]{\orgdiv{Department of Electrical and Photonics Engineering},
           \orgname{Technical University of Denmark},
           \orgaddress{\city{Kgs.\ Lyngby}, \postcode{2800}, \country{Denmark}}}

\affil[3]{\orgname{DEVCOM Army Research Laboratory},
           \orgaddress{\city{Adelphi}, \state{Maryland}, \postcode{20783}, \country{USA}}}

\abstract{\unboldmath{}Programmable photonic meshes can perform general linear transformations, but universal transformations on $N$ modes require $O(N^2)$ programmable phase shifters, which cannot simultaneously achieve low drive voltage, gigahertz speed, zero static power, lossless modulation, and a compact footprint. Signal routing does not require universal transformations, which allows considerable circuit simplifications.
We present a compact reflective circuit architecture that connects all input-output pairs across $N$ modes in $N$ steps using only $O(N)$ programmable phase shifters. The circuit comprises a single \emph{passive} $N\times N$ interferometer and a single column of programmable reflective phase shifters.
We demonstrate the architecture by implementing a $3\times 3$ router in x-cut thin-film lithium niobate with a passive multimode interferometer and phase shifters based on single-sided photonic-crystal nanobeam cavities. Our programmable devices exhibit a $1.92$~GHz/V resonance tuning efficiency via the Pockels effect, a footprint of $100~\upmu$m$^2$, sub-5-V drive voltage, and a 6.2~GHz electro-optic bandwidth.
Through voltage-space characterization and physical modeling, we identify routing states with extinction ratios exceeding 11~dB at each output and demonstrate sequential routing at an equipment-limited speed of 10~MHz.
Our results establish cavity-enhanced Pockels switching as a routing layer for classical and
quantum computing interconnects across visible and near-infrared wavelengths.}

\keywords{thin-film lithium niobate, photonic crystal, optical switch, electro-optic modulator,
          Pockels effect, photonic integrated circuit, multimode interferometer}

\maketitle

\begin{figure*}[!t]
  \centering
  \includegraphics[width=\textwidth,height=0.9\textheight,keepaspectratio]{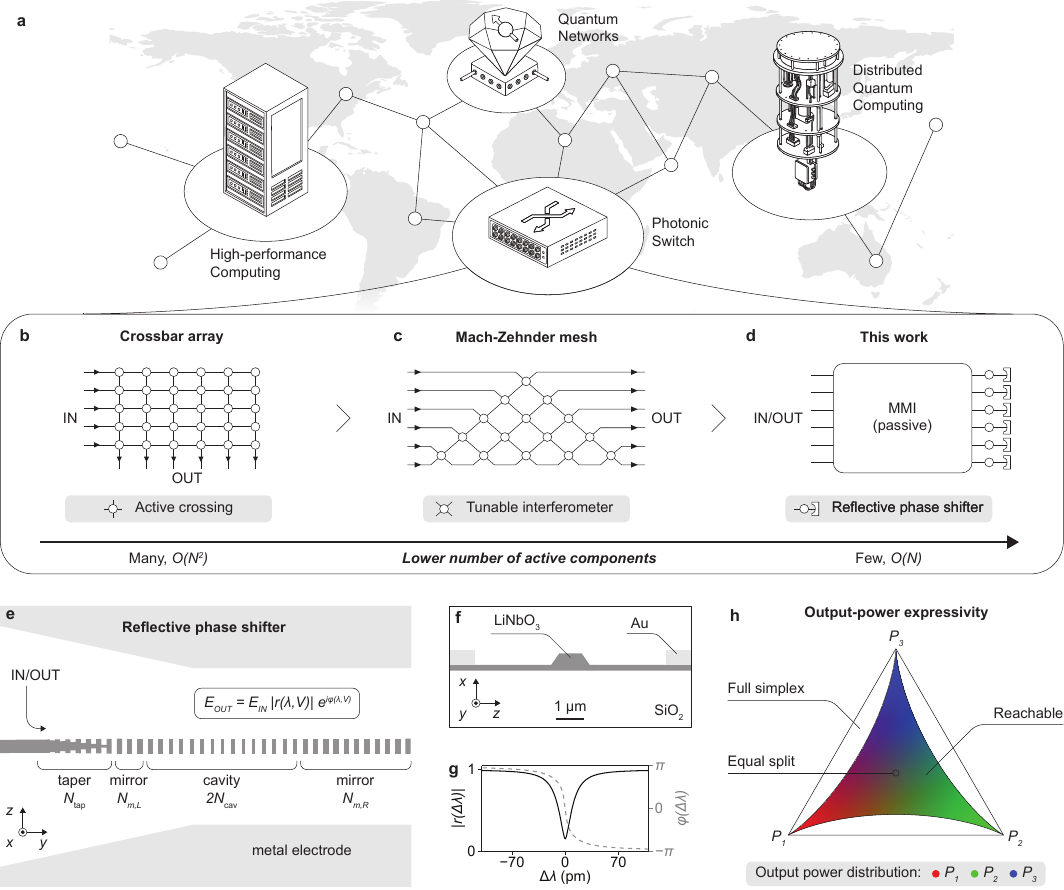}
  \caption{\textbf{Multiport photonic routing architecture and
  electro-optic cavity phase-shifter principle.}
    \textbf{a}, Network-level role of photonic switches in connecting
    high-performance computing systems, quantum networks, and distributed
    quantum processors.
    \textbf{b}, Crossbar switch with an active element at each
    input--output crossing, requiring $O(N^2)$ active elements.
    \textbf{c}, Universal Mach--Zehnder-interferometer mesh implementing
    arbitrary linear transformations with $N(N-1)/2$ tunable interferometers,
    scaling as $O(N^2)$.
    \textbf{d}, Pairwise-routing architecture demonstrated in this work: a
    passive $N\times N$ multimode interferometer (MMI) terminated by $N$
    independently controlled reflective phase shifters.
    \textbf{e}, Illustrative reflective phase-shifter design comprising a tapered
    photonic-crystal nanobeam cavity and an integrated metal electrode for
    voltage control of the reflected optical phase. The design shown differs
    from the cavities integrated in the $3\times3$ switch.
    \textbf{f}, Cross-section of the electro-optic cavity region showing
    the LiNbO$_3$ waveguide and adjacent gold electrode on an SiO$_2$ substrate.
    \textbf{g}, Calculated reflection amplitude (black) and phase (gray)
    versus wavelength detuning, with phase referenced to its resonant value.
    The linewidth uses the spectroscopic $Q_{\rm L}=5.07\times10^4$;
    $\kappa_i/\kappa_e=0.737$ is obtained from the reciprocal transfer-matrix
    fit described in Supplementary Note~4.
    \textbf{h}, Reachable output-power region of the ideal three-port
    architecture under phase-only control, shown on the $(P_1,P_2,P_3)$
    simplex; color indicates the relative power in each output port.}
  \label{fig:device}
\end{figure*}

The growing demand for computation is driving a shift toward increasingly
distributed architectures, from AI accelerator clusters to modular quantum
processors. In these architectures, communication between nodes is increasingly
constrained by the bandwidth, routing capacity, and energy cost of the
interconnect fabric~\cite{Miller2017,Shastri2021,Bogaerts2020}. AI systems
require high-throughput optical links between accelerator nodes, whereas distributed quantum processors require coherent photonic links between remote
quantum registers, with platform-specific transduction at the network
endpoints~\cite{Kimble2008,Wehner2018,Pompili2021}. Reconfigurable optical
switches address both requirements by dynamically allocating paths within the photonic network without
optical-to-electronic conversion (Fig.~\ref{fig:device}a).

Integrated photonic switches offer a chip-scale, low-latency, energy-efficient
path to reconfigurable optical routing~\cite{Chen2023}. High speed, low optical
loss, and wide wavelength coverage across the visible and near-infrared spectrum
make thin-film lithium niobate (TFLN) an ideal material platform for quantum technology,
optical interconnects, and biomedical devices~\cite{Zhu2021,Boes2023}. However, the scale and control of integrated switches are fundamentally limited by the size and number of
programmable components required~\cite{Bogaerts2020}. Crossbar arrays require
$N^2$ active crossings, while universal Mach--Zehnder
interferometer (MZI) meshes implementing arbitrary $N\times N$ linear
transformations require $N(N-1)/2$ tunable interferometers, corresponding, together with the output-phase layer, to $O(N^2)$ programmable phase controls, all scaling
quadratically with port count (Fig.~\ref{fig:device}b,c)~\cite{Reck1994,Clements2016}. Architectures for universal transformations based on layers of discrete Fourier transforms (DFTs) and programmable phase masks similarly require $O(N^2)$ tunable elements~\cite{LopezPastor2021}. Restricting the target operation from an arbitrary linear transformation to routing reduces the required programmability~\cite{Jenkins1994,Soldano1995}. Networks that realize arbitrary input--output permutations, such as Bene\v{s}
networks, require $O(N\log N)$~\cite{Benes1964,Spanke1987} tunable elements.

In this work, we present a reflective circuit architecture capable of pairwise routing among $N$ ports using only $N$ programmable phase shifters. For $N=2^k$ ($k$ an integer), each phase shifter switches between two states separated by $\pi$, and all $N(N-1)/2$ distinct input--output pairs can be addressed in the minimum possible $N-1$ configurations. The architecture consists of a single passive $N\times N$ interferometer and one reflective phase shifter per port (Fig.~\ref{fig:device}d)~\cite{Heuck2023PhotonRouting}. We demonstrate the architecture with a $3\times 3$ router using compact photonic-crystal (PhC) cavity-based phase shifters operating at telecom wavelengths.
Conventional non-resonant TFLN phase shifters require millimeter-scale interaction
lengths~\cite{Wang2018, Xu2020}, whereas resonant phase shifters convert a small
electro-optic resonance shift into a large reflected-phase change over a
capacitive interaction length of only tens of micrometers.
This enables low drive voltage, low switching energy, and static power
consumption limited by leakage currents~\cite{Li2020,Larocque2024}.
Our implementation uses a multimode interferometer (MMI) as the passive mixing element, with a multimode section measuring $(450\!\times\!15)\,\upmu\mathrm{m}$. The single-sided PhC cavity-based phase shifters measure $(100\!\times\!1)\,\upmu\mathrm{m}$ and simultaneously combine a 6.2-GHz electro-optic bandwidth with sub-5-V drive and negligible static power dissipation. We demonstrate sequential routing between three input--output pairs using 20-ns state slots, with extinction ratios above 11~dB across all three output ports. Automated characterization of 60 complete devices, comprising 180 one-sided switch cavities and 60 two-sided reference cavities, shows loaded quality factors reaching $7.1\times10^4$, close resonance matching, and tuning efficiencies up to 1.92~GHz\,V$^{-1}$.

\section{Results}\label{sec:results}

\subsection{Architecture and operating principle}\label{subsec:arch}

Light entering any input port of the MMI (Fig.~\ref{fig:device}d) is split equally among the reflector arms; each branch acquires a voltage-controlled phase shift at its reflector, and the reflected fields are mixed again on the return pass through the MMI.

The reflective phase shifters consist of single-sided PhC
nanobeam cavities (Fig.~\ref{fig:device}e), flanked by gold electrodes
(Fig.~\ref{fig:device}f). The reflection coefficient of each cavity follows from coupled-mode
theory~\cite{Suh2004},
\begin{equation}
  r(\Delta)
  = \frac{i\Delta - (\kappa_e - \kappa_i)/2}
         {i\Delta + (\kappa_e + \kappa_i)/2},
  \label{eq:r}
\end{equation}
where $\Delta = \omega_{\rm res}(V) - \omega$ is the detuning between the cavity resonance and the probe field, $\kappa_e$ is the external coupling
rate to the waveguide, $\kappa_i$ the intrinsic loss rate, and
$\omega_{\rm res}(V) = \omega_0 + 2\pi\eta V$ shifts linearly with applied voltage $V$
through the Pockels effect with a signed tuning coefficient $\eta$. Quoted tuning efficiencies denote its magnitude. In the ideal overcoupled limit
($\kappa_i/\kappa_e \to 0$), $|r| \to 1$ while $\arg(r)$ winds through a full
$2\pi$ across the resonance, controlled by the voltage $V$. Fig.~\ref{fig:device}g illustrates the resonant amplitude and phase response with $\kappa_i/\kappa_e=0.737$, showcasing a large phase swing across resonance. We thus realize a low-loss, voltage-tunable phase shifter with static power consumption limited only by leakage currents.

To describe the complete switch, we write the reflection coefficient of each
cavity as $r_j=|r_j| \exp(i\phi_j)$. Writing $\mathbf{M}$ for the unitary forward
transfer matrix of the MMI, reciprocity makes the return pass
$\mathbf{M}^{\mathsf{T}}$, giving the round-trip transfer matrix
\begin{equation}
  \mathbf{T}(\omega, \mathbf{V})
  = \mathbf{M}^{\mathsf{T}}
    \, \mathrm{diag}\bigl[r_1, r_2, r_3\bigr]
    \, \mathbf{M},
  \label{eq:T}
\end{equation}
where diag$[x,y,z]$ represents a diagonal matrix with the arguments as the ordered non-zero entries. When only port $m$ is illuminated, the power routed to output port $k$ is $P_{k} = |T_{km}|^2 P_{{\rm{in}},m}$. Independently tuning $(\phi_1,\phi_2,\phi_3)$ through the control voltages $(V_1,V_2,V_3)$ changes the interference conditions at the output ports. The resulting output-power distributions define the reachable region within the $(P_1,P_2,P_3)$ simplex (Fig.~\ref{fig:device}h). For four ports, equal-split MMIs can have different reachable power regions even though each still permits perfect pairwise routing (Supplementary Note~7).

\subsection{Routing configurations}\label{subsec:routing-configurations}

In Supplementary Notes 1 and 2, we provide several proofs that explain the possibilities and limitations of the circuit architecture in Fig.~\ref{fig:device}d. First, we show that any single pairwise connection among $N$ input--output ports is reachable if the MMI implements a DFT and the reflections are of the form $r_j=\exp(i2\pi jl/N)$, where the integer $l$ controls which ports are connected. Next, we generalize the result and show that any lossless, equal-split MMI with $|M_{nm}|=1/\sqrt{N}$ suffices for single-pair routing. Finally, for Walsh matrices of size $N=2^k$, all $N(N-1)/2$ distinct port pairs can be covered in $N-1$ configurations using only binary reflection phases $r_j=\pm1$. This means that no port is ever idle, and the phase shifters can be controlled with digital electronics. In each step, $N/2$ pairings are active, but for $N\geq8$ there is insufficient freedom to choose among the $(N-1)!!=(N-1)(N-3)\cdots 3\cdot 1$ pairing configurations (where $!!$ denotes the double factorial). This would require $O(N \log N)$ programmable elements, as in Bene\v{s} networks~\cite{Benes1964,Spanke1987}. Our reflective circuit connects all pairs in the minimum possible number of steps ($N-1$), making it ideally suited for networks with a fixed, traffic-independent switching schedule~\cite{Mellette2017}.

\subsection{Device and experimental setup}\label{subsec:device}

\begin{figure*}[!t]
  \centering
  \includegraphics[width=\textwidth,height=0.9\textheight,keepaspectratio]{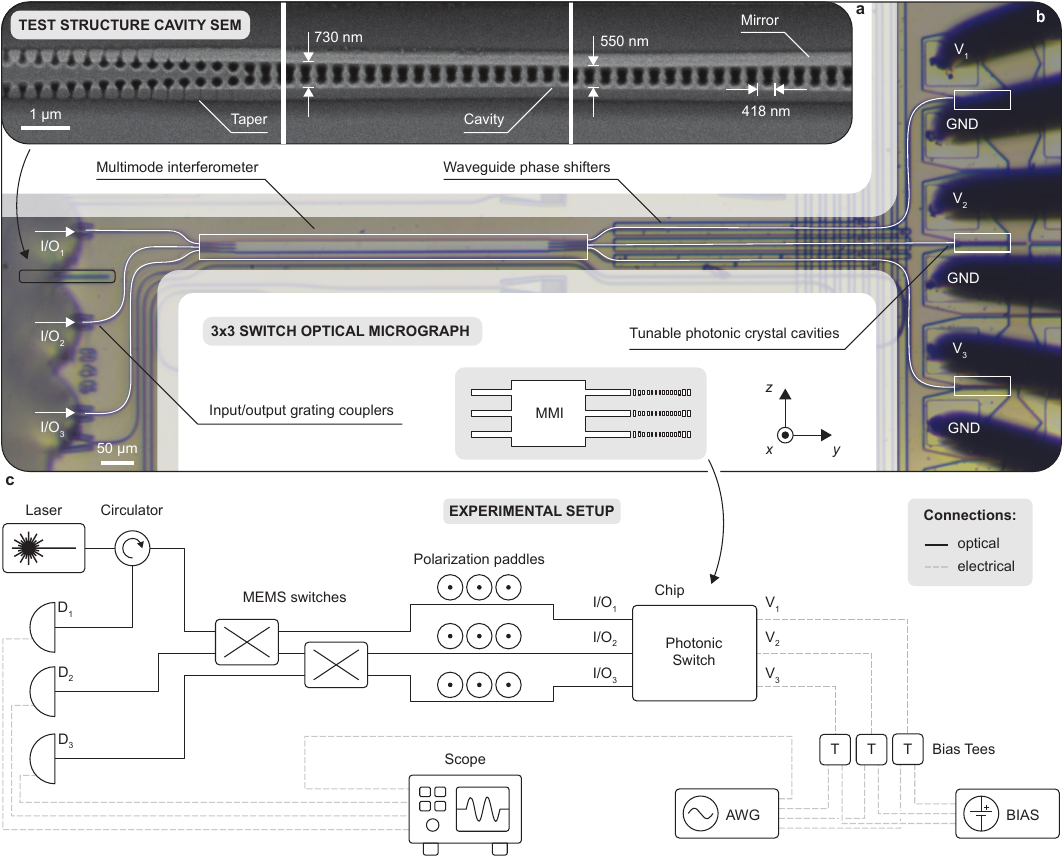}
  \caption{\textbf{Fabricated $3\times3$ photonic switch and
  experimental setup.}
    \textbf{a}, Scanning electron micrographs of a dedicated cavity test
    structure after local removal of the SiO$_2$ cladding, identifying the
    taper, cavity, and mirror sections and representative dimensions. The scale bar shown in the taper section applies to all three sections.
    \textbf{b}, Optical micrograph of the fabricated switch, showing the
    input/output grating couplers, central MMI, waveguide phase shifters, three
    tunable photonic-crystal cavities, and electrical contact region. The driven
    cavity electrodes $V_1$, $V_2$, and $V_3$, together with the intervening
    ground contacts, are identified.
    \textbf{c}, Experimental setup used to select the driven port, control the
    three cavity electrodes through bias tees, and simultaneously record the
    three reflected output powers. Black and light-gray lines denote optical
    and electrical connections, respectively.}
  \label{fig:device-setup}
\end{figure*}

The switch is fabricated on x-cut TFLN and clad with SiO$_2$ (Methods). The
crystal orientation aligns the guided TE-like mode with the in-plane DC and
radio-frequency control fields, maximizing the electro-optic response through
the dominant Pockels coefficient $r_{33}$ (Fig.~\ref{fig:device}f).

Figure~\ref{fig:device-setup}a shows a single-sided PhC nanobeam cavity divided
into taper, front-mirror, cavity, and back-mirror sections. Light traverses a
$\sim$15~$\upmu$m taper and the front mirror, which sets the external coupling
rate $\kappa_e$, enters the central defect, and is retroreflected by the back
mirror. The four sections contain $N_{\rm{tap}}$, $N_{m,L}$, $2N_{\rm{cav}}$,
and $N_{m,R}$ periods with pitch $a=418$~nm. The cavity and mirror sections are
550 and 730~nm wide, respectively. Full design details are provided in
Supplementary Note~3.

Figure~\ref{fig:device-setup}b shows the fabricated $3\times3$ switch.
Uniform TE grating couplers (I/O$_1$--I/O$_3$) couple light to and from the
three bidirectional ports of a single $3\times3$ MMI. Each MMI arm also carries
a waveguide phase shifter for optional static phase control. These phase
shifters were left unbiased in this work. Each arm then terminates in a
single-sided PhC nanobeam cavity. Coplanar gold electrodes straddle the three
cavities and are addressed by voltages $(V_1,V_2,V_3)$ referenced to a shared
ground in a $GV_1GV_2GV_3G$ pad configuration. The electrodes are separated by a
6.8~$\upmu$m gap and cover 64~$\upmu$m of the cavity's length so that the applied
electric field overlaps the guided mode. Finite-element electrostatic simulations give a terminal capacitance $C=18.6\pm0.3$~fF between each cavity drive electrode and ground (Supplementary Note~3). For a conventional non-energy-recovering driver, the resulting energy dissipated per 5-V transition is $E_{\rm tr}=C(\Delta V)^2/2=232.5\pm3.8$~fJ.

Figure~\ref{fig:device-setup}c shows the fiber-coupled routing setup. A tunable
laser, circulator, and pair of MEMS fiber switches select the driven port and
direct the three reflected outputs to detectors. Polarization paddles align the
fiber array with the grating coupler's TE mode. The outputs are recorded
simultaneously on an oscilloscope. Each cavity electrode is driven through a
bias tee combining a DC operating-point voltage with an
arbitrary-waveform-generator (AWG) signal bounded between $-5$ and $+5$~V at
the high-impedance device. The AWG synchronization signal triggers the
oscilloscope and aligns the drive waveforms with the optical responses.
Instrument details are provided in Methods.

\begin{figure*}[!t]
  \centering
  \includegraphics[width=\textwidth,height=0.9\textheight,keepaspectratio]{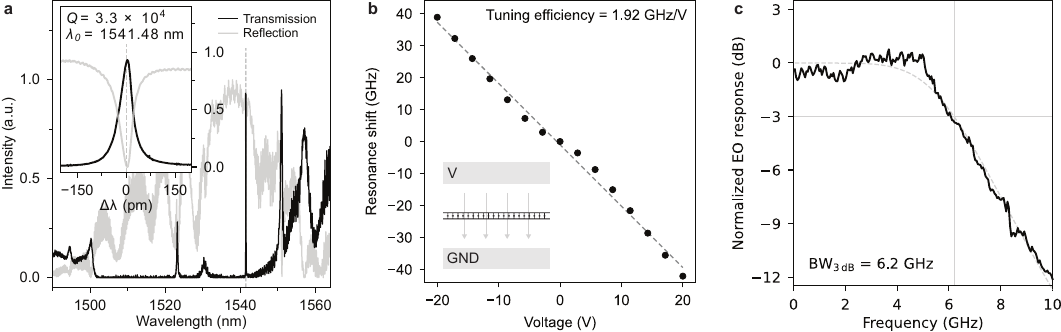}
  \caption{\textbf{Cavity spectroscopy, electro-optic tuning and
  modulation bandwidth.}
    \textbf{a}, Measured broadband transmission (black) and reflection
    (gray) spectra of a representative two-sided reference cavity, independently
    scaled for display; inset: the
    measured cavity resonance versus wavelength detuning.
    \textbf{b}, Measured resonance-frequency shift of a single-sided
    device cavity versus applied DC voltage at a $-14$-dBm laser-output
    setting (black points). The dashed gray
    line is a linear fit with an effective DC tuning rate of
    $\eta_{\rm DC}=1.92$~GHz\,V$^{-1}$; a kink around zero voltage indicates a
    deviation from linearity. Inset: schematic of the electric field between the drive and
    ground electrodes.
    \textbf{c}, Measured normalized electro-optic response of a single-sided device cavity
    versus modulation frequency (black), with a third-order low-pass fit
    (dashed gray).}
  \label{fig:bw}
\end{figure*}

\subsection{Cavity electro-optic characterization}\label{subsec:bw}

Figure~\ref{fig:bw}a shows the broadband reflection and transmission
spectra of a two-sided reference cavity. The single-sided switch cavities use
a partially transmitting front mirror with $N_{m,L}=40$ periods and a highly
reflecting back mirror with $N_{m,R}=100$ periods. The two-sided reference
cavity instead uses two symmetric 40-period mirrors,
$N_{m,L}=N_{m,R}=40$. This symmetric geometry clearly reveals the PhC bandgap
between approximately 1500 and 1560~nm and a high-$Q$ cavity resonance at
$\lambda_0=1541.48$~nm. A joint fit to the reflection and transmission
resonance gives a loaded quality factor
$Q=(3.30\pm0.02)\times10^4$, corresponding to an optical linewidth of
$\Delta\nu_{\rm cav}=\nu_0/Q=5.9$~GHz. The reference-cavity linewidth
characterizes the fabricated PhC optical quality. The full-device model
described below uses the one-sided switch-cavity parameters to determine the
phase response relevant to routing.

Figure~\ref{fig:bw}b shows the voltage-dependent resonance shift of a
single-sided device cavity. At a $-14$-dBm laser-output setting, a least-squares
fit over the full voltage
range gives an effective quasi-static tuning rate of
$\eta_{\rm DC}=1.92\pm0.06$~GHz\,V$^{-1}$. A deviation from linearity is visible
around zero voltage. At higher optical power, photo-induced space charge more
strongly screens the applied field, making the DC-tuning kink more pronounced
and reducing the effective $\eta_{\rm DC}$.

While $\eta_{\rm DC}$ describes tuning under a slowly varying bias, we measure the normalized high-speed electro-optic response as a function of modulation frequency by applying an RF tone to a cavity electrode and recording the response with a vector network analyzer (VNA).
The response stays approximately flat through the low-gigahertz range. A third-order empirical low-pass fit gives a 3-dB bandwidth of
6.2~GHz (Fig.~\ref{fig:bw}c; Supplementary Note~5).

We performed automated measurements of 60 complete devices,
comprising 180 one-sided switch cavities and 60 two-sided reference cavities,
spanning a range of designs.
Independent fits to 44 reference-cavity spectra acquired without detector saturation and with
$R^2\geq0.9$ give loaded quality factors reaching $7.1\times10^4$. Across 50
complete devices for which a visual review identified two or three resolved
switch-cavity resonances, we define the within-device resonance span
$\Delta\lambda$ as the wavelength separation between the outermost resolved
resonances among the three nominally identical, closely spaced switch cavities. The median
$\Delta\lambda$ is 0.63~nm, with an interquartile range of 0.44--0.91~nm, and
82\% have $\Delta\lambda<1$~nm (Supplementary Note~3).

\subsection{Experimental \texorpdfstring{$3\times3$}{3 x 3} routing}\label{subsec:switching}

\begin{figure*}[!t]
  \centering
  \includegraphics[width=\textwidth,height=0.9\textheight,keepaspectratio]{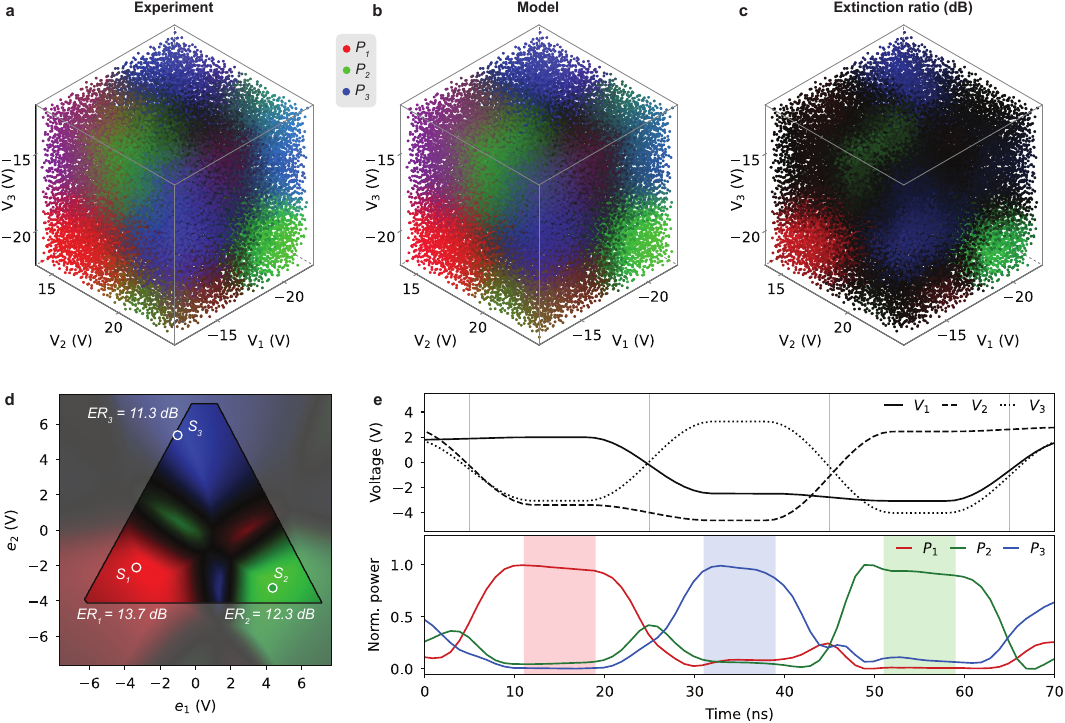}
  \caption{\textbf{Voltage-space characterization and model-guided
  time-domain routing.}
    \textbf{a}, Measured output-power distributions across the
    $(V_1,V_2,V_3)$ voltage cube sampled by the multisine drive. The RGB value
    of each point encodes the relative output powers, with red, green, and blue
    components corresponding to $P_1$, $P_2$, and $P_3$, respectively.
    \textbf{b}, Output-power distributions predicted by the fitted
    transfer-matrix model on a regular voltage grid, using the same display
    scheme and RGB encoding as in a.
    \textbf{c}, Extinction ratio computed from the measured,
    channel-normalized output signals across the voltage cube; color
    identifies the target output port, and brighter regions indicate higher
    routing contrast.
    \textbf{d}, Model-predicted extinction ratio on a slice through
    the voltage cube containing the three optimized routing points, with
    orthonormal in-plane voltage coordinates $(e_1,e_2)$, expressed in volts.
    Color identifies the target output port,
    and brighter regions indicate higher routing contrast. The gray overlay
    marks extrapolation beyond the measured voltage range; $S_1$, $S_2$, and
    $S_3$ mark routing from input port 1 to output ports 1, 2, and 3,
    respectively. The colors show the calibrated model prediction, whereas the
    extinction ratios annotated next to the three states are the corresponding
    experimental values from the dynamic measurement in e.
    \textbf{e}, Time-domain sequential routing. Top: programmed AWG voltage
    waveforms $V_1$ (solid), $V_2$ (dashed), and $V_3$ (dotted). Bottom:
    simultaneously recorded normalized output powers $P_1$ (red), $P_2$ (green)
    and $P_3$ (blue); shaded windows mark the stable routing intervals for each
    target output port.}
  \label{fig:switching}
\end{figure*}

To map the switch across its accessible voltage space, we first align the three cavity resonances using static DC biases. We then explore the
three-dimensional voltage space by driving the cavity electrodes
$(V_1,V_2,V_3)$ with orthogonal multisine waveforms, allowing the voltage cube
to be sampled densely in a single acquisition. We map the output-power distributions accessible to the switch by recording $P_1$, $P_2$, and
$P_3$ as the drive traverses the cube (Fig.~\ref{fig:switching}a).

We fit the measured point cloud using the transfer-matrix model of
\eqref{T}, with cavity parameters initialized from reference-cavity measurements.
The fitted model reproduces the measured output
distribution (Fig.~\ref{fig:switching}b) and provides a continuous mapping from
electrode voltages to output powers, allowing optimal routing points to be
identified numerically.

To identify the routing state for each target output $k$, we define the
worst-case extinction ratio
\begin{equation}
  \mathrm{ER}_k(\mathbf{V}) = 10\log_{10}\!\!\left[\frac{\widetilde{P}_k}{ \max\limits_{j \neq k}\!\big(\widetilde{P}_j\big)}\right],
  \label{eq:ER}
\end{equation}
which compares the target-port signal with the strongest off-target signal.
Here, $\widetilde{P}_k$ denotes the independently normalized output power
(Methods).
Figure~\ref{fig:switching}c plots the measured extinction ratio across the
voltage cube. Maximizing $\mathrm{ER}_k$ in the fitted model identifies the
optimal voltage tuple for each routing state; Fig.~\ref{fig:switching}d shows a
model-predicted slice through the three optimized points. The corresponding
model-reconstructed cavity spectra are discussed in Supplementary Note~4.

Applying the optimized voltage tuples sequentially demonstrates time-domain
routing (Fig.~\ref{fig:switching}e). We use a 100-ns waveform divided into
five 20-ns slots, corresponding to a 10-MHz repetition rate and a 50-MHz slot
rate. Three slots apply the voltage tuples for routing states $S_1$, $S_2$, and
$S_3$, which route light from input port 1 to output ports 1, 2, and 3,
respectively. Because the bias tees block DC and frequencies below 10~MHz,
each AWG waveform must have zero average over a complete cycle. We therefore
add two compensation slots whose voltages cancel the average of the three
routing slots. Within each slot, the voltage is held constant for 6~ns and
changes during a 14-ns raised-cosine transition. The demonstrated timing was
set by the experimental drive and acquisition setup, including the 10-MHz
bias-tee lower cutoff and 200-MHz oscilloscope bandwidth, rather than by the
intrinsic device response. The 14-ns transitions were chosen conservatively
to preserve waveform fidelity and suppress electrical ringing. During the
three routing slots, the switch directs light to output ports 1, 2, and 3 with
maximum instantaneous, normalized-signal extinction ratios of 13.7, 12.3,
and 11.3~dB, respectively, relative to the strongest off-target port, using
AWG excursions below 5~V about the static cavity biases. The variation among
output ports is consistent with differences in cavity coupling across the
three arms.

An independent power-budget measurement at the 1541.66-nm switching wavelength
gives a calibration-corrected summed-output excess loss of 3.65~dB relative to
a same-session grating-reference path. For routing state $S_1$, the
calibrated model attributes approximately 2.33~dB to cavity dissipation. The
estimated remaining excess of approximately 1.3--1.6~dB is comparable to that
measured over the off-resonant scan interval (Supplementary Note~4). This can
include two MMI passages, routing-waveguide loss, and residual path mismatch,
so it cannot be assigned to the MMI alone.

\section{Discussion}\label{sec:discussion}

The key performance metrics of our compact router architecture are insertion
loss, extinction ratio, speed, control voltage, and dynamic energy
consumption. Below, we discuss ways to improve these metrics in future
demonstrations.

For the circuit architecture requiring only two phase settings, we can choose
an operating point so the two voltages correspond to detunings
$\pm\Delta_\pi=\pm\tfrac{1}{2}\sqrt{\kappa_e^2-\kappa_i^2}$. From~\eqref{r},
the corresponding phase difference is
$\arg\{r(\Delta_\pi)\}-\arg\{r(-\Delta_\pi)\}=\pi$, and the reflected powers
are
$|r(\Delta_\pi)|^2=|r(-\Delta_\pi)|^2=(\kappa_e-\kappa_i)/(\kappa_e+\kappa_i)$.
Such an operating point requires overcoupled cavities, $\kappa_e>\kappa_i$.
If all cavities have identical coupling rates, perfect extinction remains
possible. An insertion loss of 10\% requires $\kappa_e/\kappa_i=19$, and 1\%
corresponds to $\kappa_e/\kappa_i=199$.

For lossy and non-identical cavities, the extinction ratio cannot be perfect
in general. Unequal arm lengths between the MMI and the cavities change the
required operating points because the cavity phase and amplitude responses are
coupled. Adding waveguide-based phase shifters in front of the cavities
decouples these responses and enables operating-point optimization. In
Supplementary Note~4, we investigate how device imperfections affect the
achievable extinction. In an optical-only sensitivity model retaining the
fitted cavity and MMI parameters but excluding empirical detector backgrounds
and calibration, optimizing the two static relative operating-point phases
increases the worst of the three separately optimized route extinctions from
16.0 to 24.0~dB and the minimum route-specific fraction of the voltage cube
above 10~dB from 1.49\% to 3.93\%. Optimizing instead for local 10-dB
voltage-error tolerance increases the worst-route half-width from
approximately $\pm0.70$ to $\pm1.13$~V per electrode. Designs with matched arm
optical path lengths reduce the
required operating-point phase-shifter length. Alternatively, millimeter-scale
waveguide-based phase shifters could implement both DC offsets and dynamic
switching, enabling broadband operation with an interaction length comparable
to that of a push--pull MZI.

Larger extinctions than the 11.3--13.7~dB shown in
Fig.~\ref{fig:switching}e are possible for single input--output pairs. We
achieved 20.8--22.6~dB with repeated returns to the same waveform phase, as
described in Supplementary Note~6. The contrast available at a fixed operating
point therefore exceeds that maintained throughout the full three-state
sequence. At the highest-contrast point, the off-target signals approach the
measurement repeatability floor, whereas noise and slow drift displace the
device from its optimized operating points during extended sequences. Faster
feedback and improved thermal stabilization should reduce this difference.

Photo-induced space charge screens the applied electric field, making the DC
response dependent on optical power and reducing the effective tuning
efficiency $\eta_{\rm DC}$ at higher power. Related electro-optic relaxation
and photorefractive resonance shifts are widely observed in TFLN modulators
and resonators~\cite{Holzgrafe2024,Xu2021,Buzzi2026Photorefractive}. This instability complicates
long-term bias control but may be reduced through surface treatment, improved
charge management, and air-clad device structures.

The speed, control voltage, and energy consumption are coupled in cavity-based
phase shifters because their modulation bandwidth is limited by the linewidth,
consistent with the gigahertz response in Fig.~\ref{fig:bw}c. The required
switching voltage satisfies
$2\pi\eta_{\rm RF}\Delta V=2\Delta_\pi=\sqrt{\kappa_e^2-\kappa_i^2}$,
where $\eta_{\rm RF}$ is the resonance-frequency tuning efficiency at the
modulation frequency. The switching energy, $C(\Delta V)^2/2$, is also coupled
to the linewidth and speed. Improving the speed--energy trade-off requires increasing
$\eta_{\rm RF}$ and reducing the device capacitance $C$. Future cavity designs
should therefore be optimized for $\eta_{\rm RF}$, $\kappa_i$, and size. In
Supplementary Note~3, we show that our nanobeam PhC cavity design enables
control of $\kappa_e$ without affecting $\kappa_i$, allowing the linewidth to
be adjusted to meet specified control-voltage and speed requirements.

\section{Conclusion}\label{sec:conclusion}

We have theoretically shown that pairwise routing requires only $O(N)$
programmable phase shifters in a compact reflective circuit architecture. Our
experimental demonstration of a $3\times3$ router in TFLN illustrates the
feasibility of the approach, providing a path toward dense, low-energy,
electronically addressable routing layers for classical photonic
interconnects. While our demonstration used TFLN, the reflective architecture
could also be implemented on other platforms, including silicon, which
benefits from mature foundry processes~\cite{Rahim2018}. There, a
ring-resonator phase shifter~\cite{Li2014,Huang2020} or a waveguide phase
shifter could be placed before a passive Sagnac-loop mirror to provide the
required controllable reflection phase.

Circuits described by permutation matrices also enable the routing of
individual photons distributed among the input ports. The high speed,
cryogenic compatibility~\cite{Lomonte2021,Thiele2020}, and broad
transparency~\cite{Desiatov2019,Boes2023} make these circuits attractive for
quantum photonics. They could enable cryogenic routing of photons between
spin--photon interfaces in distributed matter-qubit
modules~\cite{Pompili2021,Choi2019}, control beams for trapped ions, neutral
atoms, and solid-state emitters~\cite{Mehta2020,Christen2025,Wan2020,Buzzi2025},
and provide essential building blocks in recently proposed recirculating
quantum photonic networks~\cite{Grovn2026}.

Cavity-enhanced Pockels switching therefore provides a versatile path toward
compact routing layers for both classical and quantum photonic interconnects.

\section{Methods}\label{sec:methods}

\subsection{Device fabrication}\label{subsec:fab-methods}

Devices were fabricated on CSEM's x-cut TFLN platform, comprising a
600~nm LiNbO$_3$ device layer on buried SiO$_2$ and silicon.
The 1550-nm waveguides and PhC cavities were patterned in the LiNbO$_3$
device layer and etched to a depth of 400~nm, leaving a 200~nm slab. The
routing waveguides had a nominal lithographic top width of 800~nm. The
nominal exterior sidewall angle was $55^\circ$ from the substrate plane
($35^\circ$ from vertical). For the 400-nm etch,
this sidewall angle made the fabricated waveguide base 560~nm wider than the
lithographic top width, giving a nominal base width of 1.36~$\upmu$m for an
800-nm lithographic top width. Gold electrodes were formed using the
600-nm-thick platform metallization layer. The devices were clad with silicon
dioxide. To expose the cavity for the SEM in
Fig.~\ref{fig:device-setup}a, the silicon-dioxide cladding was locally removed
from a dedicated test structure while the switch cavities remained oxide-clad.

\subsection{Optical and electrical measurement setup}
\label{subsec:optical-methods}

For optical measurements, the chip was mounted on a Thorlabs M30XY
translation stage and coupled through the grating couplers using a fiber array
mounted at an $8^\circ$ angle on a Thorlabs NanoMax stage. The optical
source was a Santec TSL-570 tunable laser with a tuning range of
1480--1640~nm. For the switching measurements in Fig.~\ref{fig:switching},
the laser output was set to 0~dBm. The laser light passed through a fiber-optic
circulator and a fiber-switching network comprising two Agiltron $2\times2$
MEMS switches that selected the driven and detected ports. Polarization paddles
aligned the launch to the TE mode before coupling to the chip. Three Thorlabs
APD431C InGaAs avalanche photodetectors (400-MHz nominal bandwidth) recorded
$P_1$, $P_2$, and $P_3$ simultaneously. The three cavity
electrodes were contacted with a FormFactor Cascade Multi~$|Z|$ RF probe
(K20-MZN-9B01-100) and driven through Mini-Circuits ZX85-12G-S+ bias tees
(10-MHz--12-GHz passband)
that combined static DC biases from three Keithley 2400 source-measure units
(100~$\upmu$A current compliance) with AC waveforms from a three-channel
arbitrary waveform generator (Zurich Instruments HDAWG8, 2.4~GSa/s,
$\pm5$~V high-impedance output span, or 10~V peak-to-peak).
For the broadband insertion-loss budget, a plain grating-reference path and
the device were measured on the same wavelength grid using mutually calibrated
detector channels. The traces were median-filtered before comparison. Because
the reference includes both grating couplers, external fiber
components, polarization mismatch, and reference-waveguide propagation, the
device values are reported as excess loss relative to this path rather than
as independently calibrated component losses (Supplementary Note~4).

\subsection{Cavity spectroscopy, DC electro-optic tuning, and device statistics}
\label{subsec:spectroscopy-methods}

For cavity spectroscopy, the laser wavelength was swept with a nominal
step of 1~pm over 1480--1640~nm, and the reflected power was normalized to the
laser output. The spectrum of the two-sided reference cavity was fitted over a 0.30-nm window centered
on the detected resonance. The reflection dip and transmission peak were fitted
simultaneously with a shared Lorentzian center and linewidth, independent
amplitudes, and independent linear backgrounds. This simultaneous Lorentzian
fit extracted the resonance wavelength, linewidth, and loaded quality factor
$Q_{\rm total}$. The reported uncertainty in $Q_{\rm total}$ is the
one-standard-error value propagated from the least-squares covariance.
For the DC tuning measurement, the voltage was stepped sequentially, and
a wavelength scan was acquired at each voltage using a $-14$-dBm laser output
setting. At each voltage, the three detector spectra were fitted independently
of the other voltage steps with a shared resonance center and linewidth and
independent linear backgrounds and Lorentzian/dispersive amplitudes.
The extracted resonance wavelengths were converted to frequency shifts and
fitted by ordinary least-squares over all 15 points from $-20$ to $+20$~V,
giving an effective quasi-static tuning rate
$\eta_{\rm DC}=1.92\pm0.06$~GHz\,V$^{-1}$. The formal slope standard error is
0.044~GHz\,V$^{-1}$; grouping two to four neighboring voltage settings to
allow correlated residuals gives descriptive standard errors up to
0.060~GHz\,V$^{-1}$, which sets the reported uncertainty. A deviation from
linearity remains near zero voltage.

Device statistics were acquired with an automated measurement routine
that measured all input--output port configurations for 60 complete devices,
comprising 180 one-sided switch cavities and 60 two-sided reference cavities.
The Thorlabs M30XY translation
stage was used to optimize the fiber coupling before each wavelength sweep.
After short translations, automatic alignment used three iterations of a
climb-ladder search along $x$ followed by $y$. After larger translations, a
coarse raster scan first located the grating couplers, followed by the same
climb-ladder procedure for fine alignment. For each complete device, the
recorded sequence comprised a reference-cavity sweep, three MMI sweeps acquired
under different programmed permutations of the MEMS fiber-switch matrix, and a
final calibration configuration. In a representative
uninterrupted block covering 27 device positions, successive device
measurements began at a median interval of 6.5~min. The five recorded sweeps
within each complete sequence spanned a median of 3.9~min; the remaining cycle
time included translation to the next device and optimization of the fiber
coupling.

\subsection{Electro-optic bandwidth measurement}\label{subsec:rf-methods}

The electro-optic bandwidth of a single cavity was measured from its
$S_{21}$ response over 50~MHz--10~GHz using a four-port Rohde \& Schwarz ZNA67
vector network analyzer connected to the cavity electrodes through one GSG
pair of a GSGSGSG probe. VNA port~1 drove the contacted electrodes
directly through the probe, without a bias tee or applied DC bias. The laser
was tuned to a point of maximum absolute slope on the cavity reflection
lineshape. The modulated
optical output was detected by a Discovery Semiconductors DSC30S InGaAs
photodetector (18-GHz nominal bandwidth), whose electrical output returned
directly to VNA port~2. The response shown in Fig.~\ref{fig:bw}c was
acquired with a nominal electrical VNA source power of 12~dBm and an optical
power setting of 13~dBm at the laser output. Accounting for the measured
1.39~dB input-path loss, an estimated
5.4--6.1~dB input-grating loss, and ideal three-way MMI splitting gives an
estimated optical power of 0.7--1.4~dBm in the addressed cavity arm, before
uncalibrated MMI excess loss.
A third-order low-pass model with a free vertical offset was fitted to
the measured $S_{21}$ magnitude, yielding a 3-dB bandwidth of 6.2~GHz.
Additional cavity-channel measurements showed similar bandwidth responses.
Supplementary Note~5 provides the fitting procedure and model comparison.

\subsection{Multisine voltage-space acquisition and delay correction}
\label{subsec:voltage-space-methods}

To map the router response over its three-dimensional control space, we
simultaneously varied the voltages applied to the three cavity electrodes while
recording the optical power at all three output ports.
The three voltage waveforms were designed to sample this space broadly
while minimizing correlations between the drive channels.
Each drive channel carried a 25~$\upmu$s-period waveform synthesized at
2.4~GSa/s from 1001 equally spaced tones between 10 and 50~MHz, with equal
initial amplitudes and pseudorandom phases.
The three waveforms were orthogonalized by QR decomposition so that their
discrete pairwise inner products over one period satisfied
$\sum_n V_i[n]V_j[n]=0$ for $i\ne j$. They were then scaled to equal peak range
and rank-remapped to give each channel a uniform amplitude distribution over
$[-1,1]$. The nonlinear remapping and delay reconstruction introduced small
residual correlations, and for the trajectories in Fig.~\ref{fig:switching}, the
largest pairwise correlation magnitude was $0.018$.
The 10--50~MHz range refers to the initial multisine synthesis band
before rank remapping.
The resulting waveforms spanned $\pm5$~V and were added to static cavity biases of
$(-17,+19,-17)$~V.
With the laser set to 1541.66~nm, four waveform periods were acquired
over 100~$\upmu$s at 1~GSa/s with 16 waveform averages.

Signal-propagation delays
between each AWG drive channel and oscilloscope acquisition channel were
calibrated using a common linearly swept chirp spanning 10--150~MHz with a
raised-cosine taper. The chirp occupied 15~$\upmu$s of each 25~$\upmu$s
period, and four periods were recorded together with a synchronous reference
tap. For each drive--detector pair, the traces were filtered over the
chirp band and cross-correlated, and the Hilbert-transform envelope was used
to estimate the delay. The nine delays were jointly fitted as
$\tau_{ij}=d_{\mathrm{in},i}+d_{\mathrm{out},j}$, with one input delay fixed
as the reference, and used to align the recorded traces. Synthetic
tests recovered injected delays within 0.2~ns, while repeated measurements
gave a relative-delay standard deviation of 1.4~ns.

For the color representation in Fig.~\ref{fig:switching}a, each detector
trace was independently min--max normalized over the acquisition.
For visualization, samples were grouped into a $15^3$ voltage grid.
Each occupied voxel was assigned its samples' mean color, and display points
were distributed uniformly within occupied voxels.
The extinction-ratio color maps in Fig.~\ref{fig:switching}c,d evaluate
the definition above using the min--max-normalized output power
$\widetilde{P}_k$ and a numerical offset $\epsilon=10^{-6}$ added to each normalized power.

For visualization, negative extinction ratios were set to zero.

\subsection{Transfer-matrix model and parameter estimation}
\label{subsec:tm-methods}

At the fixed probe frequency, the transfer-matrix model predicts the
three output powers from the applied voltages, using the voltage-dependent
cavity reflection coefficients $r_j(V_j)$ defined in \eqref{r}.
For each cavity, the numerical fit was parameterized by the zero-voltage resonance detuning $\Delta\nu_0=(\omega_0-\omega)/(2\pi)$,
external-to-internal coupling ratio, $\kappa_e/\kappa_i$, electro-optic tuning efficiency, $\eta$, and
loaded quality factor, $Q_{\rm L}=\omega_0/(\kappa_e+\kappa_i)$. Here, $\omega$ is the fixed probe angular frequency and $\nu_0=\omega_0/(2\pi)$ is the zero-voltage resonance frequency.
At the fixed probe frequency, however, the cavity response constrains the
products $Q_{\rm L}\Delta\nu_0/\nu_0$ and $Q_{\rm L}\eta/\nu_0$, rather than
$Q_{\rm L}$ and $\eta$ independently. Equivalently, it determines the voltage
required to tune through one loaded linewidth,
$V_{\rm linewidth}=\nu_0/(Q_{\rm L}\eta)$.
The nonlinear cavity parameters were constrained to the following bounds:
$-60$ to $60$~pm for the resonance wavelength offset,
$0.2$ to $20$ for the external-to-internal coupling ratio,
$0.95$ to $3.0$~GHz/V for the tuning efficiency, and $2.5\times10^4$ to
$6.0\times10^4$ for the loaded quality factor.
This loaded-quality-factor range was used as a prior informed by full-device
spectroscopy, including the reference-cavity measurements, rather than as an
upper bound on quality factors observed across the device array.
The MMI was modeled using a fixed phase convention based on the
discrete Fourier transform (DFT)~\cite{Bachmann1994,Soldano1995}. Only two relative amplitude factors were
fitted, each constrained to $0.85$--$1.15$; thus, the fit did not use an
unrestricted MMI matrix $\mathbf{M}$.
Reciprocal return propagation was evaluated with the transpose of the
forward MMI matrix, giving the round-trip transformation
$\mathbf{M}^{\mathsf{T}}\mathbf{R}\mathbf{M}$.
The fit also included two relative arm phases constrained to
$[-\pi,\pi]$, a common voltage-dependent phase correction constrained to
$\pm0.15$~rad/V, and a coherent background field at each detector with
relative amplitude constrained to $0$--$0.35$.
A gain and offset for each detector were determined by linear least
squares at every model evaluation.

For parameter estimation, the delay-aligned samples were binned into a
$20\times20\times20$ voltage grid, and each detector signal was averaged within
each occupied voxel.
The residual from each voxel was weighted by the square root of its
sample count.
The nonlinear parameters were fitted by minimizing the weighted
root-mean-square residual using bounded multistart optimization with a
differential-evolution fallback and L-BFGS-B refinement. A weak dimensionless
regularization term (overall scale $0.015$) favored similar parameters among
the three cavities and broad reference values of $1.7$~GHz/V for tuning
efficiency, $3.6\times10^4$ for loaded quality factor, and $5$ for the mean
external-to-internal coupling ratio. It also weakly favored MMI amplitude
factors near unity, a small voltage-dependent phase correction, and small
coherent background fields.
The reciprocal fit achieved $R^2=0.85$--$0.96$ across the three detector
signals.

\subsection{Routing-point optimization and time-domain measurement}
\label{subsec:time-routing-methods}

Routing operating points were selected by maximizing
$\mathrm{ER}_k(\mathbf{V})$ on the voltage grid used for model fitting.
For the representative device used in Figs.~\ref{fig:bw} and
\ref{fig:switching} (device A46), the absolute electrode-voltage tuples
$(V_1,V_2,V_3)$ used at 1541.66~nm were
\[
\begin{aligned}
 \mathbf{V}(S_1)&=(-15.01,15.61,-20.07)~\mathrm{V},\\
 \mathbf{V}(S_2)&=(-20.08,21.44,-21.04)~\mathrm{V},\\
 \mathbf{V}(S_3)&=(-19.48,14.38,-13.75)~\mathrm{V},\\
 \mathbf{V}_{\rm bias}&=(-17,19,-17)~\mathrm{V}.
\end{aligned}
\]
The listed absolute voltages include the static bias shown above. The
corresponding AWG excursions were
\[
\begin{aligned}
 \Delta\mathbf{V}(S_1)&=(1.99,-3.39,-3.07)~\mathrm{V},\\
 \Delta\mathbf{V}(S_2)&=(-3.08,2.44,-4.04)~\mathrm{V},\\
 \Delta\mathbf{V}(S_3)&=(-2.48,-4.62,3.25)~\mathrm{V}.
\end{aligned}
\]
The tuples are device- and wavelength-specific
calibrated operating points rather than universal switching voltages.
For time-domain routing, the AWG generated a 100-ns periodic waveform
containing three routing slots and two compensation slots. Each 20-ns slot
comprised a 6-ns constant-voltage interval followed by a 14-ns raised-cosine
transition. Because the bias tees block DC and frequencies below 10~MHz, each
channel's waveform was made zero-mean over the complete cycle. Both
compensation slots were therefore set to minus half the sum of the three
routing-slot voltages in that channel.

The three detector signals and a synchronous reference were acquired
on a Tektronix MSO2024B oscilloscope (200-MHz nominal bandwidth, 1~GSa/s)
using the acquisition settings and delay correction described above. The
demonstrated timing was set by these drive and acquisition constraints rather
than by the intrinsic electro-optic response.
For each target output, the reported extinction ratio was the maximum
instantaneous value over a 25~$\upmu$s analysis interval.

\subsection{Theoretical analysis}
\label{subsec:proof-methods}

Under the ideal assumptions of a lossless, equal-split MMI and
unit-magnitude reflector coefficients with phases independently tunable over
the full $2\pi$ range, every three-port coupler admits the
factorization
$\mathbf{M}=\mathbf{D}_{\rm out}\mathbf{P}_{\rm out}\mathbf{F}
\mathbf{P}_{\rm in}\mathbf{D}_{\rm in}$, where $\mathbf{F}$ is the three-point
Fourier-transform matrix, the $\mathbf{D}$ matrices are diagonal phases, and
the $\mathbf{P}$ matrices are permutations. Substituting this factorization
into the reciprocal round-trip transformation
$\mathbf{T}=\mathbf{M}^{\mathsf{T}}\mathbf{R}\mathbf{M}$ shows that the internal
phases can be absorbed into the controllable reflector phases, while
external phases leave output powers unchanged, and permutations only relabel
ports. Every such three-port coupler therefore has the same reachable power
region as the Fourier-transform model, up to port relabeling.

For any $N$, these ideal assumptions allow unit incident power at any
input to be routed entirely to any chosen output, yielding $P_k=1$ and
$P_{j\neq k}=0$. Full proofs and the analytic three-port
boundary are provided in Supplementary Notes~1 and~2. The routing theorems,
the three-port reachable-region equivalence theorem, and supporting algebraic
identities were
machine-checked in Lean~4 using Mathlib; the formal source is maintained in
the GitHub repository cited under Code availability.

\subsection{AI assistance}\label{subsec:ai-assistance}

\begingroup\raggedright
Generative AI tools assisted with portions of experimental control-code
development, data analysis, mathematical analysis, Lean verification, and
manuscript review. The authors reviewed and validated the AI-assisted
work. They retain full responsibility for the scientific content, results,
and conclusions.
\par\endgroup

\section*{Data availability}

\begingroup\raggedright
The source data for the main-text numerical panels are publicly available
in Zenodo at \url{https://doi.org/10.5281/zenodo.23194529}.
\par\endgroup

\section*{Code availability}

\begingroup\raggedright
The code for waveform generation, transfer-matrix modeling, routing
optimization, figure generation, and formal verification is maintained at
\url{https://github.com/sandrobuzzi/Multiport-Photonic-Routing}.
\par\endgroup

\section*{Acknowledgements}

This work was supported by the DEVCOM Army Research Laboratory under
Cooperative Agreement No.~W911NF-22-2-0127. A.B.\ acknowledges support from the Air Force Research
Laboratory under Award No.~FA8750-25-2-0500 and from\
the Defense Advanced Research Projects Agency
(DARPA) through the NanoWatt Platforms for Sensing, Analysis and Computation
(NaPSAC) program as part of the TOPCHIP project under subcontract No.~SCON-00005835. M. H. acknowledges support from The Novo Nordisk Foundation (REPEAT, grant no. NNF24OC0096351) and Innovation Fund Denmark (EQUAL, grant no. 4356-00007B). The authors thank Steven Nagle and the
T.~J. Rodgers RLE Laboratory for their continued support in developing the
experimental setup and characterizing the electro-optic bandwidth.
The authors thank CSEM, in particular Homa Zarebidaki, Amir H. Ghadimi,
and Hamed Sattari, for fabricating the chips.

\section*{Author contributions}

A.B.\ built the characterization setup, designed the control system, and measured the devices with support from B.A.F.\ and D.L.P.V.\ A.B., D.L.P.V., and M.H.\ analyzed the data. M.H.\ conceived the circuit concept, designed the devices, performed the optical simulations, and provided the DFT-based proof for pairwise routing. A.B.\ performed the electrostatic simulations and developed the Lean formalization, which D.R.E.\ reviewed. A.B.\ wrote the manuscript with support from M.H.\ and D.L.P.V.\ All authors reviewed the manuscript. D.L.P.V., M.H., and D.R.E.\ supervised the work.

\section*{Competing interests}

The authors declare no competing interests.

\bibliography{references}

\clearpage
\onecolumn
\setcounter{section}{0}
\setcounter{secnumdepth}{2}
\renewcommand{\thesubsection}{S\arabic{section}.\arabic{subsection}}
\makeatletter
\def\@seccntformat#1{\expandafter\ifx\csname the#1\endcsname\thesection
  \else
    \csname the#1\endcsname\hskip.5em\fi}
\makeatother
\setcounter{figure}{0}
\renewcommand{\thefigure}{S\arabic{figure}}
\renewcommand{\theHfigure}{S\arabic{figure}}
\renewcommand{\figurename}{Supplementary Figure}
\renewcommand{\figurecaptionfont}{\normalfont\rmfamily\fontsize{9}{12}\selectfont}
\setlength{\abovecaptionskip}{4.5pt}
\makeatletter
\long\def\@figurecaption#1#2{{\figurecaptionfont\raggedright
   {\bfseries #1}\hskip.7em#2\par}}
\makeatother
\setcounter{equation}{0}
\renewcommand{\theequation}{S\arabic{equation}}
\renewcommand{\theHequation}{S\arabic{equation}}
\setcounter{table}{0}
\renewcommand{\tablename}{Supplementary Table}
\renewcommand{\thetable}{\arabic{table}}
\renewcommand{\theHtable}{S\arabic{table}}

\begin{center}
  {\LARGE\bfseries Supplementary Information}\\[5mm]
  {\large\bfseries Multiport photonic routing with a single
   cavity-based electro-optic phase shifter per port}\\[5mm]
  Alessandro Buzzi$^{1,*}$, Bjarke A. Frederiksen$^{1,2}$,
  Dashiell L.\,P. Vitullo$^{1,3}$, Mikkel Heuck$^{2}$,
  and Dirk R. Englund$^{1}$\\[3mm]
  {\small\itshape
  $^1$Research Laboratory of Electronics, Massachusetts Institute of
  Technology, Cambridge, MA 02139, USA\\
  $^2$Department of Electrical and Photonics Engineering, Technical
  University of Denmark, Kgs.\ Lyngby 2800, Denmark\\
  $^3$DEVCOM Army Research Laboratory, Adelphi, Maryland 20783, USA}\\[2mm]
  {\small $^*$abuzzi@mit.edu}
\end{center}

\vspace{6mm}
\noindent\textbf{Contents}\\[1mm]
\hspace*{4mm}Supplementary Note 1: Pairwise routing with a DFT MMI\\
\hspace*{4mm}Supplementary Note 2: General routing theorems and three-port reachable-region equivalence\\
\hspace*{4mm}Supplementary Note 3: Device design, electrostatics and cavity-array characterization\\
\hspace*{4mm}Supplementary Note 4: Cavity response, insertion loss and routing-performance analysis\\
\hspace*{4mm}Supplementary Note 5: Electro-optic bandwidth extraction\\
\hspace*{4mm}Supplementary Note 6: Single-channel high-extinction routing\\
\hspace*{4mm}Supplementary Note 7: Four-port reachable-region extension
\vspace{6mm}

\section{Supplementary Note 1: Pairwise routing with a DFT MMI}
\subsection{Solution to the pairwise routing problem}
\begin{figure}[!ht]
\centering
  \includegraphics[width = 0.45\textwidth] {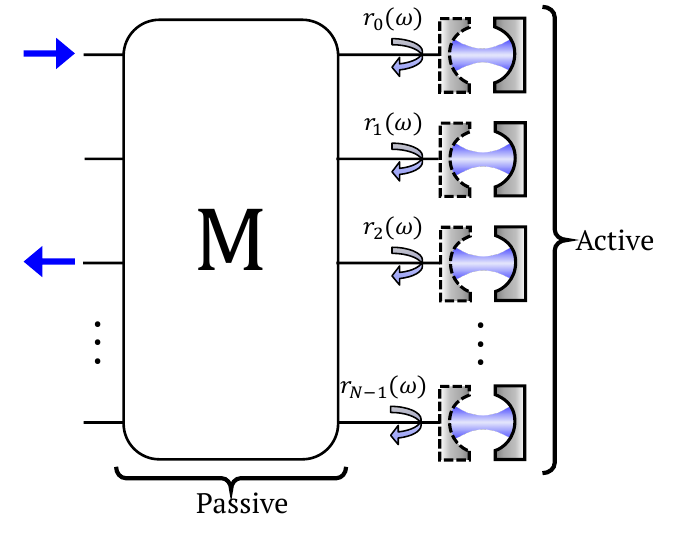}
\caption{Sketch of an $N$-port pairwise router with a passive MMI described by
a transfer matrix $\mathbf{M}$ and $N$ single-sided cavity phase shifters with
complex field-reflection coefficients $r_j(\omega)$. Only $N-1$ relative
reflector phases are independent because a common phase does not affect the
output powers.}
\figlab{Concept_Fig}  
\end{figure}  
We want to show that any input mode $n$ can be routed to any output mode $m$ of
a reflection-based $N\times N$ interferometer using $N$ cavity-based phase
shifters, of which only $N-1$ relative phases are independent, as shown
in~\figref{Concept_Fig}.
We represent incident light in mode $n$ by an $N\!\times\!1$ vector, $\mathbf{x}$, which is zero everywhere except $x_n = 1$. The forward transfer matrix of the passive interferometer is denoted by $\mathbf{M}$. Reciprocity makes the reverse transfer matrix $\mathbf{M}^{\mathsf{T}}$, so the round-trip input-output transformation is
\begin{align}\eqlab{IO vector definition}
    \mathbf{z} = \mathbf{T}(\mathbf{R})\mathbf{x} = \mathbf{M}^{\mathsf{T}} \mathbf{R} \mathbf{M} \mathbf{x},
\end{align}
where the transmission matrix, $\mathbf{T}(\mathbf{R})$, is a function of the diagonal matrix containing the reflection coefficients of the cavity phase shifters, $\mathbf{R} = {\rm{diag}}[r_k(\omega)]$. Below, we show that the pairwise routing problem may be solved by choosing $\mathbf{M}$ to be the discrete Fourier transform (DFT) $\mathbf{F}=\mathbf{F}^{\mathsf{T}}$,
\begin{align}\eqlab{Uft definition}
    \mathbf{y} = \mathbf{F}\mathbf{x}, ~~\text{with} ~~ y_k = \frac{1}{\sqrt{N}}\sum_{j=0}^{N-1} e^{-i\frac{2\pi}{N}kj}x_j.
\end{align}

First, we show that if all the phase shifts are zero, corresponding to $\mathbf{R}=\mathbf{I}$, the transformation $\mathbf{T}(\mathbf{I})$ is a particular permutation matrix. Then, we use the shifting property of the DFT to provide solutions for $\mathbf{R}$ that result in $N$ cyclic permutations of the rows of $\mathbf{T}(\mathbf{I})$. Taken together, these $N$ configurations cover every ordered input--output pair: for each selected input $n$ and output $m$, one configuration gives $T_{mn}=1$, with all other elements in column $n$ equal to zero. Thus, any selected pair can be routed perfectly, although different pairs generally require different reflector settings.

For an input with light in only the $n$th mode, only the $n$th term in the sum in~\eqref{Uft definition} is non-zero
\begin{align}\eqlab{y_k}
    y_k = \frac{1}{\sqrt{N}}e^{-i\frac{2\pi}{N}kn}.
\end{align}
If $\mathbf{R}=\mathbf{I}$, the output is
\begin{align}\eqlab{z_m}
    z_m = \frac{1}{\sqrt{N}} \sum_{k=0}^{N-1}e^{-i\frac{2\pi}{N}km}y_k =  \frac{1}{N} \sum_{k=0}^{N-1}e^{-i\frac{2\pi}{N}k(m+n)}.
\end{align}
If we define the phase $\theta_{mn}=2\pi(m+n)/N$, we can write~\eqref{z_m} as
\begin{align}\eqlab{z_m theta_mn}
    z_m = \frac{1}{N} \sum_{k=0}^{N-1}e^{-ik\theta_{mn}} = \frac{1}{N} \left(\frac{1-e^{-iN\theta_{mn}}}{1-e^{-i\theta_{mn}}}\right).
\end{align}
Since $m$ and $n$ are integers, we see that $N\theta_{mn}=2\pi p$ with $p\in\mathbb{N}_0$. This means that $z_m=0$ unless $\theta_{mn}=2\pi p$, which results in the requirement that $m+n=0$ or $m+n=N$. When this requirement is fulfilled, we have from~\eqref{z_m} that $z_m=1$. This means that an input in mode $n$ is routed to an output in mode $m$, or, equivalently, the transmission matrix is a permutation matrix

\begin{align}\eqlab{T permutation matrix}
    T_{mn}(\mathbf{I}) = \delta_{m,(-n)\bmod N} ~~\text{with}~~ \left\{ \begin{array}{lll} m=0 & ~\text{if} & n=0\\ {}&{}&{}\\ m=N-n & ~\text{if} & n\neq 0\end{array} \right . .
\end{align}
The Kronecker delta introduced in~\eqref{T permutation matrix} is 1 when its two indices agree and zero otherwise, with indices understood modulo $N$. The relation in~\eqref{T permutation matrix} shows that for $\mathbf{R}=\mathbf{I}$, each input mode $n$ is routed to a \emph{particular} output mode $m$. To show an input mode $n$ may be routed to \emph{any} output mode, we consider the shifting property of the Fourier transform
\begin{align}\eqlab{FT shifting}
    z_{m-l} = \frac{1}{\sqrt{N}}\sum_{k=0}^{N-1} e^{-i\frac{2\pi}{N}km} \left(y_k e^{i\frac{2\pi}{N}kl}\right).
\end{align}
In this context,~\eqref{FT shifting} means that the output light will shift from mode $m$ to mode $m+l$ (modulo $N$) if we choose the reflection coefficients as
\begin{align}\eqlab{r_k solution}
    r_k(\omega) = e^{i\frac{2\pi}{N}kl}.
\end{align}
This also becomes clear by inserting~\eqref{y_k} into~\eqref{FT shifting} and defining a new phase $\theta_{mln}=(m-l+n)2\pi/N$. Following the arguments above, this results in
\begin{align}\eqlab{T permutation matrix R}
    T_{mn}\!\left(\mathbf{R}\right) = \delta_{m,(l-n)\bmod N} ~~\text{with}~ \left\{ \begin{array}{lll} m\!=\!l\!-\!n & ~\text{if} & n\!\leq\! l\\ {}&{}&{}\\ m\!=\!N\!-\!(n\!-\!l) & ~\text{if} & n\!>\!l\end{array} \right . \!\!.
\end{align}
With the $N$ different choices of the phase in~\eqref{r_k solution}, corresponding to $l\in\{0,1,2,\ldots,N-1\}$, we get $N$ cyclic permutations of the rows of $\mathbf{T}(\mathbf{I})$. This means that for any choice of $n\in\{0,1,2,\ldots,N-1\}$, we can choose an $l$, such that we get any $m\in\{0,1,2,\ldots,N-1\}$, which solves the routing problem. Note that $r_0(\omega)=1$ independently of our choice of $l$, so we only need $N-1$ tunable cavities.

At this point, we also note that only $N$ out of the $N!$ possible $N\times N$ permutation matrices may be realized by the solution provided here. We have solved the pairwise routing problem, but not the realization of arbitrary permutation matrices.

\section{Supplementary Note 2: General routing theorems and three-port reachable-region equivalence}
\label{sec:general-routing-theorems}

This note gives three structural results and one quantitative consequence.
First, fixed reciprocal phase delays between an MMI and its reflectors do not
change its reachable set of output-power distributions. Second, every lossless
equal-split $N\times N$ MMI supports unit-efficiency routing between any selected
input--output pair, without requiring the MMI to be a DFT. Third, for
$3\times3$ devices, every lossless equal-split MMI has the same reachable
output-power region as the DFT, up to a fixed relabeling of ports. Finally, we
derive this common three-port region analytically, including the sharp
$(4/9,4/9,1/9)$ balanced two-output limit.

All theorems in this note, together with their supporting algebraic lemmas,
have been formalized and machine-checked in Lean~4 against Mathlib. The formal
development is maintained in the GitHub repository cited under Code
availability and includes checks of the
explicit three-port field identities and the sharp balanced-splitting result
derived below.

Throughout this note, external ports are indexed by $n,m,q$ and the internal
reflector arms by $j$.  The forward MMI matrix is $\mathbf{M}$, reciprocity
makes the reverse matrix $\mathbf{M}^{\mathsf{T}}$, and
\begin{align}
 \mathbf{D}&=\operatorname{diag}(d_0,\ldots,d_{N-1}),
 &d_j&=e^{i\delta_j},\notag\\
 \mathbf{R}&=\operatorname{diag}(r_0,\ldots,r_{N-1}),
 &|r_j|&=1
 \eqlab{general D and R}
\end{align}
denote the fixed arm phases and the controllable reflector phases, respectively.
With ideal components, the round-trip matrix is
\begin{align}
 \mathbf{T}_{\mathbf{M},\mathbf{D}}(\mathbf{R})
 =\mathbf{M}^{\mathsf{T}}\mathbf{D}^{\mathsf{T}}
  \mathbf{R}\mathbf{D}\mathbf{M}.
 \eqlab{general round trip}
\end{align}
For a fixed input port $n$, the basis vector $\mathbf{e}_n$ represents unit
excitation at that port. For a chosen reflector configuration $\mathbf{R}$,
the product $\mathbf{T}_{\mathbf{M},\mathbf{D}}(\mathbf{R})\mathbf{e}_n$ is the
complex output-field vector. Its $q$th component is the field amplitude at
output port $q$, so the corresponding output power is
\begin{align}
 p_q(\mathbf{R};n)
 &:=\Big|[\mathbf{T}_{\mathbf{M},\mathbf{D}}(\mathbf{R})\mathbf{e}_n]_q\Big|^2,
 &q&=0,\ldots,N-1.
\end{align}
Collecting these components gives the output-power vector
$\mathbf{p}(\mathbf{R};n)=(p_0(\mathbf{R};n),\ldots,p_{N-1}(\mathbf{R};n))$.
The reachable power set is
\begin{align}
 \mathcal{P}_{\mathbf{M},\mathbf{D}}(n)
 :=\left\{\mathbf{p}(\mathbf{R};n):
 \mathbf{R}\ \text{diagonal and unitary}\right\}.
 \eqlab{general reachable set}
\end{align}
Thus, $\mathcal{P}_{\mathbf{M},\mathbf{D}}(n)$ contains every output-power
distribution obtainable by varying the independently controllable reflector
phases while holding the input port $n$ fixed.

\subsection{Fixed diagonal phases preserve the reachable set}

\begin{theorem}[Fixed-phase invariance]
\label{thm:fixed-phase-invariance}
Let $\mathbf{M}$ be any $N\times N$ matrix and let $\mathbf{D}$ be any fixed
diagonal unitary matrix.  Then, for every input port $n$,
\begin{align}
 \mathcal{P}_{\mathbf{M},\mathbf{D}}(n)
 =\mathcal{P}_{\mathbf{M},\mathbf{I}}(n).
 \eqlab{fixed phase reachable equality}
\end{align}
In fact, the sets of reachable round-trip matrices are identical.
No equal-splitting or DFT assumption is required for this invariance.
\end{theorem}

\begin{proof}
Because $\mathbf{D}$ is diagonal,
$\mathbf{D}^{\mathsf{T}}=\mathbf{D}$, and because both $\mathbf{D}$ and
$\mathbf{R}$ are diagonal they commute.  Consequently
\begin{align}
 \mathbf{D}^{\mathsf{T}}\mathbf{R}\mathbf{D}
 =\mathbf{D}\mathbf{R}\mathbf{D}
 =\mathbf{D}^2\mathbf{R}.
 \eqlab{D squared R}
\end{align}
Introduce the effective reflector matrix
\begin{align}
 \mathbf{R}'=\mathbf{D}^2\mathbf{R}.
 \eqlab{effective reflector general}
\end{align}
It is diagonal and unitary.  Moreover, the map
$\mathbf{R}\mapsto\mathbf{R}'$ is a bijection on the set of diagonal unitary
matrices: for any desired $\mathbf{R}'$, its unique preimage is
$\mathbf{R}=\mathbf{D}^{-2}\mathbf{R}'$.  Hence
\begin{align}
 \left\{
 \mathbf{M}^{\mathsf{T}}\mathbf{D}^{\mathsf{T}}
 \mathbf{R}\mathbf{D}\mathbf{M}:\mathbf{R}\ \text{diagonal unitary}
 \right\}
 =
 \left\{
 \mathbf{M}^{\mathsf{T}}\mathbf{R}'\mathbf{M}:
 \mathbf{R}'\ \text{diagonal unitary}
 \right\}.
 \eqlab{round trip set equality}
\end{align}
Acting on $\mathbf{e}_n$ and taking componentwise squared magnitudes gives
\eqref{fixed phase reachable equality}.
\end{proof}

The theorem assumes that all $N$ diagonal phases can be varied.  Only $N-1$
independent controls are needed for powers.  Indeed, multiplying every
reflector coefficient by the same $e^{i\vartheta}$ multiplies the complete
output field by $e^{i\vartheta}$ and leaves every output power unchanged.
Every diagonal unitary can therefore be represented, up to this irrelevant
common phase, by one whose reference entry is unity.

\subsection{Pairwise routing for any equal-split unitary and any \texorpdfstring{$N$}{N}}

Supplementary Note~1 establishes pairwise routing for a DFT MMI. We now show
that the DFT structure is unnecessary: equal splitting and unitarity alone
suffice.

By the fixed-phase-invariance theorem, the transformation
$\mathbf{R}_0=\mathbf{D}^{\mathsf{T}}\mathbf{R}\mathbf{D}$ is a bijection on
the set of diagonal unitary reflector matrices. We may therefore set
$\mathbf{D}=\mathbf{I}$ while deriving the routing solution. For a device with
fixed arm phases $\mathbf{D}$, the corresponding physical reflector matrix is
$\mathbf{R}=(\mathbf{D}^{\mathsf{T}})^{-1}\mathbf{R}_0\mathbf{D}^{-1}
=\mathbf{D}^{-2}\mathbf{R}_0$.

\begin{theorem}[General pairwise routing]
\label{thm:general-pairwise-routing}
Let $\mathbf{M}\in U(N)$ be lossless and equal-split,
\begin{align}
 |M_{jk}|=\frac{1}{\sqrt{N}},\qquad j,k=0,\ldots,N-1.
 \eqlab{general equal split}
\end{align}
For every input port $n$ and target output port $m$, there is a diagonal
phase-only reflector matrix $\mathbf{R}_0$ such that
\begin{align}
 \mathbf{M}^{\mathsf{T}}\mathbf{R}_0\mathbf{M}\mathbf{e}_n
 =e^{i\theta}\mathbf{e}_m,
 \eqlab{general exact route}
\end{align}
where $\theta$ is an arbitrary common phase. Thus, all $N(N+1)/2$ distinct
reciprocal routing connections are individually reachable with unit efficiency
in the ideal case: $N(N-1)/2$ connect different pairs of ports, and the
remaining $N$ are self-reflections.
\end{theorem}

\begin{proof}
Follow the input field through the device:
\begin{align}
 \mathbf{e}_n
 \xrightarrow{\mathbf{M}}\mathbf{M}\mathbf{e}_n
 \xrightarrow{\mathbf{R}_0}\mathbf{R}_0\mathbf{M}\mathbf{e}_n
 \xrightarrow{\mathbf{M}^{\mathsf{T}}}
 \mathbf{M}^{\mathsf{T}}\mathbf{R}_0\mathbf{M}\mathbf{e}_n.
 \eqlab{general propagation chain}
\end{align}
The amplitude at an arbitrary output port $q$ is
\begin{align}
 z_q
 :=[\mathbf{M}^{\mathsf{T}}\mathbf{R}_0\mathbf{M}\mathbf{e}_n]_q
 =\sum_{j=0}^{N-1}M_{jq}\,r_{0,j}\,M_{jn}.
 \eqlab{general output component}
\end{align}
Choose the reflector coefficients arm by arm as
\begin{align}
 r_{0,j}=e^{i\theta}\frac{\overline{M_{jm}}}{M_{jn}},
 \eqlab{general reflector solution}
\end{align}
where the overline denotes complex conjugation.
The division is well-defined because equal splitting makes every $M_{jn}$
non-zero.  More importantly, \eqref{general reflector solution} is
phase-only:
\begin{align}
 |r_{0,j}|=\frac{|M_{jm}|}{|M_{jn}|}
 =\frac{1/\sqrt{N}}{1/\sqrt{N}}=1.
 \eqlab{general reflector unit modulus}
\end{align}
Substitution into \eqref{general output component} gives
\begin{align}
 z_q
 &=e^{i\theta}\sum_{j=0}^{N-1}
 M_{jq}\overline{M_{jm}}\notag\\
 &=e^{i\theta}\delta_{qm}.
 \eqlab{general routing verification}
\end{align}
The final identity is the orthogonality of columns $q$ and $m$ of the unitary
matrix $\mathbf{M}$:
$\sum_jM_{jq}\overline{M_{jm}}=\delta_{qm}$. Therefore, the desired output has
unit amplitude and every other output has zero amplitude, proving
\eqref{general exact route}.
\end{proof}

For exact phase-only routing between every input--output pair, the relevant
column magnitudes must satisfy
\begin{align}
 |M_{jn}|=|M_{jm}|\qquad\text{for all }j,n,m.
\end{align}
Thus, for each row $j$, all $N$ matrix elements have the same magnitude.
Because $\mathbf{M}$ is unitary, each row has unit norm, so
\begin{align}
 \sum_{k=0}^{N-1}|M_{jk}|^2=1
\end{align}
implies
\begin{align}
 |M_{jk}|=\frac{1}{\sqrt{N}}\qquad\text{for all }j,k.
\end{align}
Equal splitting is therefore necessary for exact phase-only routing between
every pair of ports, and the construction above shows that it is also
sufficient. The arbitrary common phase $\theta$ may be chosen so that one
selected $r_{0,j}$ is unity, leaving $N-1$ independently controlled reflector
phases.

For the DFT convention of Supplementary Note~1,
$M_{jn}=N^{-1/2}e^{-i2\pi jn/N}$, the general solution reduces to
\begin{align}
 r_{0,j}=e^{i\theta}e^{i2\pi j(m+n)/N},
 \eqlab{general theorem DFT corollary}
\end{align}
which is the Fourier phase ramp after the fixed arm phases have been absorbed.

\subsection{Simultaneous port swaps, self-reflections, and routing schedules}
\label{sec:simultaneous-pairings}

Pairwise routing completeness means that any selected input--output pair can be connected. It does not require that only one
input be illuminated: a fixed phase setting acts on all inputs simultaneously, although the resulting
connections cannot generally be chosen independently. Absorbing fixed arm
phases into $\mathbf{R}$, the lossless round-trip matrix
$\mathbf{T}=\mathbf{M}^{\mathsf{T}}\mathbf{R}\mathbf{M}$ is unitary and symmetric.
We call a simultaneous routing configuration exact when every input $n$ is
routed with unit efficiency to a single output $\pi(n)$, so that
\begin{align}
 |T_{mn}|^2=\delta_{m,\pi(n)}.
\end{align}
Equivalently, the power-transfer matrix $\mathbf{P}=|\mathbf{T}|^2$ is a
permutation matrix. Reciprocity makes $\mathbf{P}$ symmetric. A symmetric
permutation matrix satisfies $\mathbf{P}^2=\mathbf{I}$, so its permutation is
an involution. An involution is a permutation that returns every port to
itself when applied twice; its cycles can therefore only have length one or
two. Thus, every exact simultaneous routing configuration consists of
reciprocal two-port swaps and single-port self-reflections, and cycles
involving three or more ports are excluded. For three ports, an exact swap
necessarily leaves the third port fully self-reflected.

\paragraph{Discrete Fourier Transform MMI.}
With the fixed arm phases absorbed into $\mathbf{R}$, define the Fourier
coefficients of the reflector settings as
\begin{align}
 c_s=\frac{1}{N}\sum_{j=0}^{N-1}r_j e^{-i2\pi js/N}.
 \eqlab{simultaneous Fourier coefficients}
\end{align}
For the DFT MMI, the round-trip matrix elements are then
\begin{align}
 T_{mn}=c_{(m+n)\bmod N}.
 \eqlab{simultaneous Fourier matrix entry}
\end{align}

From \eqref{general theorem DFT corollary}, after absorbing the fixed arm
phases, the reflector setting for routing between ports $n$ and $m$ depends
only on
\begin{align}
 l=(m+n)\bmod N.
 \eqlab{Fourier link setting}
\end{align}
The same setting routes every input $n$ to the output
\begin{align}
 m=(l-n)\bmod N.
 \eqlab{simultaneous Fourier permutations}
\end{align}
For a fixed setting $l$, if $m=n$, that port is self-reflected; otherwise,
$n$ and $m$ form a reciprocal pair.
For $N=3$, the settings $l=0,1,2$ respectively produce
$0\to0$ with $1\leftrightarrow2$, $2\to2$ with
$0\leftrightarrow1$, and $1\to1$ with $0\leftrightarrow2$. In the physical
one-based port labels, these are the three possible reciprocal swaps, each
with the remaining port self-reflected.
Thus these $N$ permutations exhaust the exact routing configurations,
up to a common phase. For odd $N$, each has one self-reflection and
$(N-1)/2$ swaps. For even $N$, odd $l$ gives $N/2$ swaps; even $l$ gives
two self-reflections and $(N-2)/2$ swaps, since fixed ports satisfy
$2n=l\pmod N$.
Because every exact reciprocal configuration is an involution, at most
$\lfloor N/2\rfloor$ swaps are possible and at least $N\bmod2$ ports must
self-reflect. The Fourier construction attains this bound for every pair
when $N$ is odd and, when $N$ is even, for opposite-parity pairs; a
same-parity pair instead gives two self-reflections, a restriction of the
Fourier family rather than a fundamental requirement. By the DFT equivalence
proved below, the same three-port result holds, up to port phases and
relabeling, for every equal-split unitary $3\times3$ MMI.
For $N=5$, the Fourier family supports five configurations, one for each
self-reflected port, out of the 15 possible configurations comprising two
swaps and one self-reflection. Once the self-reflected port is chosen, the
Fourier construction fixes the other two pairs; without that restriction,
the remaining four ports admit three pairings, giving $5\times3=15$
possibilities in total.

\paragraph{An all-to-all schedule for every $N$.}
As shown in Supplementary Note~1, the $N$ DFT reflector settings generate
permutation matrices satisfying $m=l-n\pmod N$. Hence, for any input--output
pair $(n,m)$, the unique required setting is $l=n+m\pmod N$. The $N$ settings
therefore route every input to every output exactly once. Thus, as $l$ runs
from $0$ to $N-1$, each of the $\binom{N}{2}$ possible reciprocal pairings of
distinct ports occurs exactly once.

We consider the scheduling problem of realizing all $\binom{N}{2}$ reciprocal
pairings between distinct ports using the fewest reflector configurations.
The combinatorial minimum is $N$ configurations for odd $N$ and $N-1$ for
even $N$. The DFT schedule uses $N$ configurations and is therefore optimal
for odd $N$, but uses one more configuration than the combinatorial minimum
for even $N$. For $N=2^k$, the Walsh construction discussed below attains the
$N-1$ minimum. Each port self-reflects in exactly one of the $N$ DFT settings,
namely $l=2n$.
The five settings and their combined pair coverage are illustrated in Supplementary Fig.~\ref{fig:Fourier-five-schedule}.

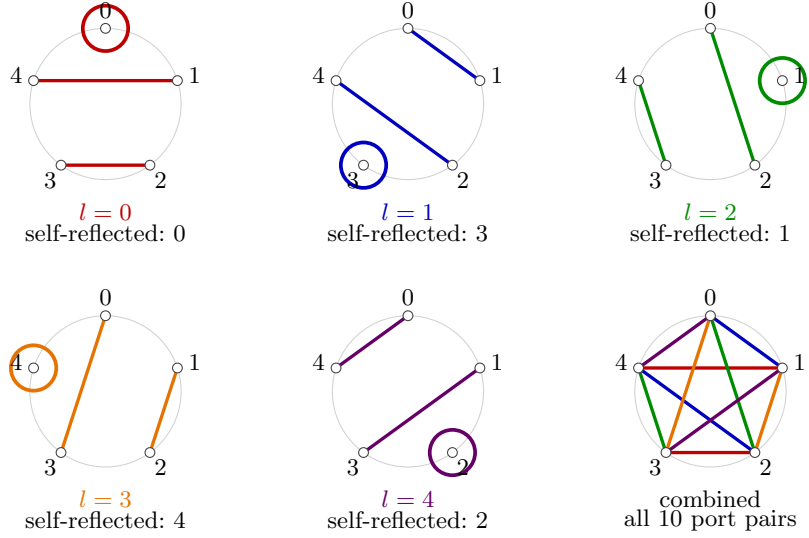
\begin{figure}[htbp]
 \centering
 \begin{tikzpicture}[
   port/.style={circle,draw=black!75,fill=white,inner sep=1.3pt},
   link/.style={line width=1.25pt},
   self/.style={circle,line width=1.4pt,minimum size=6mm}]
  \newcommand{\dftfivepanel}[6]{\begin{scope}[shift={(#1,#2)}]
    \draw[gray!35] (0,0) circle (1);
    \foreach \u/\v in {#4}
      \draw[link,#6] (90-72*\u:1) -- (90-72*\v:1);
    \node[self,draw=#6] at (90-72*#5:1) {};
    \foreach \n in {0,...,4} {
      \node[port] at (90-72*\n:1) {};
      \node[font=\scriptsize] at (90-72*\n:1.24) {\n};
    }
    \node[font=\small,text=#6] at (0,-1.42) {$l=#3$};
    \node[font=\scriptsize] at (0,-1.70) {self-reflected: $#5$};
   \end{scope}}
  \dftfivepanel{0}{2.05}{0}{1/4,2/3}{0}{red!75!black}
  \dftfivepanel{4.00}{2.05}{1}{0/1,2/4}{3}{blue!75!black}
  \dftfivepanel{8.00}{2.05}{2}{0/2,3/4}{1}{green!55!black}
  \dftfivepanel{0}{-1.75}{3}{0/3,1/2}{4}{orange!90!black}
  \dftfivepanel{4.00}{-1.75}{4}{0/4,1/3}{2}{violet!80!black}
  \begin{scope}[shift={(8.00,-1.75)}]
   \draw[gray!35] (0,0) circle (1);
   \foreach \u/\v in {1/4,2/3} \draw[link,red!75!black] (90-72*\u:1)--(90-72*\v:1);
   \foreach \u/\v in {0/1,2/4} \draw[link,blue!75!black] (90-72*\u:1)--(90-72*\v:1);
   \foreach \u/\v in {0/2,3/4} \draw[link,green!55!black] (90-72*\u:1)--(90-72*\v:1);
   \foreach \u/\v in {0/3,1/2} \draw[link,orange!90!black] (90-72*\u:1)--(90-72*\v:1);
   \foreach \u/\v in {0/4,1/3} \draw[link,violet!80!black] (90-72*\u:1)--(90-72*\v:1);
   \foreach \n in {0,...,4} {
    \node[port] at (90-72*\n:1) {};
    \node[font=\scriptsize] at (90-72*\n:1.24) {\n};
   }
   \node[font=\small] at (0,-1.42) {combined};
   \node[font=\scriptsize] at (0,-1.70) {all 10 port pairs};
  \end{scope}
 \end{tikzpicture}
 \caption{Five-setting all-to-all schedule of the five-port Fourier router.
 The first five panels are the switching configurations; the sixth panel
 combines their color-coded reciprocal pairings and is not itself one switch setting.
 Setting $l$ implements $n\mapsto l-n\pmod 5$.  Each pentagon contains two
 simultaneous reciprocal pairings and one self-reflected port (colored ring),
 which is unavoidable for odd $N$.  Across the five panels, every side and
 every diagonal of the pentagon occurs exactly once, so all ten port pairs
 are connected in the minimum possible five settings.}
 \label{fig:Fourier-five-schedule}
\end{figure}

\paragraph{Walsh interferometer.}
A Walsh interferometer implements the normalized Walsh--Hadamard transform, whose entries
have equal magnitude and phases restricted to $0$ or $\pi$. It is defined for
$N=2^k$ ports. We label each port index $n$ and arm index $j$ by the $k$-bit
binary representation of its integer index, so $n,j\in\mathbb{F}_2^k$. We
denote the Walsh matrix by $\bm{W}$, with entries
\begin{align*}
 W_{jn}=N^{-1/2}(-1)^{j\cdot n},
\end{align*}
where $j\cdot n=\sum_{t=0}^{k-1}j_tn_t\pmod 2$ is the parity of the positions
in which both binary labels contain a one. Thus $W_{jn}$ is
$+1/\sqrt{N}$ for even parity and $-1/\sqrt{N}$ for odd parity.

The symbol $\oplus$ denotes the exclusive-OR (XOR) of the binary labels,
performed independently on every bit:
$0\oplus0=1\oplus1=0$ and $0\oplus1=1\oplus0=1$. For example, in an
eight-port device, $3\oplus5=011_2\oplus101_2=110_2=6$.
Following the field from input $n$ through arm $j$ and back to output $m$ gives
\begin{align*}
 T_{mn}
 &=\sum_j W_{jm}r_jW_{jn}\\
 &=\frac{1}{N}\sum_j r_j
   (-1)^{j\cdot m}(-1)^{j\cdot n}\\
 &=\frac{1}{N}\sum_j r_j(-1)^{j\cdot(m\oplus n)}.
\end{align*}
The last equality follows because the exponent matters only modulo two:
\begin{align*}
 j\cdot m+j\cdot n
 &\equiv \sum_t j_t(m_t+n_t)
 \equiv \sum_t j_t(m_t\oplus n_t)
 =j\cdot(m\oplus n)\pmod 2.
\end{align*}
We define the Walsh-transform coefficients of the reflector settings by
\begin{align}
 T_{mn}=b_{m\oplus n},\qquad
 b_a=\frac{1}{N}\sum_j r_j(-1)^{j\cdot a}.
 \eqlab{simultaneous Walsh coefficients}
\end{align}
Like $c_s$ in the Fourier construction, $b_a$ is a coherently summed field
coefficient, not a physical component. Selecting a binary mask $a$ and setting
the reflector coefficients to
\begin{align*}
 r_j=e^{i\theta}(-1)^{j\cdot a}
\end{align*}
gives, for an arbitrary binary label $u$,
\begin{align*}
 b_u
 &=\frac{1}{N}\sum_j e^{i\theta}(-1)^{j\cdot a}(-1)^{j\cdot u}\\
 &=\frac{e^{i\theta}}{N}\sum_j(-1)^{j\cdot(a\oplus u)}.
\end{align*}
In the second line we used
$j\cdot a+j\cdot u=j\cdot(a\oplus u)\pmod 2$. If $u=a$, then
$a\oplus u=0$, so all $N$ terms in the sum equal $+1$. If $u\ne a$,
choose a bit position $t$ for which $(a\oplus u)_t=1$. Each term indexed
by $j$ then cancels the term indexed by $j\oplus\mathbf{e}_t$, where
$\mathbf{e}_t$ has a one only in position $t$. Hence
\begin{align}
 b_u&=e^{i\theta}\delta_{u,a},\notag\\
 T_{mn}&=b_{m\oplus n}
 =e^{i\theta}\delta_{m\oplus n,a}
 =e^{i\theta}\delta_{m,n\oplus a}.
 \eqlab{simultaneous Walsh permutations}
\end{align}
The last equality follows by applying XOR with $n$ to
$m\oplus n=a$, using $n\oplus n=0$. Consequently, every input $n$ exits
from the unique output
$m=n\oplus a$. For example, for $N=4$ the labels are
$00,01,10,11$; choosing $a=01$ simultaneously pairs
$00\leftrightarrow01$ and $10\leftrightarrow11$.

There are exactly $N$ configurations: identity for $a=0$, and $N-1$
complete pairings without self-reflections for $a\ne0$. For even $N$,
reciprocity permits $(N-1)!!=(N-1)(N-3)\cdots 3\cdot1$ reflection-free permutation
matrices: the lowest-index unpaired port has $N-1$ possible partners, the
lowest-index remaining port then has $N-3$, and so on. The $N-1$ Walsh
matchings form an optimal all-to-all schedule: no setting can contain more
than $N/2$ disjoint swaps, and these matchings together cover each of the
$\binom{N}{2} = N(N-1)/2$ unordered port pairs exactly once. Including the $N$ self-reflections realized by the identity setting $a=0$, the full schedule covers all $N(N+1)/2$ unordered input--output connections. Indeed, the swap
$n\leftrightarrow m$ is obtained with $a=n\oplus m\ne0$, while a fixed point
would require $n\oplus a=n$ and hence $a=0$. A Walsh $4\times4$ therefore
supports all three complete pairings, but cannot perform just one swap while
self-reflecting the other two ports. For larger devices it still selects only
a subset of complete matchings: Walsh $8\times8$ supports seven of the
$7\times5\times3=105$ possibilities, although each pair occurs in
one of those seven reflection-free matchings.

\paragraph{The $N=8$ schedule and its binary drive.}
Table~\ref{tab:Walsh-eight-schedule} writes the complete schedule explicitly.
Ports and arms are labeled by the integers $0,\ldots,7$, identified with
their three-bit binary representations. In row $a$, port $n$ is paired with
$n\oplus a$, while arm $j$ receives a relative $\pi$ phase precisely when
the binary inner product $j\mathbin{\cdot}a$ is odd. Arm 0 is used as the
phase reference and therefore remains at relative phase zero in every row.
\begin{table*}[htbp]
 \centering
 \small
 \setlength{\tabcolsep}{4.2pt}
 \begin{tabular}{@{}c l *{7}{c}@{}}
  \toprule
  & & \multicolumn{7}{c}{programmable arm $j$} \\
  \cmidrule(l){3-9}
  routing mask $a$ & simultaneous port pairs $\{n,n\oplus a\}$
    & 1 & 2 & 3 & 4 & 5 & 6 & 7 \\
  & binary arm label $j$
    & $001$ & $010$ & $011$ & $100$ & $101$ & $110$ & $111$ \\
  \midrule
  $001$ & $0$--$1$, $2$--$3$, $4$--$5$, $6$--$7$
    & $\pi$ & $0$ & $\pi$ & $0$ & $\pi$ & $0$ & $\pi$ \\
  $010$ & $0$--$2$, $1$--$3$, $4$--$6$, $5$--$7$
    & $0$ & $\pi$ & $\pi$ & $0$ & $0$ & $\pi$ & $\pi$ \\
  $011$ & $0$--$3$, $1$--$2$, $4$--$7$, $5$--$6$
    & $\pi$ & $\pi$ & $0$ & $0$ & $\pi$ & $\pi$ & $0$ \\
  $100$ & $0$--$4$, $1$--$5$, $2$--$6$, $3$--$7$
    & $0$ & $0$ & $0$ & $\pi$ & $\pi$ & $\pi$ & $\pi$ \\
  $101$ & $0$--$5$, $1$--$4$, $2$--$7$, $3$--$6$
    & $\pi$ & $0$ & $\pi$ & $\pi$ & $0$ & $\pi$ & $0$ \\
  $110$ & $0$--$6$, $1$--$7$, $2$--$4$, $3$--$5$
    & $0$ & $\pi$ & $\pi$ & $\pi$ & $\pi$ & $0$ & $0$ \\
  $111$ & $0$--$7$, $1$--$6$, $2$--$5$, $3$--$4$
    & $\pi$ & $\pi$ & $0$ & $\pi$ & $0$ & $0$ & $\pi$ \\
  \bottomrule
 \end{tabular}
 \caption{Seven-setting all-to-all schedule and binary phase drive for an
 eight-port Walsh router. Every row is a reflection-free complete matching,
 and every one of the 28 reciprocal pairings occurs in exactly one row. The entries give
 the reflector phase on each programmable arm relative to arm 0. Equivalently,
 arm $j$ is set to $\pi$ when $j\mathbin{\cdot}a$ is odd and to zero otherwise.
 The omitted setting $a=000$ is the identity and sets every arm to the same
 phase.}
 \label{tab:Walsh-eight-schedule}
\end{table*}

In the phase-reference convention of Table~\ref{tab:Walsh-eight-schedule},
moving between any two distinct switch configurations, including a transition
to or from the identity, changes exactly four of the eight arm phases, that is,
half of them. Because the zero-labeled arm is fixed as the phase reference,
all four changes occur among the seven programmable arms shown in the table.
Indeed, configurations $a$ and $a'$ differ on arm $j$ precisely when
$j\mathbin{\cdot}(a\oplus a')=1$. For $a\ne a'$, this nonzero binary linear
functional is balanced over the eight three-bit arm labels, taking the value
one for exactly four labels.
The seven nonidentity matchings and their combined pair coverage are illustrated in Supplementary Fig.~\ref{fig:Walsh-eight-schedule}.

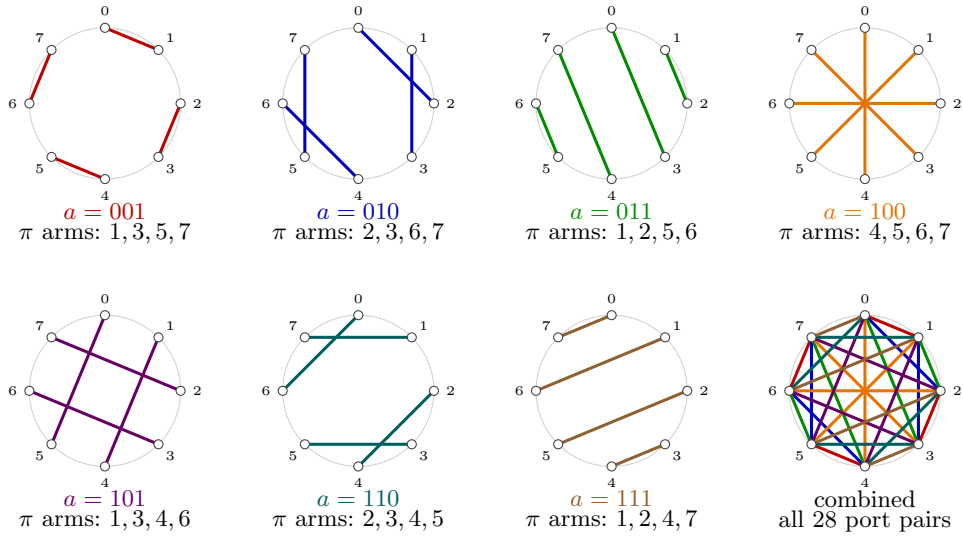
\begin{figure}[htbp]
 \centering
 \begin{tikzpicture}[
   port/.style={circle,draw=black!75,fill=white,inner sep=1.2pt},
   link/.style={line width=1.15pt}]
  \newcommand{\walshpanel}[6]{\begin{scope}[shift={(#1,#2)}]
    \draw[gray!35] (0,0) circle (1);
    \foreach \u/\v in {#4}
      \draw[link,#6] (90-45*\u:1) -- (90-45*\v:1);
    \foreach \n in {0,...,7} {
      \node[port] at (90-45*\n:1) {};
      \node[font=\tiny] at (90-45*\n:1.22) {\n};
    }
    \node[font=\small,text=#6] at (0,-1.42) {$a=#3$};
    \node[font=\scriptsize] at (0,-1.72) {$\pi$ arms: $#5$};
   \end{scope}}
  \walshpanel{0}{2.15}{001}{0/1,2/3,4/5,6/7}{1,3,5,7}{red!75!black}
  \walshpanel{3.35}{2.15}{010}{0/2,1/3,4/6,5/7}{2,3,6,7}{blue!75!black}
  \walshpanel{6.70}{2.15}{011}{0/3,1/2,4/7,5/6}{1,2,5,6}{green!55!black}
  \walshpanel{10.05}{2.15}{100}{0/4,1/5,2/6,3/7}{4,5,6,7}{orange!90!black}
  \walshpanel{0}{-1.65}{101}{0/5,1/4,2/7,3/6}{1,3,4,6}{violet!80!black}
  \walshpanel{3.35}{-1.65}{110}{0/6,1/7,2/4,3/5}{2,3,4,5}{teal!75!black}
  \walshpanel{6.70}{-1.65}{111}{0/7,1/6,2/5,3/4}{1,2,4,7}{brown!75!black}
  \begin{scope}[shift={(10.05,-1.65)}]
   \draw[gray!35] (0,0) circle (1);
   \foreach \u/\v in {0/1,2/3,4/5,6/7} \draw[link,red!75!black] (90-45*\u:1)--(90-45*\v:1);
   \foreach \u/\v in {0/2,1/3,4/6,5/7} \draw[link,blue!75!black] (90-45*\u:1)--(90-45*\v:1);
   \foreach \u/\v in {0/3,1/2,4/7,5/6} \draw[link,green!55!black] (90-45*\u:1)--(90-45*\v:1);
   \foreach \u/\v in {0/4,1/5,2/6,3/7} \draw[link,orange!90!black] (90-45*\u:1)--(90-45*\v:1);
   \foreach \u/\v in {0/5,1/4,2/7,3/6} \draw[link,violet!80!black] (90-45*\u:1)--(90-45*\v:1);
   \foreach \u/\v in {0/6,1/7,2/4,3/5} \draw[link,teal!75!black] (90-45*\u:1)--(90-45*\v:1);
   \foreach \u/\v in {0/7,1/6,2/5,3/4} \draw[link,brown!75!black] (90-45*\u:1)--(90-45*\v:1);
   \foreach \n in {0,...,7} {
    \node[port] at (90-45*\n:1) {};
    \node[font=\tiny] at (90-45*\n:1.22) {\n};
   }
   \node[font=\small] at (0,-1.42) {combined};
   \node[font=\scriptsize] at (0,-1.72) {all 28 port pairs};
  \end{scope}
 \end{tikzpicture}
 \caption{The seven nonidentity settings of the eight-port Walsh router.
 The first seven panels are the switching configurations; the eighth panel
 combines their color-coded reciprocal pairings and is not itself one switch setting.
 Each octagon shows one reflection-free complete matching
 $n\leftrightarrow n\oplus a$; taken together, the panels contain every side
 and diagonal of the eight-vertex complete graph exactly once.  The line
 below each panel lists the arms whose reflectors receive the relative
 $\pi$ phase $(-1)^{j\cdot a}=-1$.  Port and arm labels use the same binary
 encoding but do not denote a direct physical port--arm connection.}
 \label{fig:Walsh-eight-schedule}
\end{figure}

Because the Walsh transform is invertible, the reflector pattern in
\eqref{simultaneous Walsh permutations} is unique up to its common phase.
After removal of that common phase, every reflector therefore uses only the two
values $+1$ and $-1$, or relative phases $0$ and $\pi$. The zero-labeled arm
is fixed, and the remaining $N-1$ arms are parity functions of the
$k=\log_2N$ address bits.

\paragraph{Why concentrating all self-reflections in the identity requires
$N=2^k$.}
Suppose a single-layer reciprocal router realizes one identity setting and
$N-1$ reflection-free complete matchings that together cover every port
pair. These $N$ settings send any chosen input once to each output. After
taking the identity setting as a reference, their diagonal reflector
patterns multiply arm by arm, so the corresponding routing permutations commute.

The product of two scheduled permutations also commutes with every schedule member. Choose the scheduled permutation that sends a reference input to the same output as this product. Since every other port is reached from the reference input by a scheduled permutation, commutativity forces the two permutations to agree on every port. The schedule is therefore closed under composition.

Reciprocity makes every reflection-free routing permutation a product of
disjoint swaps, so applying any such operation twice gives the identity.

The independent routing operations are therefore binary choices: each is
either used once or not used, and adding one independent choice doubles the
number of configurations. With $k$ independent choices the family contains
$2^k$ configurations. Since the assumed schedule contains exactly $N$,
\begin{align}
 N=2^k.
 \eqlab{binary schedule power two}
\end{align}
For six ports, for example, the identity and five matchings would require
six configurations, whereas two binary choices give four and three give
eight; six is impossible. This is an architectural, not a combinatorial,
restriction: five complete matchings do exist for six ports, but realizing
them requires a more general mixing/control architecture. The Walsh interferometer
construction attains the binary schedule when $N=2^k$: $a=0$ is the single
all-self-reflecting identity, while every nonzero XOR mask is
reflection-free. Other port counts can instead use the $N$-setting Fourier
schedule above.

\paragraph{Selected-pair routing versus simultaneous port permutations.}
Equal splitting alone guarantees phase-only routing of one selected input to
one selected output, as proved in \eqref{general reflector solution}.
For $N>3$ it does not by itself require the remaining $N-2$ inputs to route
one-to-one to distinct outputs without power splitting; they may instead
undergo a smaller unitary mixing. The
Fourier and Walsh matrices above are explicit constructions for which the
entire matrix is a permutation in every listed setting. More generally, if
a single-layer router does possess exact one-to-one routing settings
(permutations up to phases) covering every port pair, the binary-choice
argument above constrains their simultaneous
routing and self-reflection patterns. This distinction separates complete
point-to-point accessibility from the stronger ability to prescribe a
complete simultaneous matching.

\paragraph{Planar realization with balanced $2\times2$ unitaries.}
For $N=2^k$, the Walsh transform can be implemented as a radix-two
butterfly containing $k=\log_2N$ sequential layers. Each layer contains
$N/2$ balanced $2\times2$ unitaries that act in parallel on disjoint pairs
of modes, so every mode passes through one $2\times2$ unitary in each layer.
Writing $\bm{H}=2^{-1/2}\bigl(\begin{smallmatrix}1&1\\1&-1\end{smallmatrix}\bigr)$
and denoting the $d\times d$ identity by $\bm{I}_d$, the factorization is
\begin{align}
 \bm{W}_N=\bm{H}^{\otimes k}
 =\prod_{r=0}^{k-1}
 \left(\bm{I}_{2^r}\otimes\bm{H}\otimes
 \bm{I}_{2^{k-r-1}}\right).
 \eqlab{Walsh butterfly factorization}
\end{align}
Each factor mixes the $N/2$ disjoint pairs of modes whose binary labels
differ in one bit. The complete butterfly therefore contains
$(N/2)\log_2N$ passive $2\times2$ unitaries. Conversely, after $L$ layers of
disjoint two-mode couplers, one input can reach at most $2^L$ outputs.
Equal splitting among all $N$ outputs therefore requires
$L\geq\log_2N$, so this depth is minimal. Fixed crossings or mode
permutations provide the required butterfly connectivity. The DFT uses the
same network with phase shifts between layers.
All unitaries remain at a $50{:}50$ power split; only relative
phases are changed. Thus, this passive realization does not alter the
$O(N)$ number of actively controlled cavity reflectors in the routing
architecture.

For the four-port case, let a symmetric balanced $2\times 2$ unitary have the scattering
matrix $\mathbf Q$. Two such unitaries acting in parallel on disjoint pairs of modes give the block-diagonal stage
$\mathbf Q_{\rm s}$: 
\begin{align}
 \mathbf{Q}&=\frac{1}{\sqrt{2}}
 \begin{pmatrix}1&i\\ i&1\end{pmatrix},
 &
 \mathbf{Q}_{\rm s}
 &=\operatorname{diag}(\mathbf Q,\mathbf Q)
 =\frac{1}{\sqrt{2}}
 \begin{pmatrix}
  1&i&0&0\\
  i&1&0&0\\
  0&0&1&i\\
  0&0&i&1
 \end{pmatrix},
 \eqlab{four port physical MMI stage}
 \end{align}
Let a waveguide crossing exchange the two middle rails; its permutation matrix is
\begin{align}
 \mathbf{P}=\begin{pmatrix}
 1&0&0&0\\
 0&0&1&0\\
 0&1&0&0\\
 0&0&0&1
 \end{pmatrix}.
 \eqlab{four port planar shuffle}
\end{align}
The required phase-only interstage section is
\begin{align}
 \mathbf{\Phi}(\alpha)
 =\operatorname{diag}\!\left(1,-i,-i,-e^{i\alpha}\right),
 \eqlab{four port physical MMI phases}
\end{align}
corresponding, modulo $2\pi$, to phase settings
$(0,-\pi/2,-\pi/2,\alpha+\pi)$. With
$\mathbf{L}=\operatorname{diag}(1,i,1,i)$, direct multiplication gives
\begin{align}
 \mathbf{Q}_{\rm s}\mathbf{\Phi}(\alpha)
 \mathbf{P}\mathbf{Q}_{\rm s}
 =\mathbf{L}\mathbf{P}\mathbf{M}_4(\alpha)\mathbf{P}\mathbf{L},
 \eqlab{four port physical MMI product}
\end{align}
where multiplying from the left by $\mathbf{P}^{\mathsf{T}}\mathbf{L}^{-1}$ and from the right by $\mathbf{L}^{-1}\mathbf{P}^{\mathsf{T}}$
recovers
\begin{align}
 \mathbf{M}_4(\alpha)=\frac{1}{2}
 \begin{pmatrix}
 1&1&1&1\\
 1&e^{i\alpha}&-1&-e^{i\alpha}\\
 1&-1&1&-1\\
 1&-e^{i\alpha}&-1&e^{i\alpha}
 \end{pmatrix}.
 \eqlab{four port planar MMI family}
\end{align}
Hence $\alpha=0$ gives the Walsh matrix, $\alpha=\pi/2$ gives the
four-point DFT, and intermediate values continuously connect the two while
preserving equal splitting. The external diagonal phases do not change
routed powers and can either be absorbed into the port reference planes or
implemented as fixed planar phase sections. A compact single-layer layout is
shown in Supplementary Fig.~\ref{fig:four-port-planar-mmi-network}.

\begin{figure}[H]
 \centering
 \begin{tikzpicture}[x=0.95cm,y=0.82cm,line width=0.55pt,font=\small]
  \foreach \y/\lab in {3/0,2/2,1/1,0/3}{
    \draw (-0.10,\y) -- (1.275,\y);
    \node[left] at (-0.10,\y) {$x_{\lab}$};
  }

  \draw (1.275,1.70) rectangle (3.425,3.30);
  \draw (1.275,-0.30) rectangle (3.425,1.30);
  \node[align=center] at (2.35,2.50) {$2\!\times\!2$ \\[2pt]$50{:}50$};
  \node[align=center] at (2.35,0.50) {$2\!\times\!2$ \\[2pt]$50{:}50$};
  \foreach \y in {3,2,1,0}{\draw (3.425,\y) -- (4.75,\y);}

  \draw (4.75,3) -- (6.65,3);
  \draw (4.75,0) -- (6.65,0);
  \draw (4.75,2) -- (6.65,1);
  \draw (4.75,1) -- (6.65,2);

  \foreach \y/\ph in {3/0,2/{-\pi/2},1/{-\pi/2},0/{\alpha+\pi}}{
    \draw (6.65,\y) -- (7.625,\y);
    \draw[fill=white] (7.625,\y-0.26) rectangle (9.075,\y+0.26);
    \node at (8.35,\y) {$\ph$};
    \draw (9.075,\y) -- (10.325,\y);
  }

  \draw[fill=white] (10.325,1.70) rectangle (12.475,3.30);
  \draw[fill=white] (10.325,-0.30) rectangle (12.475,1.30);
  \node[align=center] at (11.40,2.50) {$2\!\times\!2$ \\[2pt]$50{:}50$};
  \node[align=center] at (11.40,0.50) {$2\!\times\!2$ \\[2pt]$50{:}50$};

  \foreach \y/\lab in {3/0,2/2,1/1,0/3}{
    \draw (12.475,\y) -- (13.85,\y);
    \node[right] at (13.85,\y) {$y_{\lab}$};
  }
 \end{tikzpicture}
 \caption{\textbf{Planar realization of a four-port MMI with balanced $2\times2$ unitaries.}
 The external ports use the bit-reversed order $(0,2,1,3)$, so only the two
 middle internal rails are exchanged by a waveguide crossing. The four unitaries remain balanced;
 the setting of the interstage phase, $\alpha$, selects the Walsh matrix ($\alpha=0$), the DFT matrix
 ($\alpha=\pi/2$), or an intermediate equal-split matrix.}
 \label{fig:four-port-planar-mmi-network}
\end{figure}
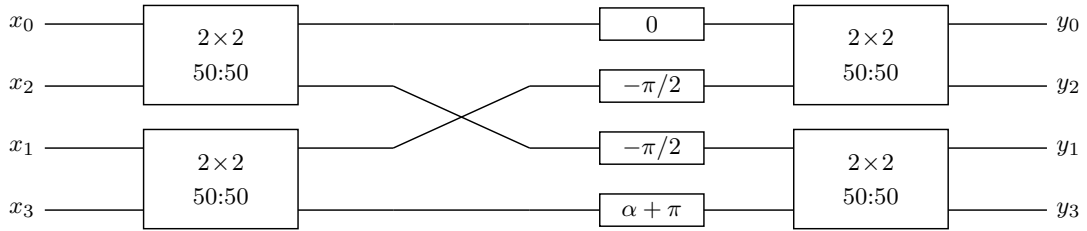

\subsection{Reachable-region equivalence for \texorpdfstring{$3\times3$}{3 x 3} equal-split MMIs}

Pairwise routing uses only the vertices of the output power simplex.  For
$3\times3$ equal-split MMIs, a stronger result holds: the complete interior reachable region is also independent of the MMI phases, up to a fixed port
permutation.

\begin{theorem}[$3\times3$ reachable-region equivalence]
\label{thm:three-port-region-universality}
Let $\mathbf{M}\in U(3)$ satisfy $|M_{jk}|=1/\sqrt{3}$, and let
$\mathbf{D}$ be any fixed diagonal unitary.  For every physical input port
$n$, there are a relabeled input index $n'\in\{0,1,2\}$ and a fixed
permutation $\Pi$ of the three output power coordinates such that
\begin{align}
 \mathcal{P}_{\mathbf{M},\mathbf{D}}(n)
 =\Pi\!\left(\mathcal{P}_{\mathbf{F},\mathbf{I}}(n')\right),
 \eqlab{three port region equality}
\end{align}
where $\mathbf{F}$ is the normalized $3\times3$ DFT matrix. Here $n'$ is the
input index in the canonical DFT representation corresponding to physical
port $n$; it is not an additional physical port. Thus fixed arm
phases and the internal phases of an arbitrary equal-split $3\times3$ MMI do
not change the shape of the reachable power region.
\end{theorem}

\begin{proof}
We first show that every equal-split $3\times3$ unitary is equivalent to the
DFT through diagonal phases and permutations.  Define
$\mathbf{H}=\sqrt{3}\,\mathbf{M}$.  Every entry of $\mathbf{H}$ has unit
modulus, and its rows and columns are mutually orthogonal.  Multiplying rows
and columns by unit-modulus constants preserves these properties.  We may
therefore multiply on the right by a diagonal phase matrix to make the first
row equal to $(1,1,1)$, and then multiply on the left by another diagonal phase
matrix to make the first column equal to $(1,1,1)^{\mathsf{T}}$.  The resulting
dephased matrix has the form
\begin{align}
 \mathbf{H}'=
 \begin{pmatrix}
  1&1&1\\
  1&a&b\\
  1&c&d
 \end{pmatrix},
 \qquad |a|=|b|=|c|=|d|=1.
 \eqlab{dephased three Hadamard}
\end{align}
Orthogonality of the first two rows gives $1+a+b=0$.  Since
$b=-(1+a)$ and $|a|=|b|=1$,
\begin{align}
 1=|b|^2=|1+a|^2=2+2\operatorname{Re}(a),
\end{align}
so $\operatorname{Re}(a)=-1/2$.  Together with $|a|=1$, this gives
$a\in\{\omega,\omega^2\}$, where $\omega=e^{i2\pi/3}$; then $b$ is the other
root.  After possibly exchanging the last two columns, the second row is
$(1,\omega,\omega^2)$.  Applying the same argument to the third row shows that
$(c,d)$ is also a permutation of $(\omega,\omega^2)$.  Orthogonality with the
second row forces the opposite ordering, so
\begin{align}
 \mathbf{H}'=
 \begin{pmatrix}
 1&1&1\\
 1&\omega&\omega^2\\
 1&\omega^2&\omega
 \end{pmatrix}
 =\sqrt{3}\,\mathbf{F},
 \eqlab{dephased matrix is Fourier}
\end{align}
up to row or column permutations associated with the DFT sign convention.
Undoing the dephasing proves the factorization
\begin{align}
 \mathbf{M}=\mathbf{D}_{\rm out}\mathbf{P}_{\rm out}
 \mathbf{F}\mathbf{P}_{\rm in}\mathbf{D}_{\rm in},
 \eqlab{three Hadamard factorization detailed}
\end{align}
where the $\mathbf{D}$ matrices are diagonal unitary and the $\mathbf{P}$
matrices are permutations.

By Theorem~\ref{thm:fixed-phase-invariance}, the separate fixed phase matrix
$\mathbf{D}$ in \eqref{general round trip} does not change the reachable
set, so it is sufficient to consider
$\mathbf{T}_{\mathbf{M},\mathbf{I}}(\mathbf{R})=
\mathbf{M}^{\mathsf{T}}\mathbf{R}\mathbf{M}$.  Substituting
\eqref{three Hadamard factorization detailed} gives
\begin{align}
 \mathbf{M}^{\mathsf{T}}\mathbf{R}\mathbf{M}
 ={}&\mathbf{D}_{\rm in}^{\mathsf{T}}
 \mathbf{P}_{\rm in}^{\mathsf{T}}\mathbf{F}^{\mathsf{T}}
 \mathbf{R}'\mathbf{F}\mathbf{P}_{\rm in}\mathbf{D}_{\rm in},
 \eqlab{three factorized round trip}
\end{align}
where
\begin{align}
 \mathbf{R}'=\mathbf{P}_{\rm out}^{\mathsf{T}}
 \mathbf{D}_{\rm out}^{\mathsf{T}}\mathbf{R}
 \mathbf{D}_{\rm out}\mathbf{P}_{\rm out}.
 \eqlab{three effective reflector}
\end{align}
The map $\mathbf{R}\mapsto\mathbf{R}'$ is again a bijection on diagonal
unitaries: diagonal phases are shifted, and the diagonal entries are reordered.

For a fixed physical input $n$, there exist a relabeled input index $n'$ and a phase
$e^{i\gamma_n}$ such that
\begin{align}
 \mathbf{P}_{\rm in}\mathbf{D}_{\rm in}\mathbf{e}_n
 =e^{i\gamma_n}\mathbf{e}_{n'}.
\end{align}
The factor $e^{i\gamma_n}$ is a common field phase.  On the output side,
$\mathbf{D}_{\rm in}^{\mathsf{T}}$ changes only the individual output phases,
and $\mathbf{P}_{\rm in}^{\mathsf{T}}$ applies a fixed permutation to the
output coordinates.  Neither operation changes the set of output powers,
apart from that permutation.  Finally, as $\mathbf{R}$ ranges over every
diagonal unitary, so does $\mathbf{R}'$.  Thus
\eqref{three factorized round trip} establishes \eqref{three port region equality}.
\end{proof}

\subsection{Analytic three-port region and the balanced-splitting limit}

The common reachable region of Theorem~\ref{thm:three-port-region-universality} can
be written explicitly. It is sufficient to use the DFT, the zero-based matrix
index $n=0$ (physical input port 1), and no fixed phases. The preceding
theorems then extend the result to every input, every fixed diagonal phase
matrix, and every equal-split $3\times3$ unitary MMI. A common reflector phase
is physically irrelevant, so write the
three effective arm phases as
\begin{align}
 (u_0,u_1,u_2)=(1,e^{i\alpha},e^{i\beta}),
 \qquad \alpha,\beta\in[0,2\pi).
 \eqlab{three effective phase coordinates}
\end{align}
With $\omega=e^{i2\pi/3}$, the three output amplitudes are
\begin{align}
 A_1(\alpha,\beta)
 &=\frac{1+e^{i\alpha}+e^{i\beta}}{3},\notag\\
 A_2(\alpha,\beta)
 &=\frac{1+\omega e^{i\alpha}+\omega^2e^{i\beta}}{3},\notag\\
 A_3(\alpha,\beta)
 &=\frac{1+\omega^2e^{i\alpha}+\omega e^{i\beta}}{3}.
 \eqlab{three DFT output amplitudes}
\end{align}
Thus, $P_k=|A_k|^2$ and the explicit two-phase power map is
\begin{align}
 P_1={}&\frac{1}{9}\left[3+2\cos\alpha+2\cos\beta
                         +2\cos(\alpha-\beta)\right],\notag\\
 P_2={}&\frac{1}{9}\left[3+2\cos\!\left(\alpha+\frac{2\pi}{3}\right)
                         +2\cos\!\left(\beta+\frac{4\pi}{3}\right)
                         +2\cos\!\left(\alpha-\beta-\frac{2\pi}{3}\right)
                   \right],\notag\\
 P_3={}&\frac{1}{9}\left[3+2\cos\!\left(\alpha+\frac{4\pi}{3}\right)
                         +2\cos\!\left(\beta+\frac{2\pi}{3}\right)
                         +2\cos\!\left(\alpha-\beta+\frac{2\pi}{3}\right)
                   \right].
 \eqlab{three DFT explicit power map}
\end{align}
Unitarity gives $P_1+P_2+P_3=1$, so the image lies in the two-dimensional
probability simplex shown in Supplementary
Fig.~\ref{fig:three-port-reachable-region}.

We next determine the boundary of this image.  Since the phase torus has no
boundary, the inverse function theorem implies that a boundary point of the
image must be the image of a critical point of the map
$(\alpha,\beta)\mapsto(P_1,P_2)$.  Direct differentiation of
\eqref{three DFT explicit power map} gives
\begin{align}
 \det\!\left[\frac{\partial(P_1,P_2)}{\partial(\alpha,\beta)}\right]
 =-\frac{2\sqrt{3}}{81}\bigl[
 &\sin(\alpha-2\beta)+\sin(\alpha+\beta)
  -\sin(2\alpha-\beta)\bigr].
 \eqlab{three DFT Jacobian}
\end{align}
The trigonometric factor has the useful factorization
\begin{align}
 &\sin(\alpha-2\beta)+\sin(\alpha+\beta)
  -\sin(2\alpha-\beta)\notag\\
 &\hspace{1cm}=2\sin\!\left(\alpha-\frac{\beta}{2}\right)
 \left[
  \cos\!\left(\frac{3\beta}{2}\right)
  -\cos\!\left(\alpha-\frac{\beta}{2}\right)
 \right].
 \eqlab{three DFT Jacobian factorization}
\end{align}
Modulo $2\pi$, the critical set is therefore the union of the three line
families
\begin{align}
 \alpha=\frac{\beta}{2}+k\pi,
 \qquad \alpha=2\beta+2k\pi,
 \qquad \alpha=-\beta+2k\pi.
 \eqlab{three DFT critical phase families}
\end{align}
Substitution into \eqref{three DFT explicit power map} shows that the three
families have the same image, differing only by cyclic reparametrization of
the power coordinates.  A single closed parametrization is
\begin{align}
 \mathbf{P}_{\partial}(t)=\frac{1}{9}\begin{pmatrix}
  [1+2\cos t]^2\\
  [1+2\cos(t-2\pi/3)]^2\\
  [1+2\cos(t+2\pi/3)]^2
 \end{pmatrix},
 \qquad 0\le t<2\pi.
 \eqlab{three DFT reachable boundary}
\end{align}
Away from these critical lines, the Jacobian is non-zero, so each regular
phase-domain component maps locally to an open subset of the simplex.  The
curve \eqref{three DFT reachable boundary} is simple between its three routing
cusps, and direct evaluation of any regular point (for example, a point near
the equal-power center) selects its bounded side. Hence, the reachable set is
exactly the closed three-arc region enclosed by
\eqref{three DFT reachable boundary}, rather than the full simplex.

\begin{figure*}[t]
 \centering
 \makebox[\textwidth][c]{\includegraphics[width=180mm]{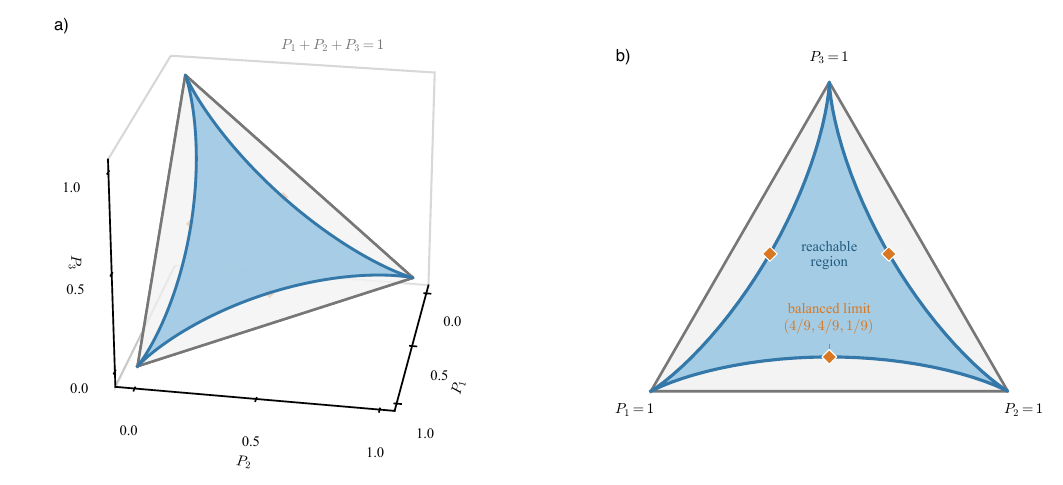}}
 \caption{\textbf{Reachable output powers of the ideal three-port router.}
 \textbf{a}, The lossless output powers lie on the conservation plane
 $P_1+P_2+P_3=1$; positivity restricts them to the triangular probability
 simplex, while phase-only control reaches the blue subset.
 \textbf{b}, Barycentric view of the simplex.  The blue curve is the exact
 analytic boundary in \eqref{three DFT reachable boundary}.  The orange
 diamonds mark the three cyclic balanced two-output optima, including
 $(4/9,4/9,1/9)$.}
 \label{fig:three-port-reachable-region}
\end{figure*}

\pagebreak
\begin{theorem}[Sharp balanced two-output limit]
\label{thm:three-port-balanced-limit}
For any reachable three-port power vector satisfying $P_1=P_2$, one has
\begin{align}
 P_3\ge\frac{1}{9},\qquad
 P_1=P_2\le\frac{4}{9}.
 \eqlab{three port balanced bound}
\end{align}
Both inequalities are attained simultaneously, so the optimum is
$(P_1,P_2,P_3)=(4/9,4/9,1/9)$.  The two other choices of balanced output pair
obey the corresponding cyclic statements.  These bounds hold for every
equal-split unitary $3\times3$ MMI, every input, and arbitrary fixed
reciprocal post-MMI phases.
\end{theorem}

\begin{proof}
Let $a=A_1$, $b=A_2$, and $c=A_3$.  Direct substitution of
\eqref{three DFT output amplitudes}, using $1+\omega+\omega^2=0$ and
$|u_j|=1$, gives the two field identities
\begin{align}
 |a|^2+|b|^2+|c|^2&=1,\notag\\
 a\overline{b}+b\overline{c}+c\overline{a}&=0.
 \eqlab{three field identities}
\end{align}
Suppose $|a|=|b|=s$.  Rearranging the second identity and applying the
triangle inequality gives
\begin{align}
 s^2=|a\overline b|
 &=|b\overline c+c\overline a|
 \le |b|\,|c|+|c|\,|a|=2s|c|.
 \eqlab{three balanced triangle bound}
\end{align}
If $s=0$, the result is immediate.  Otherwise, division by $s$ and squaring
give $s^2\le4|c|^2$.  Combining this with
$2s^2+|c|^2=1$ yields $|c|^2\ge1/9$ and $s^2\le4/9$.
The effective phase vector
\begin{align}
 (u_0,u_1,u_2)=(1,\omega,-\omega^2)
 \eqlab{three balanced witness phases}
\end{align}
produces $(|a|^2,|b|^2,|c|^2)=(4/9,4/9,1/9)$, proving sharpness for the DFT.
Fixed-phase invariance and
Theorem~\ref{thm:three-port-region-universality} transfer both the bound and
the attaining state to arbitrary fixed phases and every equal-split
$3\times3$ unitary MMI.
\end{proof}

The residual $1/9\simeq11.1\%$ is unavoidable power in the undesired output
when the other two outputs are required to be exactly balanced.  It is
therefore routing leakage (or crosstalk), not dissipative optical loss: the
three powers still sum to unity.

Theorem~\ref{thm:three-port-region-universality} is stronger than
Theorem~\ref{thm:general-pairwise-routing}, but it is special to dimensions in
which all equal-split unitaries are DFT-equivalent. This reachable-region
equivalence already fails at $N=4$, as the following explicit example shows.
Consider the two normalized equal-split unitary matrices
\begin{align}
 \mathbf{W}_4&=\frac{1}{2}
 \begin{pmatrix}
  1&1&1&1\\
  1&1&-1&-1\\
  1&-1&1&-1\\
  1&-1&-1&1
 \end{pmatrix}, &
 \mathbf{F}_4&=\frac{1}{2}
 \begin{pmatrix}
  1&1&1&1\\
  1&i&-1&-i\\
  1&-1&1&-1\\
  1&-i&-1&i
 \end{pmatrix}.
 \eqlab{four port Walsh and Fourier}
\end{align}
For the zero-based matrix index $n=0$ (physical input port 1), no fixed phases,
and reflector vector
$\mathbf{r}=(1,1,i,i)$, direct multiplication gives
\begin{align}
 \mathbf{W}_4^{\mathsf{T}}\operatorname{diag}(\mathbf{r})
 \mathbf{W}_4\mathbf{e}_0
 =\left(\frac{1+i}{2},\frac{1-i}{2},0,0\right)^{\mathsf{T}},
 \eqlab{Walsh four port witness field}
\end{align}
so the power vector $(1/2,1/2,0,0)$ is reachable with $\mathbf{W}_4$.

It is not reachable with $\mathbf{F}_4$.  Indeed, suppose a unit-modulus
reflector vector produced a field $(a,b,0,0)^{\mathsf{T}}$ through
$\mathbf{F}_4$.  Inverting the four-point DFT relation forces its four entries
to be
\begin{align}
 a+b,\qquad a-ib,\qquad a-b,\qquad a+ib.
 \eqlab{Fourier four port inverse controls}
\end{align}
Their unit moduli imply, by comparing the first with the third and the second
with the fourth,
\begin{align}
 a\overline{b}+b\overline{a}&=0,&
 a\overline{b}-b\overline{a}&=0.
\end{align}
Hence $a\overline{b}=0$.  This contradicts the required powers
$|a|^2=|b|^2=1/2$.  Thus even equal-split unitary MMIs can have different
complete reachable power regions for $N>3$.  The fixed-phase invariance theorem
shows that adding arbitrary fixed reciprocal post-MMI phases does not remove
this difference.  Theorem~\ref{thm:general-pairwise-routing} nevertheless
guarantees that every such region contains all $N$ routing vertices for every
input port. Uniform loss across all arms reduces the total output power but leaves the normalized reachable region unchanged. Arm-to-arm loss imbalance deforms the region and generally removes the exact routing vertices, thereby limiting the achievable extinction ratio.

\section{Supplementary Note 3: Device design, electrostatics and cavity-array characterization}

The MMI is designed to split any input evenly among the output channels.
Supplementary Figure~\ref{fig:app MMI Parameters} shows its port-numbering convention and
geometrical parameters.
\begin{figure*}[!ht]
  \centering
  \includegraphics[width=0.8\textwidth]{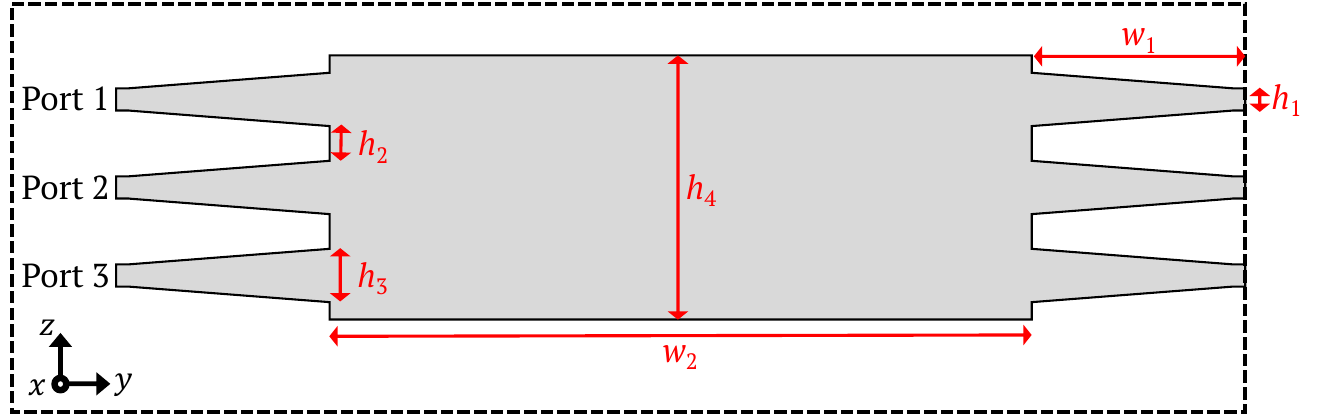}
  \caption{\textbf{Sketch of the MMI Design.} The coordinate system shown in the lower left corner corresponds to the crystal axes of the LiNbO$_3$ membrane. The parameters of the device used in the main text are: $h_1=800\,$nm, $h_2=2000\,$nm, $h_3=3150\,$nm, $h_4=15\,\upmu$m, $w_1=40\,\upmu$m, $w_2=435\,\upmu$m.}
  \figlab{app MMI Parameters}
\end{figure*}
We used the finite-difference eigenmode expansion solver in Ansys Lumerical
to compute the scattering matrix between the input and output fundamental TE-like waveguide modes. The structure was optimized to reach $|M_{mn}|^2=1/3$ for every input mode $n$ and output mode $m$. Supplementary Fig.~\ref{fig:app MMI Simulation} plots the calculated matrix elements corresponding to the optimized structure.
\begin{figure*}[!ht]
  \centering
  \includegraphics[width=0.75\textwidth]{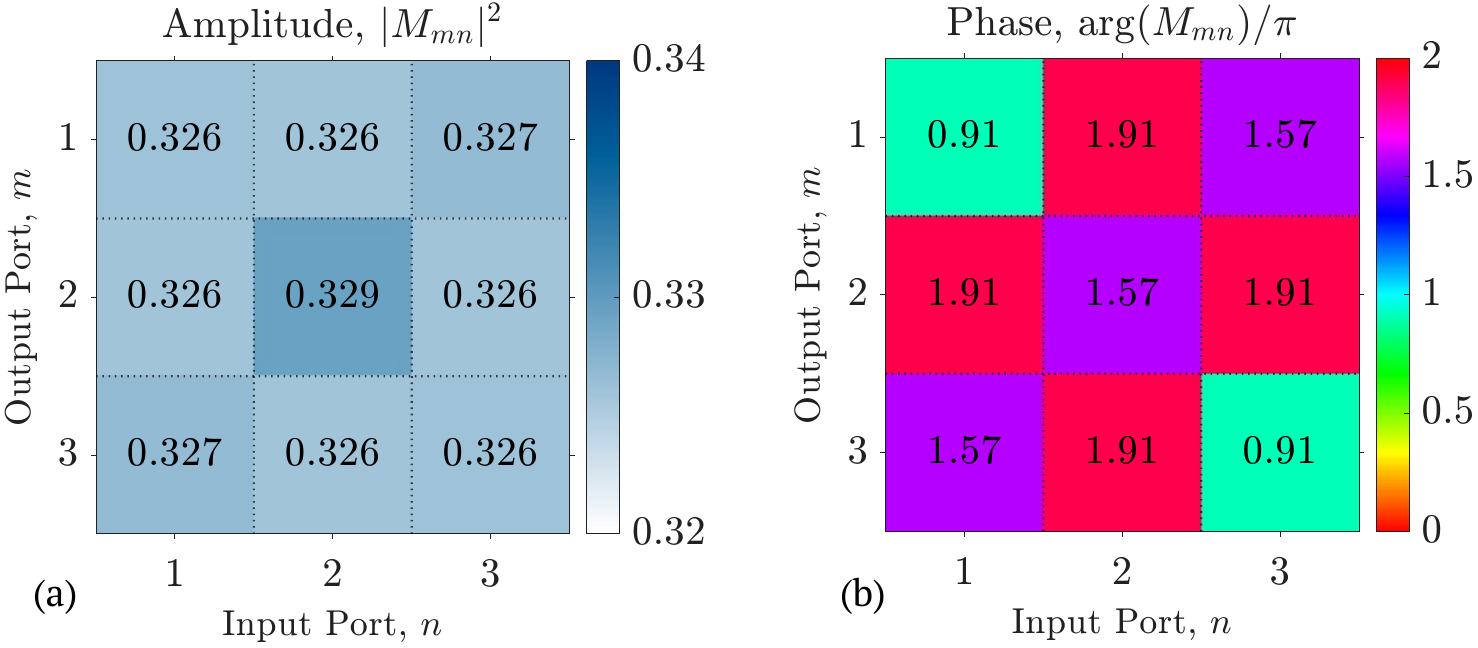}
  \caption{\textbf{Simulated Scattering Matrix of the MMI.}
    \textbf{a}, Calculated intensity transmission, $|M_{mn}|^2$.
    \textbf{b}, Calculated transmission phase, $\arg(M_{mn})$. The
    simulation-specific optimized parameters are $h_1=800\,$nm,
    $h_2=2000\,$nm, $h_3=3200\,$nm, $h_4=15\,\upmu$m,
    $w_1=40\,\upmu$m and $w_2=435.4\,\upmu$m; Supplementary
    Fig.~\ref{fig:app MMI Parameters} reports how the parameters are related to the geometrical structure of the MMI.}
  \figlab{app MMI Simulation}
\end{figure*}
The calculated transmission loss
\begin{equation}
  T_{{\rm{loss}},n} = 1 - \sum_{m=1}^{3} |M_{mn}|^2,
  \eqlab{app Tloss}
\end{equation}
is less than 2.2\% for all input ports. The deviation from the ideal splitting ratio of 1/3 is less than 1\% for all matrix elements.\\

The PhC cavity design follows Ref.~\cite{Quan2010} and uses band diagrams of
infinite periodic waveguides with unit cells parametrized as shown
in Supplementary~\figref{app-phc-parameters}(a). The structure is periodic in the
$y$-direction with lattice constant $a$, and its band diagram gives the
dispersion relation $\omega(k_y)$ for the first few modes. We calculate these
band diagrams using the open-source MIT Photonic Bands package~\cite{Johnson2001}.
The PhC cavity comprises four regions: 1) a taper of $N_{\rm tap}$ unit cells
that converts the rib waveguide smoothly into a periodic structure; 2) a
front mirror of $N_{m,L}$ identical unit cells that provides partial reflection
at the target resonance frequency $\omega_0$; 3) a cavity of $2N_{\rm cav}$
unit cells with smoothly varied geometric parameters; and 4) a back mirror of
$N_{m,R}$ identical unit cells that ideally provides 100\% reflection at
$\omega_0$.
\begin{figure*}[!ht]
  \centering
  \includegraphics[width=0.9\textwidth]{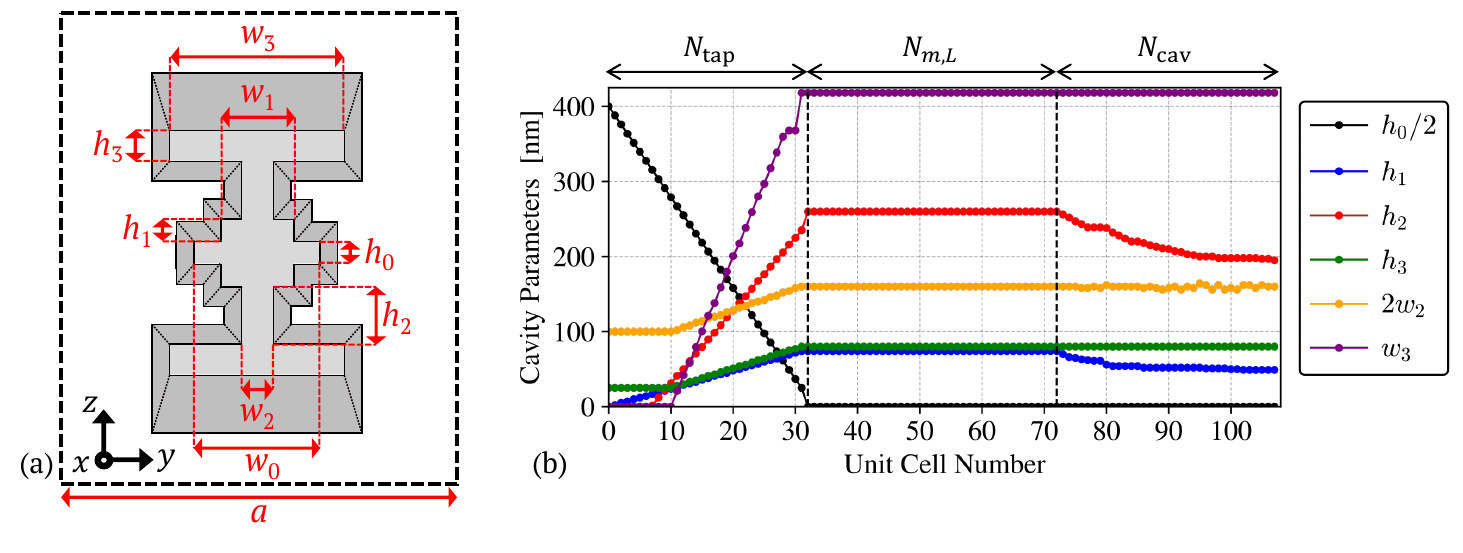}
  \caption{\textbf{Design Parameters of the PhC Cavities.}
    \textbf{a}, Sketch of a single unit cell of the PhC cavity indicating its geometrical parameters. \textbf{b}, Plot of the parameters as a function of unit cell number for the left part of the full cavity structure. The parameters plotted correspond to the device in the main text. The fixed parameters are: $a=418\,$nm, $w_0=a$, and $w_1=148\,$nm. The lengths of the regions are: $N_{\rm{tap}}=32$, $N_{m,L}=40$, and $N_{\rm{cav}}=36$. Note that the parameters plotted here differ from the design parameters by the offset parameter $\delta_{\rm{off}}=-16\,$nm, as defined in~\eqref{app offset}.
    }
  \label{fig:app-phc-parameters}
\end{figure*}
Supplementary Figure~\ref{fig:app-phc-parameters}(b) shows how the parameters are varied in the taper section (unit cell 0 to $N_{\rm{tap}}-1=31$), left mirror section (unit cell 32 to $N_{\rm{tap}}+N_{m,L}-1=71$), and the left half of the cavity section (unit cell 72 to $N_{\rm{tap}}+N_{m,L}+N_{\rm{cav}}-1=107$). The full cavity section is symmetric around its center at unit cell 107, and the right mirror contains $N_{m,R}=100$ unit cells.

Note that for the fabricated devices, we add an offset parameter, $\delta_{\rm{off}}$, to the design parameters to compensate for over- or under-etching. This offset is added to the geometrical parameters in Supplementary~\figref{app-phc-parameters}(a) as
\begin{equation}
  h_n\rightarrow h_n + \delta_{\rm{off}} \qquad \text{and} \qquad 
  \eqlab{app offset} w_n\rightarrow w_n + 2\delta_{\rm{off}}.
\end{equation}
Supplementary Figure~\ref{fig:app-phc-parameters}(b) shows values for the device used in the main text for which $\delta_{\rm{off}}=-16\,$nm. The nominal thickness of the LiNbO$_3$ membrane is $h_{\rm{mem}}=600\,$nm, which is etched by $h_{\rm{wg}}=400\,$nm to form the photonic structures. The nominal exterior sidewall angle of the PhC cavity (Fig.~\ref{fig:device-setup}a) is $55^\circ$ from the substrate plane ($35^\circ$ from vertical). The sidewalls inside the narrow PhC Bragg-mirror gaps are steeper. Following Supplementary Section~V of Ref.~\cite{Larocque2024}, which reports an inner-gap angle of approximately $17^\circ$ from vertical ($73^\circ$ from the substrate plane) based on SEM imaging and transmission modeling, our PhC design model uses $73^\circ$ from the substrate plane for these internal sidewalls. The design assumes this angle on all sidewalls, since the inner-gap angle is the most significant to the PhC design.

Inside the cavity section, the parameters $h_1$, $h_2$, and $w_2$ are varied to create a linearly decreasing mirror strength, as shown in Supplementary~\figref{app-phc-simulation-q}(a). The mirror strength is defined in terms of frequencies at the band edge of the band diagram, $k_y=\pi/a$, as~\cite{Quan2010}
\begin{equation}
  MS = \sqrt{\frac{(\omega_{\rm{air}} - \omega_{\rm{diel.}})^2}{(\omega_{\rm{air}} + \omega_{\rm{diel.}})^2} - \frac{(\omega_0 - \omega_{\rm{cen}})^2}{\omega_{\rm{cen}}^2}},
  \eqlab{app MS}
\end{equation}
where $\omega_{\rm{air}}$, $\omega_{\rm{diel.}}$, and $\omega_{0}$ are the frequencies of the air band, dielectric band, and desired resonance, while $\omega_{\rm{cen}} = (\omega_{\rm{air}}+\omega_{\rm{diel.}})/2$. For designs that are divided into sections, it is straightforward to vary the extrinsic coupling rate, $\kappa_e$, with a minimal effect on the intrinsic loss rate, $\kappa_i$. Supplementary Figure~\ref{fig:app-phc-simulation-q}(b) shows the calculated quality factors (defined as $Q_e=\omega_0/\kappa_e$ and $Q_i=\omega_0/\kappa_i$, respectively) as a function of the number of unit cells in the left mirror section, $N_{m,L}$. The extrinsic coupling $Q$ increases exponentially with $N_{m,L}$, while the intrinsic loss rate is roughly independent of $N_{m,L}$.
\begin{figure*}[!ht]
  \centering
  \includegraphics[width=0.9\textwidth]{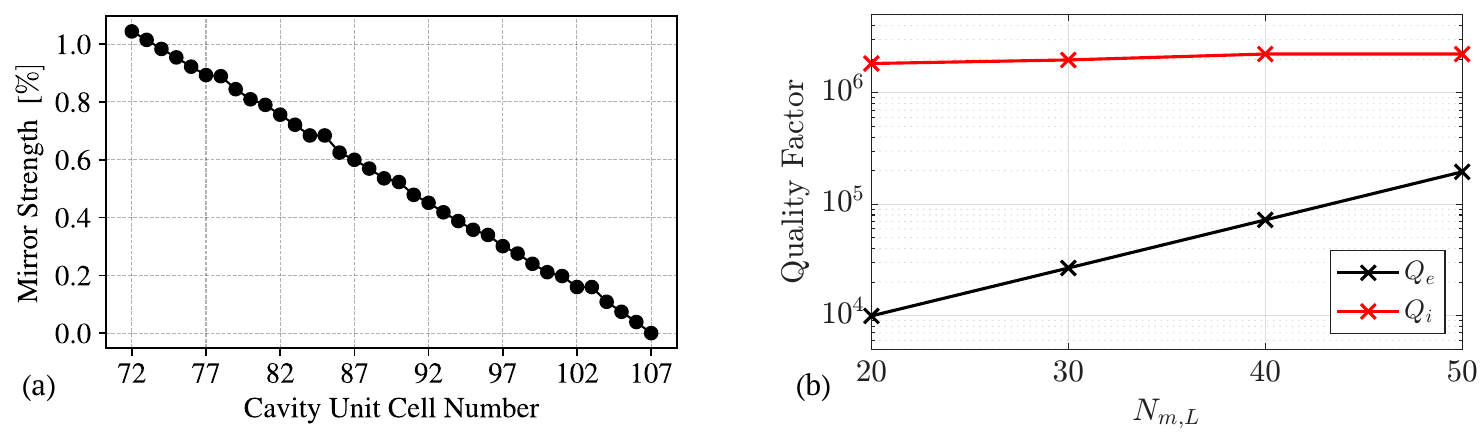}
  \caption{\textbf{Calculated Mirror Strength and Quality Factors of the Single-Sided PhC Cavity.}
    \textbf{a}, Calculated mirror strength, $MS$, from~\eqref{app MS} as a function of unit cell number in the left half of the cavity section. \textbf{b}, Calculated quality factors as a function of the length of the left mirror section, $N_{m,L}$. The parameters used for the simulations are the same as in Supplementary~\figref{app-phc-parameters}(b), except for $\delta_{\rm{off}}$, which is zero here. The length of the right mirror section is $N_{m,R}=100$.}
  \label{fig:app-phc-simulation-q}
\end{figure*}
The simulations in Supplementary~\figref{app-phc-simulation-q}(b) were performed using Ansys Lumerical's Finite-Difference Time-Domain solver. A cross-section of the field profile of the resonance corresponding to the geometry with $N_{m,L}=40$ in Supplementary~\figref{app-phc-simulation-q}(b) is plotted in Supplementary~\figref{app PhC Simulation Mode} with the structure overlaid. The colored contour corresponds to the real part of the $y$-component, normalized as
\begin{equation}
  \mathcal{E}_y(y,z) = E_y(y,z)\frac{\lambda_0}{\sqrt{\int |\bm{E}(y,z)|^2 dydz}},
  \eqlab{app Enorm}
\end{equation}
where $\lambda_0 = 1549.2\,$nm is the corresponding resonance wavelength.
\begin{figure*}[!ht]
  \centering
  \includegraphics[width=1\textwidth]{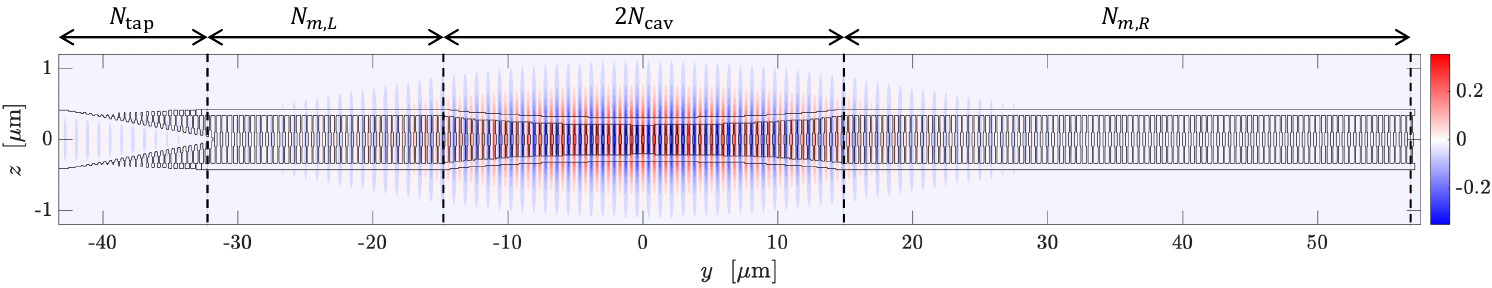}
  \caption{\textbf{Simulated Mode Profile of the Single-Sided PhC Cavity.}
    The plot shows the simulated real part of the electric field normalized as in~\eqref{app Enorm}. The black curve shows the cavity structure at the same cross-section, $x=(h_{\rm{mem}}-h_{\rm{wg}})+h_{\rm{wg}}/2$. The four regions of the cavity are indicated with $N_{\rm{tap}}=32$, $N_{m,L}=40$, $N_{\rm{cav}}=36$, and $N_{m,R}=100$. The geometrical parameters are the same as in Supplementary~\figref{app-phc-simulation-q}(b).}
  \figlab{app PhC Simulation Mode}
\end{figure*}
The field is observed to be concentrated in the cavity region, with some leakage to the left but none to the right, as required for a single-sided cavity.

Assuming a uniform field $V/g$ and unit optical overlap, the Pockels effect gives
$|\mathrm{d}\nu/\mathrm{d}V|\simeq\nu_0n_e^2r_{33}/(2g)$.
At a wavelength of 1542~nm, using $n_e=2.14$, $r_{33}=30.9$~pm/V and an
electrode gap $g=6.8$~$\upmu$m gives approximately 2.0~GHz/V.
An approximate overlap calculation using local periodic cells of the
patterned cavity and the nonuniform electrostatic field gives approximately
0.94~GHz/V. These estimates support the order of magnitude of the measured
quasi-static tuning rate.

\subsection{Electrode capacitance and switching energy}

We estimated the terminal capacitance of the phase-shifter electrode pair with
an electro-quasistatic finite-volume solution of
$\nabla\!\cdot(\boldsymbol{\epsilon}\nabla\phi)=0$. The three-dimensional model
uses the exact 24-vertex electrode footprint extracted from the GDS, including
the taper and end fringing, together with a continuous LiNbO$_3$ rib and
slab, the SiO$_2$ claddings, the dielectric and electrically
uncontacted silicon handle, air, and 600~nm thick gold electrodes. We use
$\epsilon_r=3.9$, 11.7, and 1 for SiO$_2$, Si, and air, respectively, and the
anisotropic LiNbO$_3$ values $\epsilon_{r,\parallel}=28.7$ and
$\epsilon_{r,\perp}=43.0$. Here, $x$, $y$ and $z$ denote the longitudinal,
transverse and vertical directions, respectively, corresponding to the crystal
axes $y$, $z$ and $x$. The far-field boundaries are fixed at the balanced
pair's 0.5-V common-mode potential and placed at $x=\pm1$~mm,
$y=1$~mm from the symmetry plane, and $z=\pm0.5$~mm; doubling all of these
distances changes the raw capacitance by only 0.096\%. Calibrating the
three-dimensional mesh against a
converged two-dimensional straight-section calculation gives
$C=18.6\pm0.3$~fF for the complete electrode pair. The two-dimensional
cross-section gives $0.1414$~fF/$\upmu$m. Applying this value over the full
116.70-$\upmu$m GDS electrode span gives $16.5$~fF; the remaining
approximately $2.1$~fF in the full three-dimensional result arises from the
tapered and widened regions and end fringing. We therefore use the
three-dimensional terminal capacitance for the switching-energy estimates.
The electrode geometry and transverse field are shown in Supplementary Fig.~\ref{fig:app-electrode-capacitance-field}, and the full three-dimensional field in Supplementary Fig.~\ref{fig:app-electrode-capacitance-field-3d}.
For a conventional
non-energy-recovering capacitive driver, the energy dissipated in one
transition of one phase shifter is
$E_{\rm tr}=C(\Delta V)^2/2$, corresponding to $232.5\pm3.8$~fJ for
$\Delta V=5$~V. For a closed cycle among the three demonstrated routing
states, the energy is
\begin{equation}
 E_{\rm cyc}=\frac{C}{2}\sum_{s}\sum_{j=1}^{3}
 \left(V_{j,s+1}-V_{j,s}\right)^2,
 \eqlab{switching-cycle-energy}
\end{equation}
where the cyclic sum includes the last-to-first transition. We evaluate this
expression using the measured AWG excursions listed in Methods and the closed
routing-state sequence
$P_1\rightarrow P_3\rightarrow P_2\rightarrow P_1$; the DC-balancing idle
states are excluded. The common static biases cancel from every voltage
difference. The result is
$2.096\pm0.034$~pJ per closed routing cycle across all three phase shifters. The
contributions from the $V_1$, $V_2$, and $V_3$ electrodes are
$0.429\pm0.007$, $0.795\pm0.013$, and $0.872\pm0.014$~pJ per cycle,
respectively. The average over the three routing transitions is
$0.699\pm0.011$~pJ per transition. Repeating this closed three-state cycle at
10~MHz corresponds to $21.0\pm0.3~\upmu$W. These values
describe capacitive driver dissipation and exclude driver quiescent power.
Their quoted uncertainties propagate the numerical capacitance uncertainty;
they do not include an independent voltage-calibration uncertainty.

The Pockels phase shifter is intrinsically capacitive and therefore consumes
no static power in the ideal dielectric limit. No device leakage was resolved
above the approximately 10-G$\Omega$ probe background, giving setup-limited
upper bounds of $|I|<0.5$~nA and $P_{\rm static}<2.5$~nW per phase shifter at
5~V. Establishing or changing a DC bias requires the one-time capacitive
charging energy $C(\Delta V)^2/2$; maintaining that bias does not produce
continuous intrinsic dissipation. Any continuous power is instead bounded by
leakage and by driver quiescent power, which is not included here. This 10-GHz
electro-quasistatic reference uses real permittivities and does not include
dielectric loss or finite silicon conductivity.

\begin{figure*}[!ht]
  \centering
  \includegraphics[width=\textwidth]{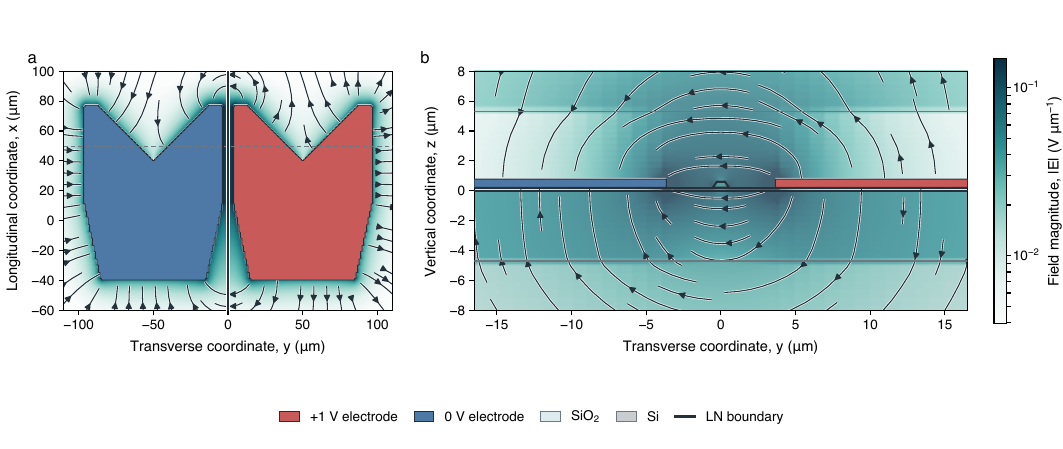}
  \caption{\textbf{Electrostatic field of the phase-shifter electrode pair.}
    \textbf{a}, Top view of the exact GDS electrode geometry, rotated so that
    the two electrodes appear on the left and right. The dashed line marks the
    transverse section used in panel~\textbf{b}. \textbf{b}, Transverse field
    distribution through the active electrode gap. Color gives the electric
    field magnitude for a 1-V differential bias on the shared logarithmic
    scale, and arrows indicate the field direction. Black outlines mark the
    LiNbO$_3$ boundary. The modeled waveguide comprises a 200-nm slab and a
    400-nm rib with a 550-nm top width and $55^\circ$ exterior sidewalls; the
    electrode gap is 6.8~$\upmu$m.}
  \figlab{app-electrode-capacitance-field}
\end{figure*}

\begin{figure*}[!ht]
  \centering
  \includegraphics[width=0.82\textwidth]{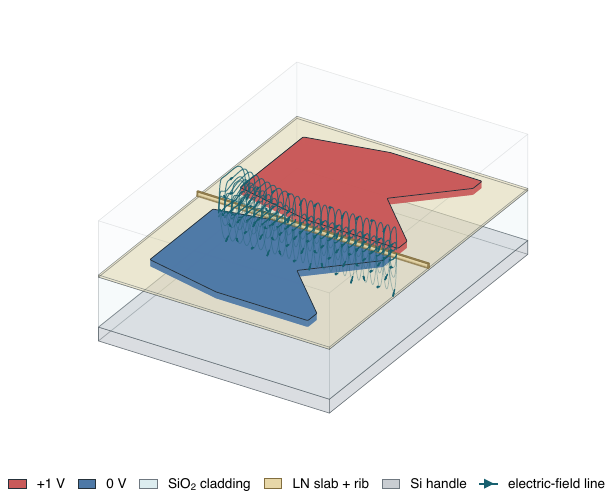}
  \caption{\textbf{Three-dimensional electrostatic field of the electrode pair.}
    Field lines are integrated through the full three-dimensional potential
    solution and arrows point from the $+1$-V electrode toward the grounded
    electrode. The exact GDS electrode footprints are shown in muted red and
    blue. The pale-gold region shows the continuous 200-nm LiNbO$_3$ slab and
    400-nm trapezoidal rib, the pale-blue transparent volumes show the upper
    and buried SiO$_2$ claddings, and the gray volume shows the
    dielectric silicon handle. The upper cladding terminates at the
    SiO$_2$--air interface. The vertical coordinate is displayed with a
    fivefold exaggeration for clarity.}
  \figlab{app-electrode-capacitance-field-3d}
\end{figure*}

\subsection{Cavity matching across complete devices}

The cavity-matching analysis used the two-sided cavity reference and two complementary
MMI transmission spectra for 60 complete three-cavity devices spanning the full
$4\times5\times3$ design set (four PhC types, five cavity offsets, and three
left-mirror counts). The external MEMS fiber-switch matrix selected the measured
optical input--output configurations, while the on-chip cavity electrodes were
left unbiased. A reference-cavity resonance was assignable in 50 of the 60
spectra and, together with the corresponding design-family trend, guided
identification of the three switch-cavity resonances in the MMI transmission
spectra. After visual review, 50 devices contained an assignable cavity group;
we refer to these as usable devices below:
34 showed three resolved resonances, and 16 showed two resolved resonances,
consistent with two cavities overlapping within the spectral resolution. For
each usable device, we define the observed resonance span as
\begin{equation}
  \Delta\lambda=\max_j(\lambda_j)-\min_j(\lambda_j),
  \eqlab{array-resonance-span}
\end{equation}
where $\lambda_j$ are the resolved resonances in the two MMI transmission
states. For devices with only two resolved resonances, $\Delta\lambda$ assumes
that the unresolved third resonance overlaps one of the observed features; the
three-resonance subset is therefore shown separately in
Supplementary Fig.~\ref{fig:app-cavity-array-matching}. Across all 50 usable
devices, the median observed span is 0.63~nm, the interquartile range is
0.44--0.91~nm, and 82\% have $\Delta\lambda<1$~nm.

For symmetric tuning of the extreme resonances toward their frequency
midpoint, the required maximum voltage magnitude is estimated as
\begin{equation}
  |V|_{\max}=\frac{\Delta\nu}{2\eta_{\rm DC}},\qquad
  \Delta\nu=c\left|\frac{1}{\lambda_{\min}}-
  \frac{1}{\lambda_{\max}}\right|,
  \eqlab{array-alignment-voltage}
\end{equation}
using the measured tuning efficiency $\eta_{\rm DC}=1.92$~GHz\,V$^{-1}$. The resulting
fractions alignable within $\pm10$, $\pm20$, $\pm30$, and $\pm40$~V are 10\%,
48\%, 76\%, and 98\%, respectively. The estimated $|V|_{\max}$ has a median
of 20.4~V and an interquartile range of 14.3--29.3~V. Thus, the approximately
30-V value corresponds to the upper quartile of the observed distribution,
rather than a voltage sufficient for every device. In particular, $\pm30$~V
aligns 38 of 50 usable devices (76\%; 95\% Wilson interval, 63--86\%). For the
stricter subset with three resolved resonances, the median is 22.5~V, the
interquartile range is 16.5--32.4~V, and 23 of 34 devices are alignable within
$\pm30$~V (68\%; 95\% Wilson interval, 51--81\%). The
$\pm0.06$~GHz\,V$^{-1}$ uncertainty of the measured tuning efficiency gives
a 3.1\% relative voltage-scale uncertainty (0.64~V at the all-device median)
and changes either $\pm30$-V device count by at most one. These estimates do not
include device-to-device variation in tuning efficiency or uncertainty in
identifying overlapping resonances; the two-resonance cases assume that the
unresolved third resonance overlaps one of the observed features.

\begin{figure*}[!ht]
  \centering
  \includegraphics[width=0.9\textwidth]{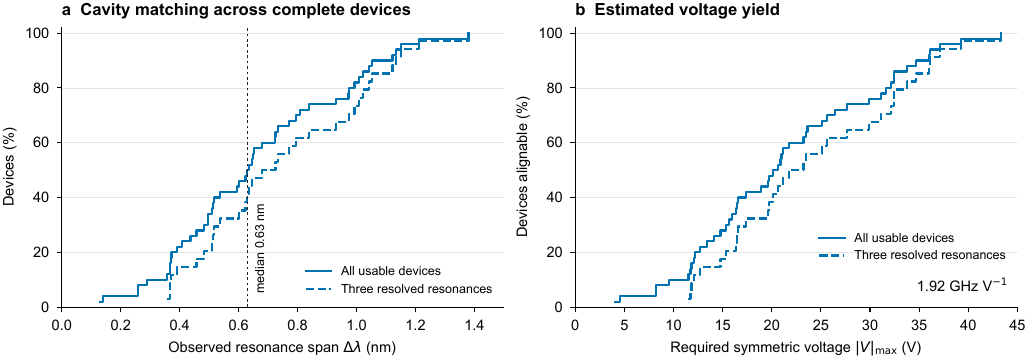}
  \caption{\textbf{Measured cavity matching and estimated alignment-voltage yield.}
    \textbf{a}, Empirical cumulative distribution of the observed wavelength
    span $\Delta\lambda$ for all 50 usable complete devices (solid) and the 34
    devices with three resolved resonances (dashed). The vertical line marks
    the all-device median of 0.63~nm. \textbf{b}, Fraction of devices whose
    extreme resolved resonances can be tuned to their frequency midpoint
    within the indicated symmetric voltage magnitude, using
    $\eta_{\rm DC}=1.92$~GHz\,V$^{-1}$.}
  \figlab{app-cavity-array-matching}
\end{figure*}

The reviewed spectra also allow the measured resonance positions to be
compared with the cavity design parameters. Within each PhC type, the mean
resolved resonance wavelength follows the lithographic cavity offset
(Supplementary Fig.~\ref{fig:app-cavity-design-trends}a). The single-sided
cavities used for the switching demonstration in device A46 belong to
type~3, with $(a,N_{\rm cav})=(418~\mathrm{nm},36)$. For the 34 devices
with three resolved resonances, type~0 has the smallest median span, although
the unequal sample counts, particularly for type~1, make this comparison
descriptive rather than conclusive. The usable-spectrum yield also varies
between the four design types (Supplementary
Fig.~\ref{fig:app-cavity-design-trends}c).

\begin{figure*}[!ht]
  \centering
  \includegraphics[width=\textwidth]{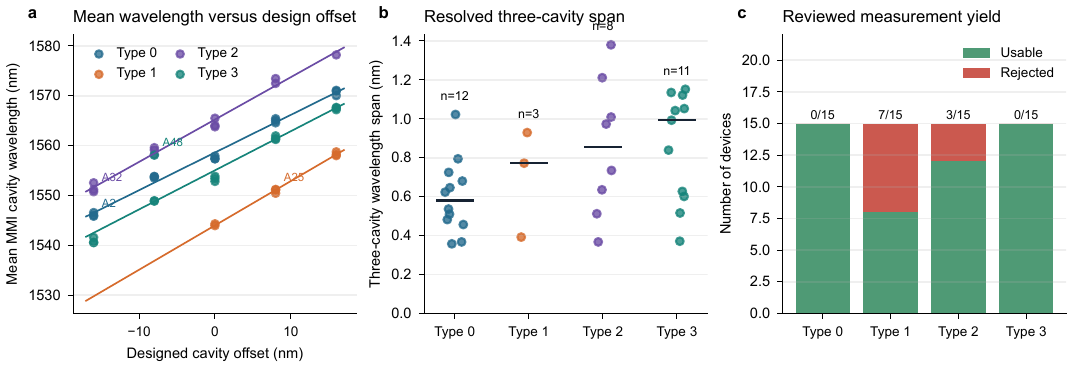}
  \caption{\textbf{Measured cavity design trends across complete devices.}
    \textbf{a}, Mean wavelength of the resolved MMI-cavity features versus the
    designed cavity offset for all 50 usable devices. Lines are linear fits
    within each PhC type. The four types use $(a,N_{\rm cav})=$
    $(424~\mathrm{nm},30)$, $(420~\mathrm{nm},30)$,
    $(422~\mathrm{nm},36)$, and $(418~\mathrm{nm},36)$, respectively.
    \textbf{b}, Max--min wavelength span for the 34 devices with three
    resolved resonances; horizontal bars indicate medians. The accepted
    devices with two resolved resonances contribute to panel~\textbf{a} but
    are excluded from panel~\textbf{b}. \textbf{c}, Numbers of usable and
    rejected spectra for each PhC type. A spectrum was rejected when the
    cavity group could not be assigned confidently.}
  \figlab{app-cavity-design-trends}
\end{figure*}

Independent fits to the reference-cavity spectra show that the loaded quality
factor increases mainly with the number of left-mirror cells, while no clear
dependence on the lithographic cavity offset is resolved
(Supplementary Fig.~\ref{fig:app-reference-q-trends}). This separation is
consistent with the mirror count controlling the external coupling and the
offset primarily selecting the resonance wavelength.

\begin{figure*}[!ht]
  \centering
  \includegraphics[width=0.92\textwidth]{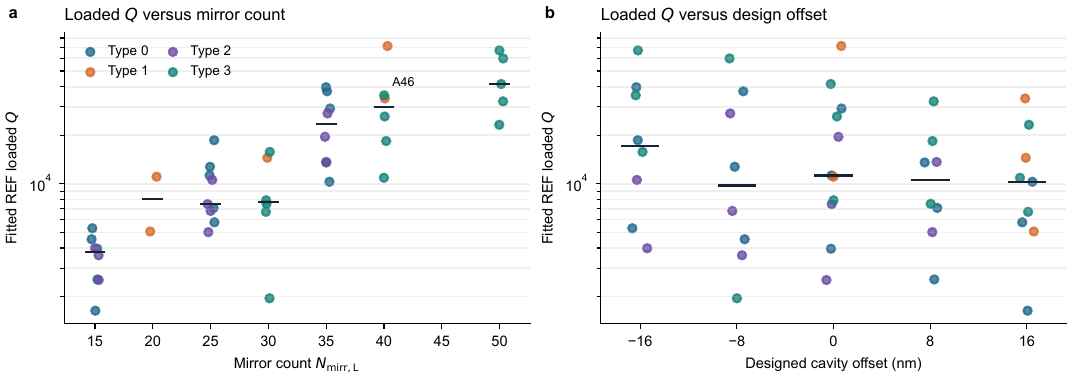}
  \caption{\textbf{Two-sided reference-cavity loaded quality-factor trends.}
    Loaded quality factors extracted from 44 two-sided reference-cavity spectra
    acquired without detector saturation and whose line-shape fits have
    $R^2\geq0.9$. \textbf{a}, Loaded $Q$
    versus the number of cells in the left mirror. \textbf{b}, Loaded $Q$
    versus the designed cavity offset. Colors identify the four PhC types;
    horizontal bars show the median at each design value.}
  \figlab{app-reference-q-trends}
\end{figure*}

\noindent\begin{minipage}{\textwidth}
For a symmetric two-sided cavity driven from one side, the field reflection
coefficient is
\begin{equation}
  r_{\rm{two-sided}}(\Delta)
  = \frac{i\Delta + \kappa_i/2}
         {i\Delta + \kappa_e + \kappa_i/2},
  \label{eq:r-two-sided}
\end{equation}
where $\Delta=\omega_0-\omega$, $\kappa_e$ is the external coupling rate
through each mirror, and $\kappa_i$ is the total intrinsic loss rate. The
loaded quality factor is therefore $Q=\omega_0/(2\kappa_e+\kappa_i)$.
\end{minipage}

The full-device model gives $\kappa_e/\kappa_i=1.09$--$1.36$ for the
one-sided switch cavities. Taking this range as an estimate of the coupling
through each 40-period mirror and assuming comparable intrinsic loss for the
two-sided reference cavity gives
$Q_i=Q(1+2\kappa_e/\kappa_i)=(1.05$--$1.23)\times10^5$. This range is a
model-based estimate rather than an independent fit of the reference-cavity
intrinsic loss.

The complete broadband measurement is shown in Supplementary
Fig.~\ref{fig:app-reference-full-spectrum}. Summing the simultaneously
measured reflection and transmission channels suppresses much of the resonant
redistribution between the two ports. Its smooth upper envelope therefore
tracks the wavelength-dependent response shared by the optical path, including
the grating couplers, but it is not an independent grating-coupler calibration
and is not used to infer an absolute coupling loss. For qualitative comparison,
a separately measured plain-grating test spectrum is vertically matched to the
long-wavelength shoulder of this envelope in the figure.

The fit quality for the representative two-sided reference cavity in device
A46 used in Fig.~\ref{fig:bw}a is shown in Supplementary
Fig.~\ref{fig:app-reference-lorentzian-fit}. The transmission peak and
reflection dip are fitted simultaneously with a shared Lorentzian center and
linewidth, independent amplitudes, and independent linear backgrounds.

\begin{figure*}[!t]
  \centering
  \includegraphics[width=\textwidth]{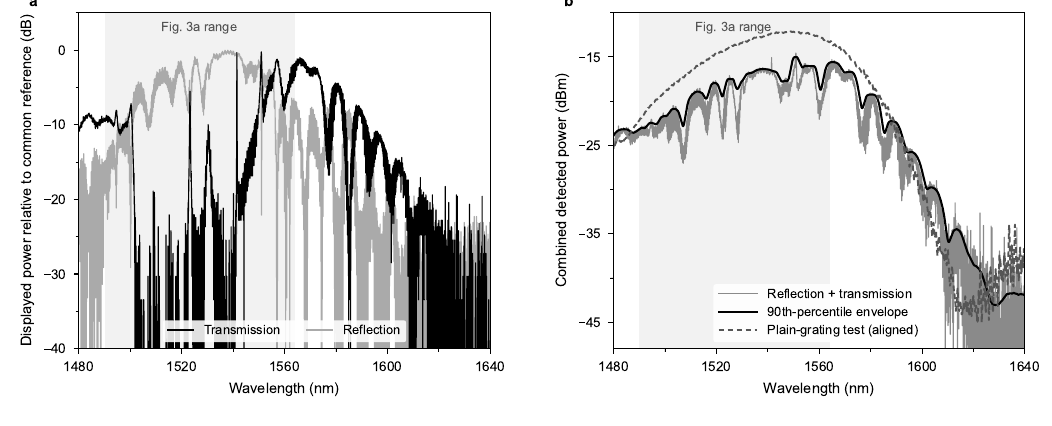}
  \caption{\textbf{Full broadband spectrum of the two-sided reference cavity.}
    \textbf{a}, Reflection and transmission channels measured over the complete
    1480--1640~nm laser sweep for device A46. Because the two detector channels
    were not absolutely calibrated relative to one another, the transmission
    trace is slightly offset for visual alignment with Fig.~\ref{fig:bw}a;
    this plotting offset has no physical significance.
    The shaded interval is the wavelength range displayed there.
    \textbf{b}, Sum of the simultaneously measured reflection and transmission
    powers and its rolling
    90th-percentile upper envelope. The dashed trace is the calibrated
    transmission of a plain-grating reference measured in a separate
    1480--1640~nm sweep. It is shown in relative decibels and
    discreetly offset to match the envelope on its long-wavelength shoulder,
    providing a qualitative comparison of spectral shape.}
  \figlab{app-reference-full-spectrum}
\end{figure*}

\begin{figure*}[!ht]
  \centering
  \includegraphics[width=0.50\textwidth]{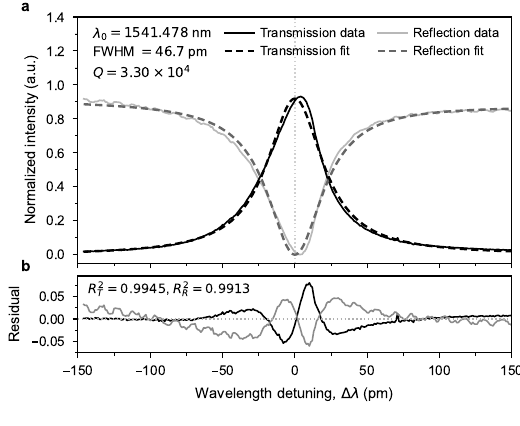}
  \caption{\textbf{Representative two-sided reference-cavity Lorentzian fit.}
    \textbf{a}, Normalized transmission (black) and reflection (light gray)
    measured around the resonance shown in Fig.~\ref{fig:bw}a. Dashed curves
    are the simultaneous Lorentzian fits. The shared fit gives
    $\lambda_0=1541.478$~nm, a full width at half maximum of 46.7~pm, and a
    loaded quality factor $Q=3.30\times10^4$; the coefficients of
    determination are $R_T^2=0.9945$ and $R_R^2=0.9913$.
    \textbf{b}, Data-minus-fit residuals for the two traces.}
  \figlab{app-reference-lorentzian-fit}
\end{figure*}

Finally, after subtracting a robust type-specific linear dependence on cavity
offset, the residual mean wavelength has no resolved gradient across the chip
(Supplementary Fig.~\ref{fig:app-cavity-spatial-residuals}). A planar fit to
the non-outlier devices gives gradients of $+0.07$ and $+0.05$~nm\,mm$^{-1}$
along $x$ and $y$, respectively, with $R^2=0.04$. Device A48 is the only
residual beyond three robust standard deviations.

\begin{figure*}[!t]
  \centering
  \includegraphics[width=0.72\textwidth]{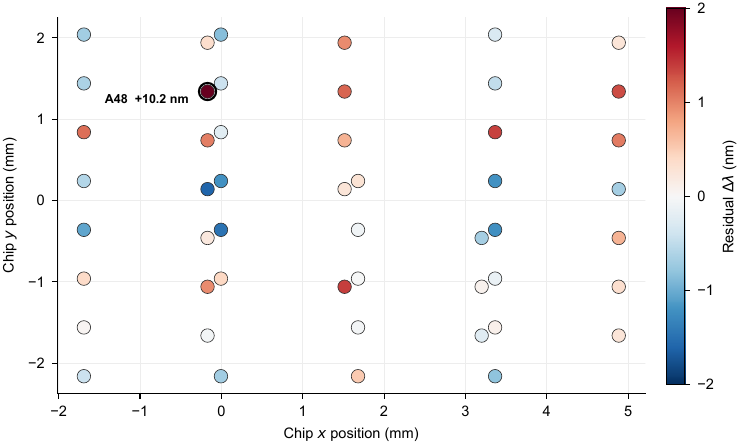}
  \caption{\textbf{Spatial residuals of the measured single-sided switch-cavity
    wavelengths.}
    Residual mean wavelength at each accepted device position after removing a
    robust, type-specific linear dependence on the designed cavity offset. The
    color scale is limited to $\pm2$~nm to show the spatial variation among the
    non-outlier devices; the outlined and labeled point A48 is the sole residual
    beyond three robust standard deviations. Excluding A48, a planar fit gives
    gradients of $+0.07$ and $+0.05$~nm\,mm$^{-1}$ along $x$ and $y$,
    respectively, with $R^2=0.04$.}
  \figlab{app-cavity-spatial-residuals}
\end{figure*}

\FloatBarrier
\section{Supplementary Note 4: Cavity response, insertion loss and routing-performance analysis}

The calibrated transfer-matrix model determines the voltage-dependent
reflection coefficient $r_j(V_j)$ of each cavity from \eqref{r}.
Supplementary Fig.~\ref{fig:app-cavity-switching-states} shows the corresponding
spectra at the three operating points obtained by maximizing the extinction
ratio for routing to $P_1$, $P_2$, and $P_3$. The three resonances move relative
to the fixed laser wavelength and thereby set the phases that interfere
constructively at the selected output after the return pass through the MMI.
To convert the normalized detunings constrained by the voltage cube to
wavelength, we set the loaded quality factor to the independently extracted
measured value $Q_{\rm L}=5.07\times10^4$ for all three nominally identical
switch cavities. This value was independently extracted from a Fano fit to
the reflection spectrum of one cavity measured in isolation on the same
fabricated router device. The resonance offsets and tuning
coefficients are rescaled together so that the modeled response at the fixed
laser wavelength is unchanged. The figure also shows the continuous cavity
reflection phases and the output powers obtained by coherently recombining the reflected fields using the
calibrated reciprocal transformation $\mathbf{M}^{\mathsf{T}}\mathbf{R}\mathbf{M}$,
including the fitted arm phases and coherent background.
The wavelength positions are therefore conditional on applying the measured
quality factor to all three cavities.

\begin{figure*}[!ht]
  \centering
  \includegraphics[width=\textwidth]{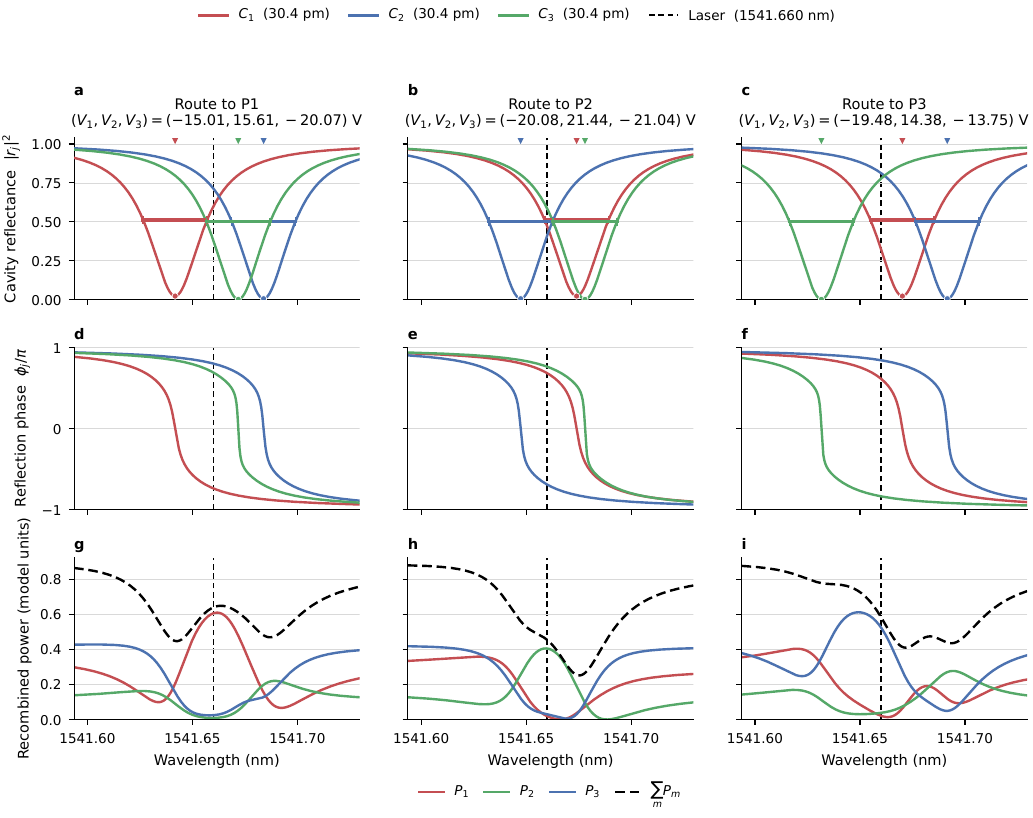}
  \caption{\textbf{Fano-Q-constrained cavity spectra, phases and output powers at the three switching
    states.}
    \textbf{a--c}, Cavity reflectance $|r_j|^2$ versus wavelength at the
    calibrated voltage tuples that route light to $P_1$, $P_2$, and $P_3$,
    respectively. The dashed vertical line marks the fixed laser wavelength,
    1541.660~nm. Triangles mark the reconstructed resonance positions, and
    horizontal bars show their loaded full widths at half maximum. A common
    $Q_{\rm L}=5.07\times10^4$, independently extracted from a Fano fit to
    one cavity measured in isolation on the same fabricated router device,
    gives linewidths of 30.4~pm. The same value is applied
    to all three nominally identical cavities; the absolute wavelength
    positions are conditional on this assumption. \textbf{d--f}, Corresponding
    continuous reflection phases $\phi_j/\pi$; only relative phases affect the
    recombined powers. \textbf{g--i}, Output powers after coherent recombination in the
    calibrated model, including the fitted arm phases, voltage-dependent
    phase correction and coherent background, before detector gain and
    offset calibration. Red, green and blue denote $P_1$, $P_2$ and $P_3$,
    respectively; the dashed black curve is their sum. Powers are in model
    units and do not represent absolute insertion-loss measurements.
    Columns correspond to routing to $P_1$, $P_2$ and $P_3$, respectively.}
  \figlab{app-cavity-switching-states}
\end{figure*}

Using the fitted voltage linewidths (4.77--4.86~V) and
external-to-internal coupling ratios (1.09--1.36), \eqref{r} predicts
cavity phase excursions of $1.67\pi$, $1.69\pi$ and $1.70\pi$
for cavities 1, 2 and 3, respectively, for a $\pm5$~V sweep
(10~V peak-to-peak) centered on the bias that brings each resonance to the
laser wavelength. These correspond to 83.5\%, 84.4\% and 85.2\% of $2\pi$.
These are model estimates of the cavity reflection phase alone, assuming
constant tuning efficiency over the sweep, rather than direct phase
measurements at the routing biases. A full $2\pi$ winding is approached
asymptotically across resonance; it is not established by the finite drive
range.

At one probe wavelength, the transformation
$Q_{\rm L}\rightarrow sQ_{\rm L}$,
$\eta\rightarrow\eta/s$ and
$\Delta\nu_0\rightarrow\Delta\nu_0/s$ leaves the modeled voltage-cube response
unchanged. The reciprocal fit therefore determines voltage linewidths of
4.78, 4.86 and 4.77~V for $C_1$, $C_2$ and $C_3$, respectively, but it does not
independently determine $Q_{\rm L}$, $\eta$ or the absolute resonance offsets
in wavelength. The separate $Q_{\rm L}$ and tuning coefficients used in the
numerical parameterization are consequently not interpreted as independent
measurements. Experimental tuning claims use the direct spectroscopic value
in Fig.~\ref{fig:bw}b.

\begin{table*}[ht]
\centering
\caption{Single-sided switch-cavity parameters constrained by the reciprocal
three-cavity voltage-cube fit and the common Fano-Q wavelength reconstruction.
Here $x=\kappa_e/\kappa_i$ and
$V_{\rm linewidth}=\nu_0/(Q_{\rm L}\eta)$. The last three columns give the
reconstructed resonance wavelengths at the states routing to $P_1$, $P_2$
and $P_3$. The coupling ratios are
model-conditioned point estimates because they can trade against the coherent
background and detector calibration. Statistical uncertainties are not quoted
because the saved regularized fit does not contain a covariance matrix or a
bootstrap ensemble.}
\label{tab:single-sided-switch-cavity-parameters}
\begin{tabular}{lrrrrr}
\toprule
Cavity & $\kappa_e/\kappa_i$ & $V_{\rm linewidth}$ (V) & $\lambda(P_1)$ (nm) & $\lambda(P_2)$ (nm) & $\lambda(P_3)$ (nm) \\
\midrule
$C_1$ & 1.357 & 4.78 & 1541.6418 & 1541.6740 & 1541.6702 \\
$C_2$ & 1.168 & 4.86 & 1541.6838 & 1541.6473 & 1541.6916 \\
$C_3$ & 1.089 & 4.77 & 1541.6718 & 1541.6780 & 1541.6316 \\
\bottomrule
\end{tabular}
\end{table*}

The A46 device uses uniform TE focusing grating couplers comprising 28
concentric circular teeth. The gratings have a constant radial period of
978~nm and a drawn tooth width of 166~nm, corresponding to a 17.0\% duty cycle
and an 812-nm gap. The innermost and outermost tooth centerlines have radii of
55.9 and 82.3~$\upmu$m, respectively. Each grating is connected through a
55-$\upmu$m-long taper that expands to a 17-$\upmu$m aperture, corresponding
to a 17.6$^\circ$ full in-plane opening angle.

\subsection{Measured insertion-loss budget}

We compared a broadband scan of device A46 with a plain grating-reference
path acquired in the same measurement session (Supplementary~\figref{app-insertion-loss-budget}). The reference and physical output~1 were
recorded on the same calibrated detector channel. The other two outputs were
placed on the output-1 scale using the independently established detector
calibration. All quoted values below are obtained after a 101-point median
filter, corresponding to 0.101~nm on the 1-pm wavelength grid.

At the fixed switching wavelength of 1541.66~nm, the grating-reference power
is $-13.63$~dBm, the physical output-1 power is $-20.28$~dBm, and the sum of
the three calibration-corrected device outputs is $-17.27$~dBm. Thus, output~1
and the summed output exhibit 6.65~dB and 3.65~dB of excess loss,
respectively, relative to the reference path. The corrected output fractions
in this scan are 50.1\%, 11.4\%, and 38.6\%. An optimized route-to-$P_1$ state
would instead concentrate most of the surviving power in $P_1$; the fitted
model predicts 94.7\%. This scan therefore does not represent the optimized
route-to-$P_1$ state. Its 6.65~dB reference-to-output-1 difference describes
this particular output distribution, rather than the insertion loss of an
optimized routing state. The summed-output excess is used below to estimate
the device-loss budget.

The reference value is itself a path measurement rather than a pure
grating-coupler calibration: it includes two grating interactions, the fiber
network, switches and connectors, polarization mismatch, and reference
waveguide propagation. If the independently verified 0-dBm launch setting
also applied to this acquisition, the measured powers correspond to total
launch-to-detector losses of 13.63~dB for the reference, 17.27~dB for the
summed device outputs, and 20.28~dB for output~1.

The fiber path from the source to the fiber array was independently measured
to have 1.39~dB loss in the output-1 configuration used here. Subtracting this
known single-sided external loss from the 13.63-dB reference-path loss leaves
12.24~dB for the two grating couplers, reference waveguide, and collection
path, corresponding to a conservative upper estimate of 6.12~dB per grating
if the remaining losses are neglected. If the collection-side external loss
is comparable to the measured launch-path loss, the inferred value is
approximately 5.43~dB per grating, before correcting for the reference-waveguide
loss. We therefore estimate a grating-coupler loss of approximately
5.4--6.1~dB per coupler.

For comparison, the fitted route-to-$P_1$ model gives cavity power survivals
of 0.600, 0.713, and 0.378 at 1541.66~nm. Weighting these by the fitted MMI arm
amplitudes gives approximately 2.33~dB of cavity dissipation, and the fitted
output distribution places 94.7\% of the surviving power in $P_1$. Subtracting
this model-conditioned cavity contribution from the measured 3.65-dB
summed-output excess gives 1.32~dB for two MMI passages, routing waveguides, and residual path mismatch. Independently, the best 1-nm-averaged
off-resonant summed-output ratio gives 1.57~dB near 1526.13~nm. Together, these measurements indicate a passive two-pass residual of approximately
1.3--1.6~dB. If assigned entirely and symmetrically to the MMI, this would
bound each MMI passage by approximately 0.7--0.8~dB; the actual MMI loss is
lower to the extent that waveguide loss or path mismatch contributes.

\begin{figure*}[ht]
  \centering
  \includegraphics[width=\textwidth]{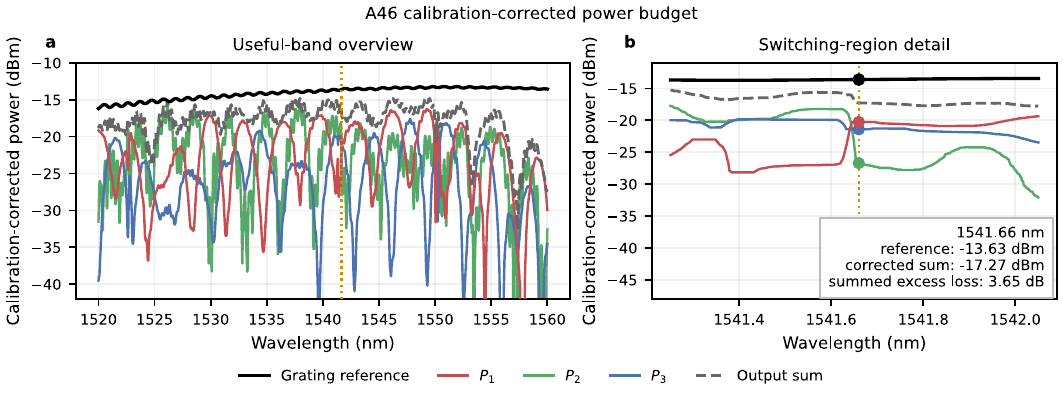}
  \caption{\textbf{Calibration-corrected insertion-loss budget for device A46.}
  \textbf{a}, Broadband grating-reference power, the three device outputs,
  and their corrected sum. Red, green and blue denote $P_1$, $P_2$ and $P_3$,
  respectively. \textbf{b}, Detail around the 1541.66-nm switching
  wavelength. Output~2 is placed on the output-1 scale using an independently
  determined detector calibration; output~1 and the reference use the same
  detector channel. The
  3.65-dB summed-output excess is relative to the complete grating-reference
  path and therefore includes cavity dissipation, two MMI passages, routing
  waveguides, and any residual path mismatch.}
  \figlab{app-insertion-loss-budget}
\end{figure*}

\subsection{Ideal sensitivity to arm-loss imbalance}

\begin{figure}[ht]
  \centering
  \includegraphics[width=0.98\textwidth]{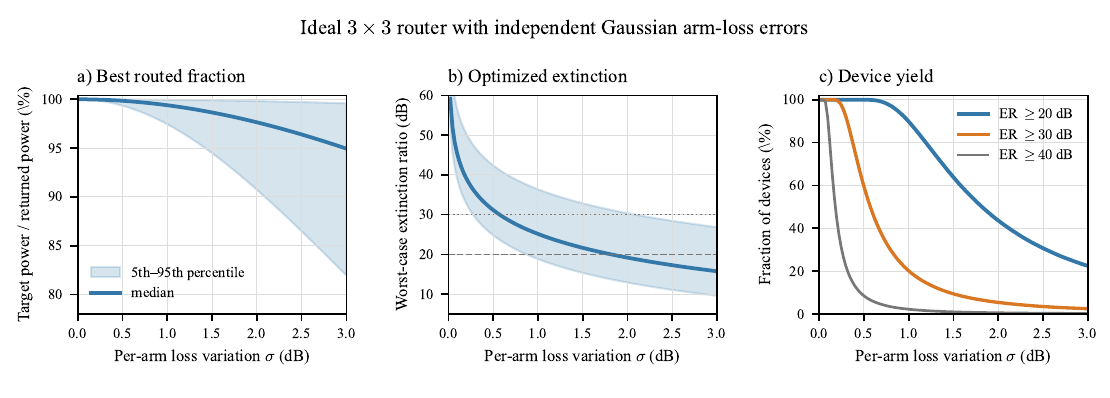}
  \caption{\textbf{Ideal routing sensitivity to independent arm-loss
  imbalance.} Monte Carlo calculation for an ideal $3\times3$ Fourier MMI
  with independently Gaussian-distributed arm losses. The common loss is
  removed from each realization, and the reflector phases are optimized for
  the target output. \textbf{a}, Maximum fraction of returned power routed to
  the target. \textbf{b}, Optimized worst-case extinction ratio. \textbf{c},
  Fraction of realizations exceeding extinction-ratio thresholds of 20, 30
  and 40~dB. Lines show medians and shaded bands show the 5th--95th
  percentiles over $3\times10^5$ realizations at each $\sigma$. This is a
  sensitivity analysis, not a measured loss distribution or fabrication-yield
  estimate.}
  \label{fig:app-random-arm-loss}
\end{figure}

To isolate the effect of differential loss, consider an ideal $3\times3$
Fourier MMI with return coefficients $r_j$ and returned arm-power fractions
$R_j=|r_j|^2$. For input power $P_{\rm in}$, the MMI sends $P_{\rm in}/3$
into each arm, so arm $j$ returns $P_{\rm in}R_j/3$. The lossless MMI
preserves their total power on the return pass, giving
$\sum_k P_k=(P_{\rm in}/3)\sum_jR_j$. Let $a_j=\sqrt{R_j}$ and
$\omega=e^{2\pi i/3}$. After choosing the reflector phases so that the
returned fields add at the target, and relabeling it as output 0, the three
output fields (up to a common phase, in units of $\sqrt{P_{\rm in}}$) are
\begin{align*}
 E_0&=\frac{a_1+a_2+a_3}{3},\\
 E_1&=\frac{a_1+\omega a_2+\omega^2a_3}{3},\\
 E_2&=\frac{a_1+\omega^2a_2+\omega a_3}{3}=E_1^*.
\end{align*}
Because the $a_j$ are real, the two unwanted fields are complex conjugates
and therefore have equal power. The power at each output is then
\begin{align*}
 P_{\rm target}&=\frac{P_{\rm in}}{9}\left(\sum_j\sqrt{R_j}\right)^2, &
 P_{\rm off,1}=P_{\rm off,2}
 &=\frac{1}{2}\left(\frac{P_{\rm in}}{3}\sum_jR_j-P_{\rm target}\right).
\end{align*}
The maximum fraction of returned output power routed to the target and the
corresponding target-to-strongest-unwanted extinction ratio are therefore
\begin{align}
 \eta_{\max}&=\frac{P_{\rm target}}{\sum_kP_k}
 =\frac{\left(\sum_j\sqrt{R_j}\right)^2}{3\sum_jR_j},\notag\\
 \mathrm{ER}_{\max}
 &=10\log_{10}\!\left[\frac{P_{\rm target}}
 {\max(P_{\rm off,1},P_{\rm off,2})}\right]
 =10\log_{10}\!\left(\frac{2\eta_{\max}}{1-\eta_{\max}}\right).
\end{align}
Here $\eta_{\max}=P_{\rm target}/\sum_kP_k$ measures routing of the power that
returns, rather than transmission relative to the input. The factor of two in
$\mathrm{ER}_{\max}$ follows because each unwanted port receives half of the
remaining power. The extinction-ratio definition has the same
target-to-strongest-unwanted form as \eqref{ER}, but uses ideal optical powers
here rather than the independently normalized outputs of the main-text
measurement.
Uniform loss therefore increases insertion loss but does not change the
maximum extinction ratio, whereas unequal arm losses prevent exact
cancellation at the unwanted outputs. Supplementary Fig.~\ref{fig:app-random-arm-loss}
quantifies this sensitivity for independent Gaussian arm-loss variations: at
$\sigma=0.5$ and 1~dB, the median routed fractions are 99.85\% and 99.39\%,
and the median extinction ratios are 31.2 and 25.2~dB, respectively.

\subsection{Arm-phase and cavity-parameter effects on routing performance}

In the fitted optical model, adjusting the static arm phases makes routing
less sensitive to voltage errors. Optimizing those phases together with the
routing voltages places all three states within a 2.46-V range on each
electrode. Higher cavity quality factor reduces the required drive but narrows
the bandwidth, while stronger overcoupling reduces cavity loss.

The fabricated device includes phase shifters in the three paths between the
MMI arms and the one-sided cavities. To investigate how their operating points
affect routing, we retained the fitted cavity and MMI parameters while varying
the two independent relative round-trip arm phases. This optical-only
sensitivity model excludes the empirical coherent detector backgrounds and
detector gain and offset calibration used in the Fig.~\ref{fig:switching} fit.
For each phase pair, we swept the three cavity voltages independently over the
$\pm5$~V cube and optimized each route separately. We recorded both the largest
extinction ratio available for each route and the fraction of sampled
voltage-cube points above a specified threshold. The reported region fraction
is the smallest of the three route-specific fractions and therefore requires
useful regions for every output port. It counts all qualifying sampled points
regardless of connectivity and is not the voltage tolerance around a selected
routing bias.

\begin{figure}[ht]
  \centering
  \includegraphics[width=0.96\textwidth]{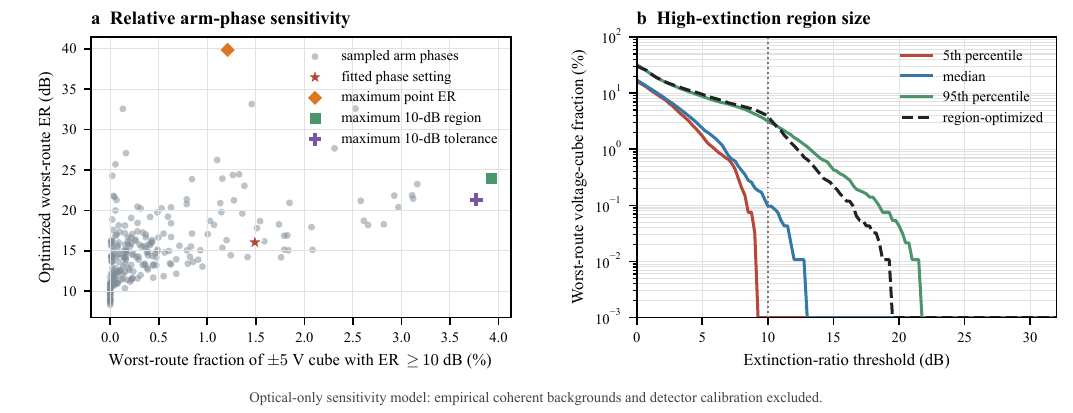}
  \caption{\textbf{Optical-only sensitivity to operating-point phases.}
  Fitted cavity and MMI parameters are retained, while empirical coherent
  detector backgrounds and detector calibration are omitted. The extinctions
  shown are not Fig.~\ref{fig:switching} model predictions. \textbf{a},
  Optimized worst-route extinction ratio and fraction of
  the $\pm5$~V voltage cube above 10~dB for 240 uniformly sampled pairs of
  relative round-trip arm phases (gray). The fitted phase setting, the phase
  setting maximizing the point extinction ratio, the setting maximizing
  the 10-dB region volume, and the setting optimizing local 10-dB voltage-error
  tolerance are highlighted. \textbf{b}, Worst-route
  voltage-cube fraction as a function of extinction-ratio threshold for
  representative 5th-, 50th-, and 95th-percentile phase settings selected by
  optimized worst-route extinction, together with the region-optimized
  setting, which is optimized specifically for the fraction above 10~dB.
  The percentile labels identify representative sampled settings rather than
  uncertainty bounds. These phase sweeps are a sensitivity analysis, not a
  measured phase distribution or a fabrication-yield estimate.}
  \label{fig:app-arm-phase-robustness}
\end{figure}

Supplementary Fig.~\ref{fig:app-arm-phase-robustness}a shows 240 uniformly
sampled operating-point phase pairs. The 5th, median, and 95th percentiles of the
optimized worst-route extinction ratio are 9.9, 14.6, and 23.3~dB,
respectively. Independently, the 5th, median, and 95th percentiles of the
voltage-cube fraction above 10~dB are 0.00\%, 0.21\%, and 2.07\%. At the fitted
operating-point phases, the optical-only model gives an optimized worst-route
extinction of 16.0~dB and a 10-dB region fraction of 1.49\%. This 16.0-dB
value is the smallest of three separately optimized route maxima, not the
calibrated Fig.~\ref{fig:switching} prediction at its displayed routing states.
Optimizing solely for the highest extinction point
produces a model value of 39.9~dB, but this solution relies on a narrow
coherent cancellation. A more robust optimization that maximizes the volume
of the 10-dB regions gives a 24.0-dB worst-route optimum and a 3.93\%
worst-route region fraction. These optical-only sensitivity values are distinct
from the fully calibrated model and measured routing states in
Fig.~\ref{fig:switching}.

To quantify local voltage-error tolerance, we find for each route the largest
axis-aligned cube centered on a route-specific voltage bias and contained in
the $\pm5$~V domain for which the modeled extinction remains at least 10~dB
throughout. We use one shared pair of operating-point phases across the three
routes and report the smallest of their cube half-widths. This half-width is
the allowed independent error on each electrode, including simultaneous
errors on all three. Multistart numerical optimization gives approximately
$\pm0.70$~V per electrode at the fitted phases and $\pm1.13$~V when the
shared phases are optimized directly for this tolerance. At the latter
setting, the optimized worst-route peak is 21.3~dB and the worst-route 10-dB
region fraction is 3.77\%. The cube interiors were checked numerically; these
values are model estimates, not certified global optima or measurements of
voltage-source stability. The region fraction and local tolerance thus
describe different aspects of robustness.

For Supplementary Fig.~\ref{fig:app-arm-phase-robustness}b, we select the
actual sampled phase pairs whose optimized worst-route extinction ratios are
closest to the 5th, 50th, and 95th percentiles. These labels therefore identify
three representative phase settings; they are not uncertainty bounds on the
curves. The region-optimized setting is selected using a different objective:
the voltage-cube fraction above 10~dB. It gives the largest fraction at that
design threshold, marked by the vertical dotted line, but need not remain
above the 95th-percentile representative at higher thresholds.

\begin{figure}[ht]
  \centering
  \includegraphics[width=0.99\textwidth]{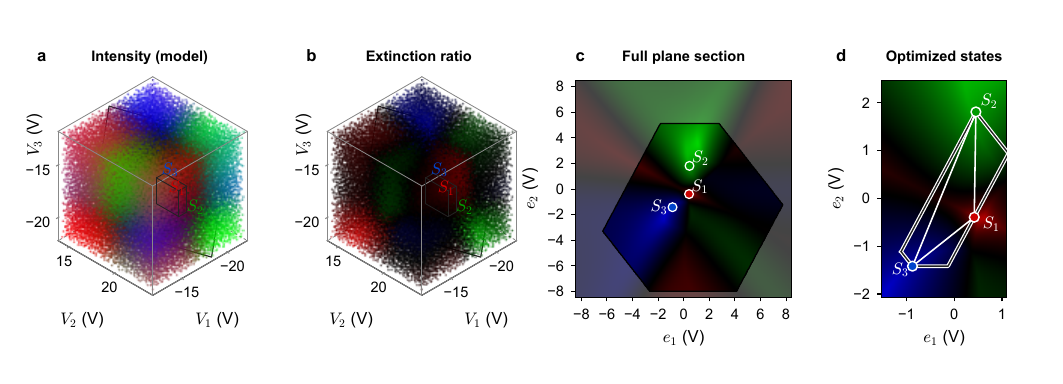}
  \caption{\textbf{Voltage-space geometry of the jointly optimized routing
  states.} The fitted reciprocal model is evaluated over the complete
  fitted-baseline $\pm5$~V cube using one shared optimized pair of relative
  round-trip arm phases. Red, green and blue correspond to routing toward
  $P_1$, $P_2$ and $P_3$, respectively. \textbf{a}, Model output powers,
  independently normalized per output over the cube and RGB encoded; color
  brightness is not an absolute-power calibration. \textbf{b}, The same
  worst-case extinction ratio $\mathrm{ER}_k$ defined in the main text,
  evaluated from the model powers and clipped at zero for display, then
  normalized over the
  cube. This winner-only encoding is primary-colored by construction.
  \textbf{c}, Extinction-ratio map in the plane through the three
  optimized states. The black hexagon is the intersection of this plane with
  the full $\pm5$~V cube; the gray exterior is outside the evaluated control
  range. The plotted $(e_1,e_2)$ coordinates describe positions within this
  plane, not individual electrode voltages; their ranges can therefore exceed
  $\pm5$~V while each electrode remains within its $\pm5$-V range.
  \textbf{d}, Enlargement of the optimized states. The white/black
  polygon is the plane section of their smallest axis-aligned bounding cube,
  whose side is $\Delta V_{\max}=2.46$~V. Lines connecting the states are
  guides to the eye. The phase offsets and voltage tuples are numerical model
  optima and are not independent measurements.}
  \label{fig:app-joint-drive-geometry}
\end{figure}

We jointly optimized the three routing voltage settings and one shared pair of
operating-point phases, requiring at least 15~dB extinction for each route.
The objective was the side length of the smallest axis-aligned cube containing
all three voltage settings, equivalently the largest inter-state voltage span
on any electrode. The smallest feasible value found by the numerical search
was 2.46~V.
Taking arm~3 as the reference, the additional round-trip operating-point phase
offsets are 0.605~rad for arm~2 and 0.925~rad for arm~1. We then reevaluated
the reciprocal model over the full fitted-baseline $\pm5$~V cube at this phase
setting to visualize the solution
(Supplementary Fig.~\ref{fig:app-joint-drive-geometry}). These are static
relative arm phases, not the global phase of the illuminated input. This
inter-state span excludes
the common DC voltages used to align the cavity resonances.

The intensity cube in Supplementary
Fig.~\ref{fig:app-joint-drive-geometry}a encodes the three model output powers
as red, green and blue after separately normalizing each output over the full
cube. Consequently, its brightness and hue do not represent absolute
throughput and should not be compared quantitatively with the measured cube
in Fig.~\ref{fig:switching}. Panel b instead evaluates the same worst-case
extinction ratio $\mathrm{ER}_k$ defined in the main text, here using the model
powers $P_k$, and displays only its positive part,
$\max\{\mathrm{ER}_k,0\}$. At most one route can have a positive extinction
ratio at a given voltage tuple, so this representation
is intentionally less color-mixed than the intensity representation. The
optimized phases relocate the interference lobes and bring the three selected
high-extinction regions close together, but the brighter or purer colors must
not be interpreted as a phase-induced increase in absolute optical power.

The three selected voltage settings do not lie on a single straight line, so
they define a plane in the three-voltage space. Its intersection with the full
voltage cube is shown in Supplementary
Fig.~\ref{fig:app-joint-drive-geometry}c, and panel d enlarges the optimized
triangle and the cross-section of its 2.46-V bounding cube. The optimization
did not impose planarity; the plane follows from the selected settings.
The construction therefore provides a convenient two-coordinate description
of transitions among these three states, rather than evidence that the full
routing landscape is intrinsically two-dimensional or that only two physical
electrodes are required.

\begin{figure}[!t]
  \centering
  \includegraphics[width=0.96\textwidth]{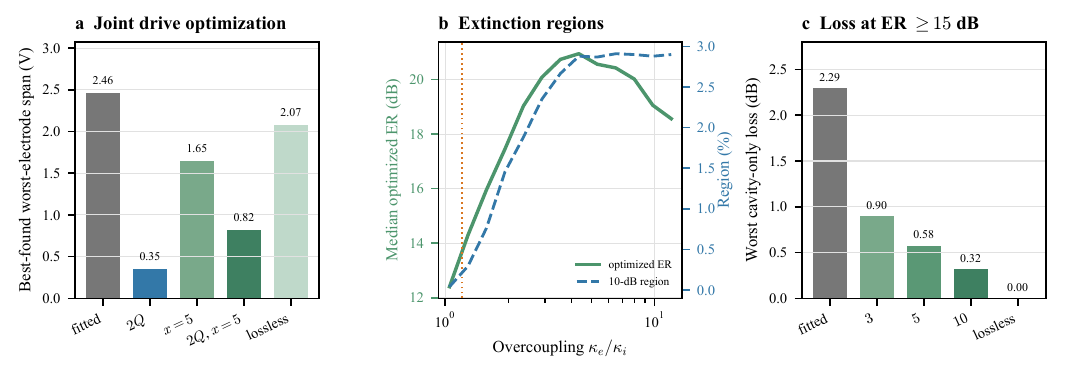}
  \caption{\textbf{Model-derived cavity-design trade-offs.} Here
  $x=\kappa_e/\kappa_i$ is the external-to-internal coupling ratio.
  \textbf{a}, Best-found maximum electrode span across the three routing
  states, obtained by joint numerical optimization subject to
  $\mathrm{ER}\geq15$~dB for every route. One shared pair of static arm
  phases and all three voltage tuples are optimized for the fitted cavities,
  doubled-loaded $Q$, common overcoupling ratio $x=5$, their combination, and
  the lossless-cavity limit. \textbf{b}, Median optimized worst-route
  extinction ratio (green) and median fraction of the $\pm5$~V cube above
  10~dB (blue dashed) versus the common overcoupling ratio, evaluated over
  random relative arm phases at fixed loaded $Q$. The vertical dotted line
  marks the geometric mean of the three fitted coupling ratios. \textbf{c},
  Worst cavity-only loss after maximizing cavity survival subject to
  $\mathrm{ER}\geq15$~dB for all routes and optimizing one shared pair of
  static arm phases. The first bar uses the fitted heterogeneous cavities;
  the following bars use the indicated common $\kappa_e/\kappa_i$ values and
  the lossless single-sided limit.}
  \label{fig:app-cavity-design-tradeoffs}
\end{figure}

Next, we varied the loaded quality factor and overcoupling in controlled sweeps
(Supplementary Fig.~\ref{fig:app-cavity-design-tradeoffs}). Scaling the loaded
quality factors while holding the tuning efficiency (in GHz\,V$^{-1}$)
and $\kappa_e/\kappa_i$ fixed isolates the linewidth effect. The voltage
required to tune through one linewidth scales as
\begin{align}
V_{\rm linewidth}=\frac{\nu_0}{Q_{\rm L}\eta},
\end{align}
where $\eta$ is the tuning efficiency. Consistent with this
scaling, we numerically optimized one shared pair of static operating-point phases and the
three routing voltage tuples while requiring every target/off-target ratio to
exceed 15~dB. For three routing states, we define the largest inter-state
electrode-voltage excursion as
\begin{align}
\Delta V_{\max}=\max_j\left[\max_s V_j^{(s)}-\min_s V_j^{(s)}\right],
\end{align}
where $s$ indexes the three routing states. The best solution found with
the fitted cavity parameters gives
$\Delta V_{\max}=2.46$~V. For comparison, the same constrained span is
0.42, 0.48, and 0.54~V at thresholds of 10, 11, and 12~dB, respectively.
Doubling all loaded quality factors reduces the 15-dB value to
0.35~V, while setting all three coupling ratios to $\kappa_e/\kappa_i=5$ at
the fitted loaded $Q$ gives 1.65~V. Combining both changes gives 0.82~V.
Higher $Q_{\rm L}$ therefore reduces the required drive, but it also reduces
the cavity-lifetime-limited bandwidth as $1/Q_{\rm L}$ and narrows the useful
voltage regions, increasing sensitivity to voltage noise and drift.

The overcoupling sweep holds $Q_{\rm L}$ fixed and assigns a common ratio
$x=\kappa_e/\kappa_i$ to the three cavities. At resonance, an overcoupled
single-sided cavity has power reflectance
\begin{align}
|r(0)|^2=\frac{(x-1)^2}{(x+1)^2}.
\end{align}
Increasing $x$ therefore suppresses the dissipative reflection dip and makes
the cavity more nearly an all-pass phase shifter. In the random-phase
ensemble, increasing $x$ between the sampled sweep points 1.58 and 2.90 raises the median optimized
worst-route extinction ratio from 16.0 to 20.1~dB and expands the median
10-dB region fraction from 0.76\% to 2.35\%. The improvement largely
saturates near $x\sim5$ in this fitted model because the remaining MMI
amplitude imbalance and finite control range become limiting.
The strictly lossless-cavity limit gives a best-found span of 2.07~V in the
same constrained optimization. The nonmonotonic ordering across the five
cases reflects both linewidth narrowing and changes in the cavity amplitude
responses. In particular, the heterogeneous fitted responses can partially
compensate the fitted MMI imbalance, so removing absorption does not
necessarily minimize the voltage span for this specific nonideal model. This
does not imply that absorption is generally beneficial: for a balanced MMI,
the lossless all-pass limit remains preferable.

Maximizing extinction alone can place a cavity close to its absorption dip,
so we separately maximized the worst-route cavity survival subject to a
minimum extinction ratio of 15~dB. After optimizing one shared pair of static
arm phases, the resulting worst cavity-only loss is 2.29~dB for the fitted
heterogeneous cavities, 0.90~dB for $x=3$, 0.58~dB for $x=5$, and 0.32~dB
for $x=10$, approaching zero in the lossless single-sided limit
(Supplementary Fig.~\ref{fig:app-cavity-design-tradeoffs}c). These are
model-derived cavity losses evaluated before the return through the MMI; they
are not total switch or fiber-to-fiber insertion losses. The $Q_{\rm L}$ and
overcoupling sweeps are varied independently to expose the trade-offs. In a
physical redesign, changing intrinsic or external coupling can change both
quantities simultaneously.

Taken together, the model identifies separate design levers: static arm-phase
tuning can improve voltage tolerance, higher loaded quality factor reduces the
required voltage span at the cost of bandwidth and narrower useful voltage
regions, and stronger overcoupling reduces cavity-only loss while improving
extinction until the remaining MMI imbalance and finite control range become
limiting. We optimized voltage span and cavity loss separately, so their
reported best values should not be attributed to the same operating settings.

\FloatBarrier
\section{Supplementary Note 5: Electro-optic bandwidth extraction}

The electro-optic transfer-function measurement contains 201 frequency points
from 50~MHz to 10~GHz. The laser output was set to 13~dBm. We correct the VNA
transmission trace with a separately measured
through-reference trace by subtracting the reference magnitude in decibels at
each device frequency, interpolating the reference where needed. This calibration
removes the cable response only; the probe, contact pads and photodetector
remain part of the measured response. Denoting the
corrected values by $d_i$, we fit all 201 points by nonlinear least squares to
the Butterworth-like low-pass family~\cite{Butterworth1930}
\begin{align}
d_n^{\mathrm{fit}}(f)=b_n-10\log_{10}\left[1+
\left(\frac{f}{f_{\mathrm{3dB}}}\right)^{2n}\right].
\eqlab{bandwidth-low-pass-family}
\end{align}
For each fixed order $n$, the two fitted parameters are the low-frequency level
$b_n$ (the model value at $f=0$) and the frequency $f_{\mathrm{3dB}}$ at which
the model falls 3.01~dB below that level. We compare $n=1,2,3,4$ on the same
samples; the model comparison below selects $n=3$. For the normalized plots,
we subtract this third-order fit's low-frequency level from both the measured
trace and the model curves, setting that level to 0~dB. The fit gives a 3-dB
bandwidth of 6.20~GHz. The measured trace itself crosses $-3$~dB at 6.07~GHz,
found by interpolating between the two surrounding measurement points.
Because every fixed-order candidate has the same two fitted parameters, we
compare their root-mean-square residual errors on the same 201 points. We also
allow $n$ to vary as a third parameter.

\begin{table}[ht]
\centering
\caption{Comparison of fits to the through-corrected electro-optic magnitude
response. The fixed-order model with the lowest RMSE is highlighted in bold.}
\label{tab:bandwidth-model-comparison}
\begin{tabular}{lrrrr}
\toprule
Model & Order & $f_{\mathrm{3dB}}$ (GHz) & RMSE (dB) & $R^2$ \\
\midrule
One-pole & 1 & 3.628 & 2.082 & 0.734 \\
Low-pass & 2 & 5.354 & 1.022 & 0.936 \\
\textbf{Low-pass} & \textbf{3} & \textbf{6.20} & \textbf{0.543} &
\textbf{0.982} \\
Low-pass & 4 & 6.765 & 0.818 & 0.959 \\
Free-order low-pass & 3.028 & 6.218 & 0.542 & 0.982 \\
\bottomrule
\end{tabular}
\end{table}

\begin{figure}[ht]
  \centering
  \includegraphics[width=0.88\textwidth]{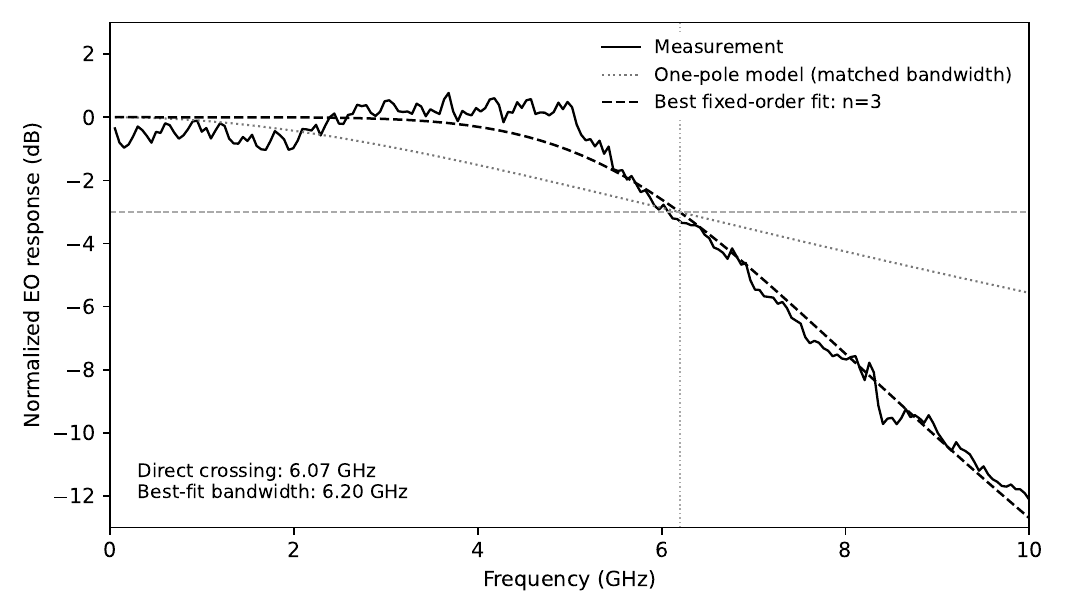}
  \caption{\textbf{Electro-optic bandwidth-model comparison.} The black curve
  is the response measured at a 13-dBm laser-output setting after through
  correction and subtraction of the fitted third-order low-frequency offset.
  The dotted gray curve is an order-one comparison constrained
  to the same low-frequency offset and 3-dB bandwidth as the best fixed-order
  model; it is not independently fitted. The dashed black curve is the best
  fixed-order model ($n=3$). This construction emphasizes the models' different
  roll-offs after their common corner. The same fitted offset is subtracted
  from the measured trace and both model curves for display. The model curves
  cross $-3.01$~dB at $f_{\mathrm{3dB}}=6.20$~GHz, marked by the horizontal
  and vertical guide lines. The directly interpolated measurement crossing
  at $-3$~dB is 6.07~GHz.}
  \label{fig:bandwidth-fit-comparison}
\end{figure}

The third-order model has the lowest fixed-order RMSE, with
$f_{\mathrm{3dB}}=6.20$~GHz, a root-mean-square residual of 0.543~dB, and
$R^2=0.982$ (Supplementary Fig.~\ref{fig:bandwidth-fit-comparison}). The
free-order fit gives $n=3.028$, close to three, with nearly the same RMSE
(0.542~dB). This supports the third-order empirical fit, from which we report
the bandwidth as 6.20~GHz. The covariance matrix gives a formal standard
error of 0.025~GHz for the fixed third-order bandwidth, but this does not
include uncertainty from the through-reference measurement, model selection,
or correlated ripple and is not quoted as a total measurement uncertainty.

Finally, the magnitude response for $n=1$ in
\eqref{bandwidth-low-pass-family} is both the magnitude of a single-pole
RC transfer function and the square root of a Lorentzian power response.
Consequently, magnitude-only $S_{21}$ data cannot distinguish those two
physical labels. The empirical preference for a third-order roll-off does not
by itself identify the physical processes that limit the measured response.
The analysis code is included in the repository listed under Code availability.

\section{Supplementary Note 6: Single-channel high-extinction routing}
To distinguish the attainable contrast of an individually optimized routing
state from the stability required to hold all three states in a switching
sequence, we performed a targeted voltage-space acquisition around the
high-contrast operating regions. The three detector traces were delay aligned
using the same calibration as Fig.~\ref{fig:switching} and independently
min--max normalized over the full acquisition, consistent with the extinction
ratio definition used for Fig.~\ref{fig:switching}c.

\begin{figure*}[!ht]
  \centering
  \includegraphics[width=\textwidth]{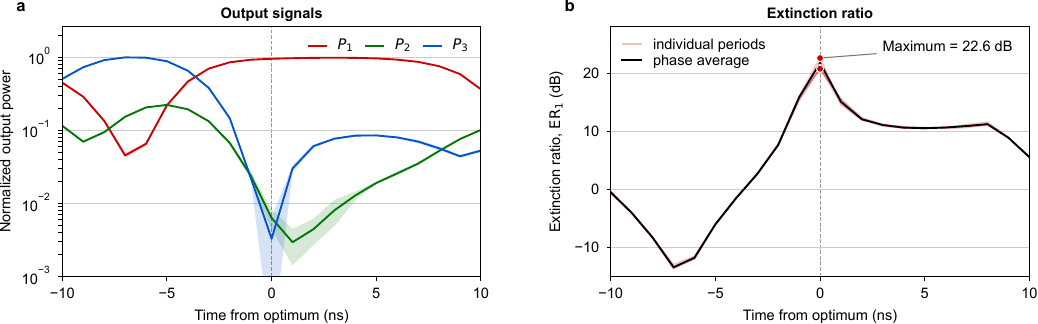}
  \caption{\textbf{Targeted single-channel high-extinction routing.}
  \textbf{a}, Normalized measured detector signals in a 20-ns window centered on the
  highest-extinction sample. Red, green and blue denote $P_1$, $P_2$ and $P_3$,
  respectively. Solid curves show the mean over four repeated
  25-$\upmu$s periods and shaded bands show one sample standard deviation; the
  dashed vertical line marks the optimum waveform phase.
  \textbf{b}, Extinction ratio for routing to $P_1$. Pale red curves show the
  four individual periods, the black curve is evaluated from the
  phase-averaged detector signals, and red markers give the four
  individual-period values at the optimum. The maximum instantaneous
  extinction is 22.6~dB.}
  \label{fig:app-single-channel-extinction}
\end{figure*}

For routing from input port 1 to output port 1, the maximum instantaneous
extinction ratio was $\mathrm{ER}_1=22.63$~dB
(Supplementary Fig.~\ref{fig:app-single-channel-extinction}). The periodic
drive revisited the same waveform phase once in each 25~$\upmu$s period. The
four visits gave 20.76, 20.98, 22.63, and 20.78~dB, corresponding to
$21.29\pm0.90$~dB (mean $\pm$ sample standard deviation). Averaging the three
detector signals across the four visits before evaluating their ratio gives
21.79~dB.

At the 22.63-dB sample, the normalized detector signals were
$(\widetilde P_1,\widetilde P_2,\widetilde P_3)
=(0.9542,0.00520,0.00237)$. The same-phase one-standard-deviation
repeatabilities of the two off-target channels were 0.00197 and 0.00319,
respectively. Thus, the $\widetilde P_3$ residual was below its measured
repeatability and the $\widetilde P_2$ residual was only 2.6 times its
repeatability. Using the larger off-target repeatability as the denominator
gives an approximate one-standard-deviation resolution scale of 24.8~dB.
The measured maximum is therefore close to the resolution of the present
setup. Because the detector channels are normalized independently and the
capture contains no separate dark/background calibration, this measurement
does not determine absolute insertion loss or absolute optical crosstalk.

\FloatBarrier
\clearpage
\section{Supplementary Note 7: Four-port reachable-region extension}
\label{sec:four-port-reachable-regions}

Supplementary Note~2 shows that every ideal equal-split unitary supports
perfect routing between any selected input and output, but that the complete
reachable power region need not be universal for $N>3$. Here, we describe this
distinction geometrically for three $4\times4$ examples: the real Walsh
matrix, the four-point DFT, and an intermediate equal-split unitary. We assume
lossless propagation and independently controllable reflector phases with
$|r_j|=1$. The plots describe power distributions for one excited input, not
simultaneous arbitrary routing of several inputs.

\subsection{A continuous family of equal-split four-port matrices}

Consider the normalized family
\begin{align}
 \mathbf{M}(\alpha)=\frac12
 \begin{pmatrix}
  1&1&1&1\\
  1&e^{i\alpha}&-1&-e^{i\alpha}\\
  1&-1&1&-1\\
  1&-e^{i\alpha}&-1&e^{i\alpha}
 \end{pmatrix}.
 \eqlab{four port continuous family}
\end{align}
Every member is unitary and all entries have modulus $1/2$. At $\alpha=0$
it equals $\mathbf{W}_4$, and at $\alpha=\pi/2$ it equals $\mathbf{F}_4$
in \eqref{four port Walsh and Fourier}. Thus the intermediate example
$\alpha=\pi/4$ is not a different splitting-ratio design: its single-pass
power splitting remains exactly equal, while its internal phases differ.

The classification of order-four complex Hadamard matrices shows that every
equal-split $4\times4$ unitary belongs to this one-parameter family up to row
and column phases and permutations~\cite{Tadej2006}. These operations change
the reachable power set only through a fixed output-port relabeling.
Complex conjugation also preserves the power set, since the controls can be
conjugated together with the matrix. In particular, $\alpha\mapsto\alpha+\pi$
exchanges rows two and four, while $\alpha\mapsto-\alpha$ conjugates the
matrix. For comparing power-region shapes, it therefore suffices to use
$0\le\alpha\le\pi/2$. There is a continuous range of possibilities, rather
than a finite list of three matrices. The three examples below illustrate
that range.

Theorem~\ref{thm:fixed-phase-invariance} allows any fixed reciprocal arm
phases $\mathbf{D}$ to be absorbed into the controllable reflector phases.
For the zero-based matrix index $n=0$ (physical input port 1),
$\mathbf{M}(\alpha)\mathbf{e}_0=(1,1,1,1)^{\mathsf{T}}/2$, so the output field
can be written
\begin{align}
 \mathbf{z}=\frac12\mathbf{M}(\alpha)^{\mathsf{T}}\mathbf{u},
 \qquad |u_j|=1,
 \qquad P_k=|z_{k-1}|^2,\quad k=1,\ldots,4.
 \eqlab{four port effective fields}
\end{align}
Here $\mathbf{u}$ contains the effective reflector phases after absorbing
$\mathbf{D}$. The same power set is obtained for every other input: the
entries of any column of an equal-split unitary have modulus $1/2$, and their
phases can likewise be absorbed into $\mathbf{u}$. Therefore the figures
apply to every input and to arbitrary fixed reciprocal post-MMI phases.

\subsection{Coordinates and cross-section construction}

Unitarity gives $P_1+P_2+P_3+P_4=1$. Consequently the four nonnegative
powers occupy a three-dimensional tetrahedron, the probability simplex. Its
four vertices are perfect single-output routing states. To inspect the
interior without overlap between front and back surfaces, define
\begin{align}
 s=P_1+P_3,\qquad x=P_1-P_3,\qquad y=P_2-P_4.
 \eqlab{four port section coordinates}
\end{align}
The height $s$ sets the total power in outputs 1 and 3; the other pair carries
$1-s$. Within a constant-$s$ plane, $x$ and $y$ specify the imbalances inside
the two pairs. These coordinates retain all four powers, since
\begin{align}
 (P_1,P_2,P_3,P_4)
 =\frac12(s+x,\,1-s+y,\,s-x,\,1-s-y).
 \eqlab{four port powers from section coordinates}
\end{align}
Positivity restricts a full simplex section to the rectangle
\begin{align}
 0\le s\le1,\qquad |x|\le s,\qquad |y|\le1-s.
 \eqlab{four port simplex section bounds}
\end{align}
In these linear coordinates the tetrahedron has vertices
$(1,0,1)$, $(0,1,0)$, $(-1,0,1)$ and $(0,-1,0)$, with coordinates ordered
as $(x,y,s)$. Its nonregular appearance in the sliced three-dimensional plot
is only a consequence of the coordinate choice.

Removing one common, physically irrelevant phase, every control vector can
be parametrized as
\begin{align}
 \mathbf{u}=
 (e^{i\phi},\,e^{i(\chi+\psi)},\,e^{-i\phi},\,e^{i(\chi-\psi)})^{\mathsf{T}}.
 \eqlab{four port three phase parametrization}
\end{align}
Substitution into \eqref{four port effective fields} gives the power map
\begin{align}
 s&=\frac{\cos^2\phi+\cos^2\psi}{2},\notag\\
 x&=\cos\phi\cos\psi\cos\chi,\notag\\
 y&=\sin\phi\sin\psi\cos(\chi+\alpha).
 \eqlab{four port explicit power map}
\end{align}
To construct a section at fixed $s$, set
$k=\cos^2\phi\cos^2\psi$ and $\ell=1-2s+k$. The allowed interval follows
from fixing the sum of $\cos^2\phi$ and $\cos^2\psi$:
\begin{align}
 \max(0,2s-1)\le k\le s^2,\qquad
 \ell=\sin^2\phi\sin^2\psi.
 \eqlab{four port section k interval}
\end{align}
After absorbing signs into $\chi$, the section is the union of the curves
\begin{align}
 x=\sqrt{k}\cos\chi,\qquad
 y=\pm\sqrt{\ell}\cos(\chi+\alpha).
 \eqlab{four port section ellipse curves}
\end{align}
For $0<\alpha\le\pi/2$ and $k\ell>0$, each curve is an ellipse. The complete
section equals the union of their filled elliptical disks. To see why filling
does not introduce extra states, decrease $k$ continuously from any disk
containing the point to its lower endpoint. At that endpoint the ellipse
degenerates to an already reachable line segment. A point either lies on
that segment or crosses an ellipse boundary at an intermediate $k$; the
boundary is attained by \eqref{four port section ellipse curves}.
This construction preserves nonconvex sections and is not a convex-hull
approximation. The intermediate sections in Supplementary Figs.~\ref{fig:four-port-region-sliced3d} and~\ref{fig:four-port-region-sections} are evaluated by a
numerically converged envelope over the interval
\eqref{four port section k interval}; the Walsh and DFT sections use the
analytic expressions below.

\begin{figure}[!t]
 \centering
 \makebox[\textwidth][c]{\includegraphics[width=180mm]{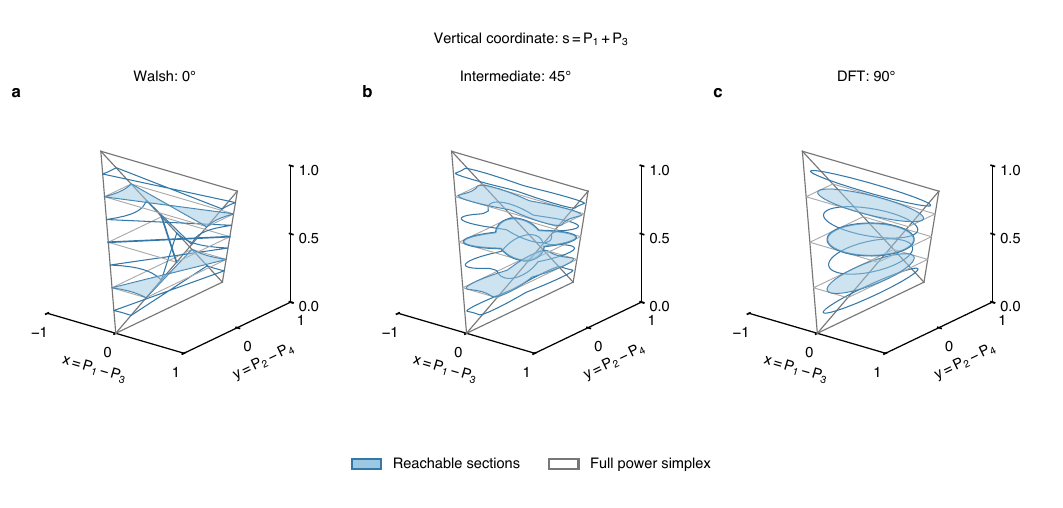}}
 \caption{\textbf{Sliced three-dimensional views of the ideal four-port regions.}
 \textbf{a}, Walsh. \textbf{b}, Intermediate matrix, $\alpha=\pi/4$.
 \textbf{c}, DFT. All panels use the same coordinates
 $x=P_1-P_3$, $y=P_2-P_4$ and $s=P_1+P_3$, and the same view.
 Gray edges outline the full power tetrahedron. Blue curves are reachable
 section boundaries at $s=1/8,1/4,3/8,1/2,5/8,3/4,7/8$; sections at
 $s=1/4,1/2,3/4$ are shaded where they have nonzero area. Their gray
 rectangular outlines show the corresponding full simplex sections.
 The sampled heights aid visualization only: all intermediate heights are
 included in the actual reachable region. Each tetrahedron vertex is a
 perfect routing state, while the different section shapes expose the
 nonequivalence of the interior reachable regions.}
 \label{fig:four-port-region-sliced3d}
\end{figure}

\subsection{Walsh, intermediate and DFT regions}

The DFT corresponds to $\alpha=\pi/2$. The elliptical disks grow with $k$,
so the largest disk, $k=s^2$ and $\ell=(1-s)^2$, gives the complete section:
\begin{align}
 \frac{x^2}{s^2}+\frac{y^2}{(1-s)^2}\le1,
 \qquad 0<s<1.
 \eqlab{four port DFT full region}
\end{align}
Thus every nondegenerate constant-$s$ section is a filled ellipse. At
$s=1/2$ it is the circle $x^2+y^2\le1/4$ inscribed in the square simplex
section. At $s=0$ and $s=1$ the sections are the complete opposite simplex
edges joining outputs $(2,4)$ and $(1,3)$, respectively.

For Walsh, $\alpha=0$, the section curves instead satisfy
$x=\sqrt{k}\cos\chi$ and $y=\pm\sqrt{\ell}\cos\chi$. For $0<s<1/2$,
varying $k$ gives the exact section
\begin{align}
 y^2\ge\frac{(1-s)^2}{s^2}x^2,
 \qquad y^2-x^2\le1-2s,
 \eqlab{four port Walsh lower sections}
\end{align}
together with \eqref{four port simplex section bounds}. Sections at
$1/2<s<1$ follow by exchanging $x$ and $y$ and replacing $s$ by $1-s$.
The central section degenerates to the two segments $y=x$ and $y=-x$
within $[-1/2,1/2]^2$. This zero-area slice is a geometric restriction,
not an absence of reachable states.

For the whole family, the central section has the particularly simple form
\begin{align}
 x^2+y^2-2\cos\alpha\,|xy|
 \le\frac{\sin^2\alpha}{4},
 \qquad |x|,|y|\le\frac12.
 \eqlab{four port central section family}
\end{align}
This follows from \eqref{four port section ellipse curves} with $\ell=k$:
the disks scale with $\sqrt{k}$ and the largest one has $k=1/4$.
At $\alpha=\pi/4$ the section is the union of two crossed elliptical disks,
whose filled union is shown in Supplementary
Fig.~\ref{fig:four-port-region-sections}(e). At the endpoints, the same
expression yields the Walsh diagonals and DFT circle. The maximum $|x|$
on the line $y=0$ is $\tfrac12\sin\alpha$, illustrating that the section
changes continuously with the MMI phase parameter.

\begin{figure}[!t]
 \centering
 \makebox[\textwidth][c]{\includegraphics[width=150mm]{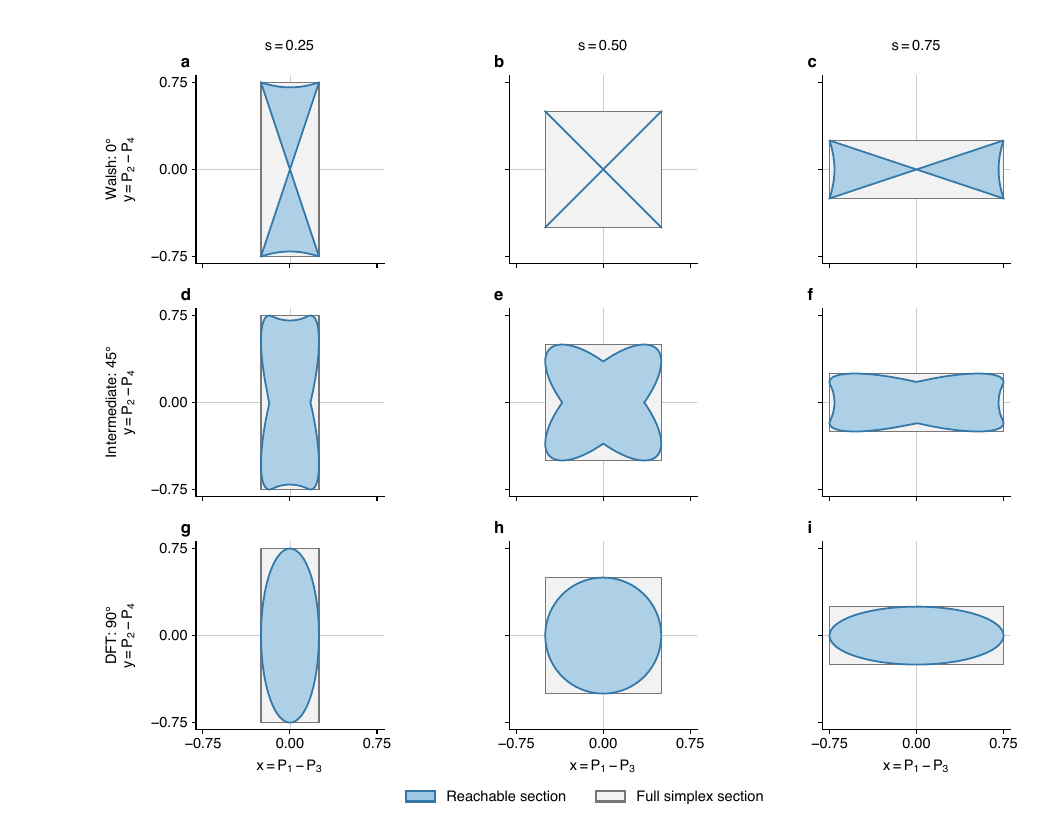}}
 \caption{\textbf{Cross-sections of three ideal four-port reachable power regions.}
 Rows show the Walsh matrix ($\alpha=0$; \textbf{a--c}), the intermediate
 matrix ($\alpha=\pi/4$; \textbf{d--f}), and the DFT
 ($\alpha=\pi/2$; \textbf{g--i}). Columns fix
 $s=P_1+P_3$ at $1/4$, $1/2$, and $3/4$, respectively.
 Blue marks reachable states; the gray rectangle is the full allowed simplex
 section, $|x|\le s$, $|y|\le1-s$. In the central column, the reachable
 section is two diagonal segments for Walsh, a union of two elliptical disks
 for the intermediate matrix, and a circular disk for DFT. The axes give
 normalized power differences, $x=P_1-P_3$ and $y=P_2-P_4$.
 Walsh and DFT curves are analytic; the intermediate noncentral boundaries
 use the converged ellipse envelope described in the text.}
 \label{fig:four-port-region-sections}
\end{figure}
\FloatBarrier

Finally, two different meanings of a ``pair'' must be distinguished.
\emph{Input--output pairwise routing} means sending all power from a
selected input $n$ to a selected output $m$. Every matrix considered here
can do this, by Theorem~\ref{thm:general-pairwise-routing}; all four routing
vertices remain reachable. \emph{Two-output splitting} means distributing
one input between two chosen outputs with zero power in the other two.
Walsh reaches every point of all six simplex edges: choosing the two nonzero
output amplitudes in quadrature makes all inverse-control entries have unit
modulus. For
$0<\alpha\le\pi/2$, only the opposite edges $(1,3)$ and $(2,4)$ have
reachable interiors; the other four edges contain only the routing vertices.
The obstruction is the same phase-matching condition used in the four-port
counterexample of Supplementary Note~2: for the other output pairs, the
relative column phases include both $\pm1$ and $\pm e^{i\alpha}$, so two
nonzero output amplitudes cannot give unit-modulus controls when
$\sin\alpha\ne0$. These edge restrictions do not contradict any-to-any
single-output routing, and fixed post-MMI phases do not change them.

The figure-generation code is included in the repository listed under Code
availability.

\end{document}